\documentclass{ws-ijmpe}

\usepackage{graphicx} 

\begin{document}

\markboth{N. Armesto}{Predictions for the heavy-ion programme at the Large Hadron Collider}

\catchline{}{}{}{}{}

\title{PREDICTIONS FOR THE HEAVY-ION PROGRAMME\\
AT THE LARGE HADRON COLLIDER}

\author{\footnotesize N\'ESTOR ARMESTO}

\address{Departamento de F\'{\i}sica de Part\'{\i}culas and Instituto Galego de F\'{\i}sica de Altas Enerx\'{\i}as,\\
Universidade de Santiago de Compostela, 15706 Santiago de Compostela, Spain\\
nestor@fpaxp1.usc.es}
%
%

\maketitle

\begin{history}
\received{(received date)}
\revised{(revised date)}
\end{history}

\begin{abstract}
I review the main predictions for the heavy-ion programme at the Large Hadron Collider (LHC) at CERN, as available in early April 2009. I begin by remembering the standard claims made in view of the experimental data measured at the Super Proton Synchrotron (SPS) at CERN and at the Relativistic Heavy Ion Collider (RHIC) at the BNL. These claims will be used for later discussion of the new opportunities at the LHC. Next I review the generic, qualitative expectations for the LHC. Then I turn to quantitative predictions: First I analyze observables which characterize directly the medium produced in the collisions - bulk observables or soft probes -: multiplicities, collective flow, hadrochemistry at low transverse momentum, correlations and fluctuations. Second, I move to calibrated probes of the medium i.e. typically those whose expectation in the absence of any medium can be described in Quantum Chromodynamics (QCD) using perturbative techniques (pQCD), usually called hard probes. I discuss particle production at large transverse momentum and jets, heavy-quark and quarkonium production, and photons and dileptons. Finally, after a brief review of pA collisions, I end with a summary and a discussion about the potentiality of the measurements at the LHC - particularly those made during the first run - to further substantiate or, on the contrary, disproof the picture of the medium that has arisen from the confrontation between the SPS and RHIC data, and theoretical models.
\end{abstract}

\section{Introduction}
\label{intro}

The experimental programme for the study of ultra-relativistic heavy-ion collisions started in 1986 at the Super Proton Synchrotron (SPS) at CERN. It accelerated protons and ions (up to Pb), at $p_{lab}\le 158$ GeV per nucleon in the case of Pb\footnote{Natural units $\hbar=c=1$, and $k_B=1$ will be used throughout this manuscript.}. The next step was the Relativistic Heavy Ion Collider (RHIC) at the BNL, which began in 2000, accelerating protons and ions up to AuAu collisions at $\sqrt{s_{NN}}=200$ GeV. Both experimental programmes have allowed for the extraction of important conclusions about the properties of the strongly interacting matter produced in such collisions \cite{Heinz:2000bk,rhic,Back:2004je,Arsene:2004fa,Adams:2005dq}.

The next step in the near future, apart from RHIC upgrades \cite{rhic2} and the energy and collision species scan at the SPS \cite{sps2}, is the heavy-ion programme at the Large Hadron Collider (LHC) at CERN \cite{Jowett:2008hb}. It will accelerate ions as heavy as Pb ($A=208$, $Z=82$), with energies
\begin{equation}
\sqrt{s_{NN}}=2\,\frac{Z}{A}\times 7 \ \ {\rm TeV}\simeq 5.5\ \ {\rm TeV\ \  for\ \  PbPb},
\label{lhcener}
\end{equation}
with a total center-of-mass energy of 1.15 PeV.
The nominal peak luminosity will be ${\cal L}_0=10^{27}$ cm$^{-2}$s$^{-1}$, with $\langle{\cal L}\rangle/{\cal L}_0=0.5$ and a estimated running time $10^6$ s/year\footnote{These numbers are to be compared with those for pp collisions: ${\cal L}_0=10^{34}$ cm$^{-2}$s$^{-1}$, with $\langle{\cal L}\rangle/{\cal L}_0$ closer to 1 and a estimated running time $8\cdot 10^6$ s/year.}. Collisions of other ions and asymmetric collisions like pPb \cite{Accardi:2004be} are possible, the latter with a shift in the center-of-mass rapidity with respect to the rapidity in the laboratory given by
\begin{equation}
\delta y=\frac{1}{2}\ln\frac{Z_1A_2}{Z_2 A_1}
\label{rapshift}
\end{equation}
for $^{A_1}_{Z_1}$A$\,^{A_2}_{Z_2}$B collisions. While the first proton beams circulated along the LHC ring in September 2008 and the first pp collisions are expected for autumn 2009, the first PbPb collisions are only expected for the second half of 2010.

Three out of the four large experiments at the LHC: ALICE, ATLAS and CMS, will measure PbPb collisions \cite{lhc,Alessandro:2006yt,D'Enterria:2007xr,Steinberg:2007nm}. While ALICE is a dedicated experiment to nucleus-nucleus collisions, both ATLAS and CMS will offer detector capabilities complementary to each other and to ALICE. They will provide a wide range of measurements covering all the main relevant observables in heavy-ion collisions. Measurements in pp and nucleus-nucleus collisions at roughly the same unexplored top energy will be, for the first time, performed using the same accelerator and detectors.

\begin{figure}[htb]
\begin{center}
\includegraphics[width=5.5cm]{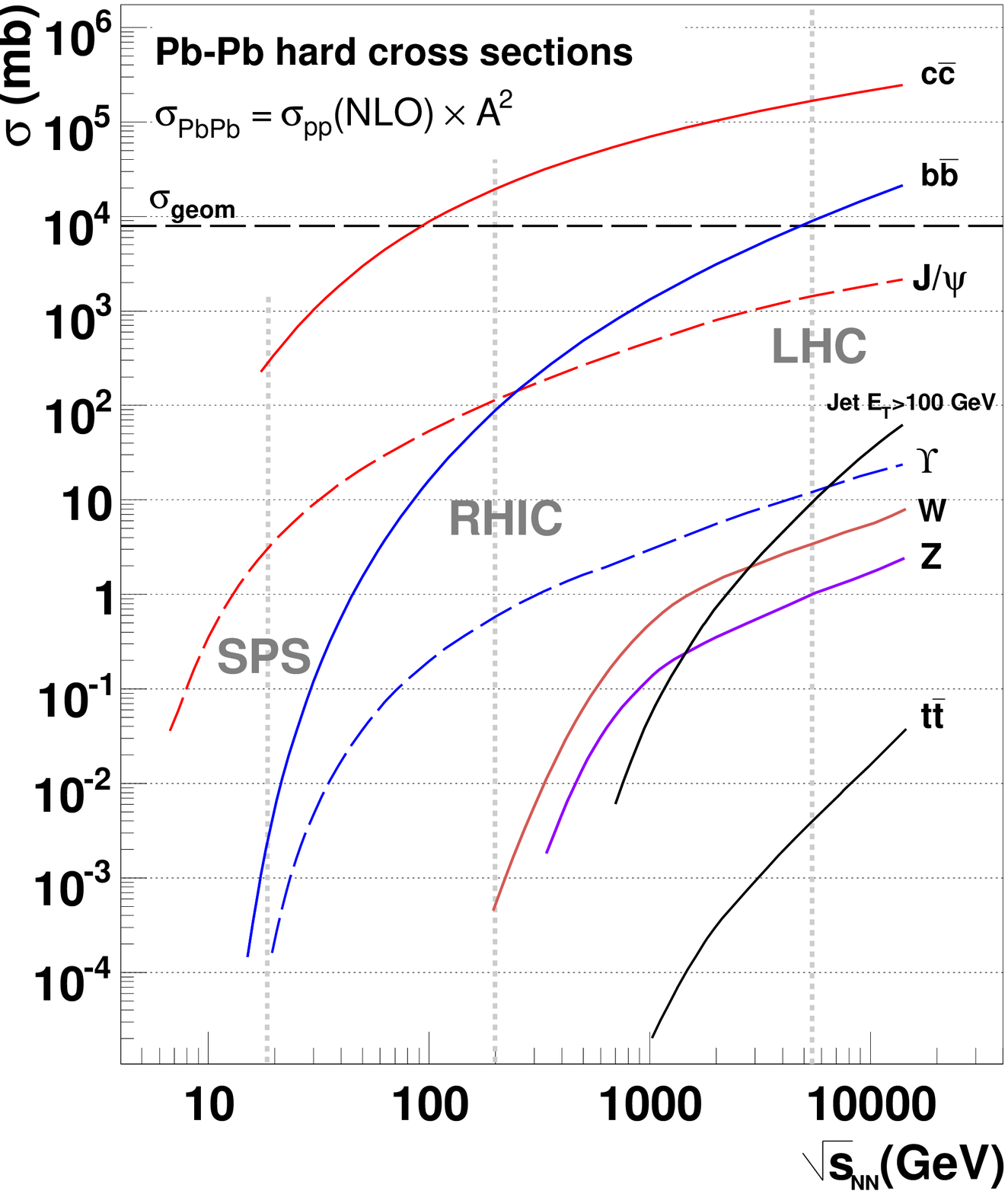}\hfill \includegraphics[width=5.6cm]{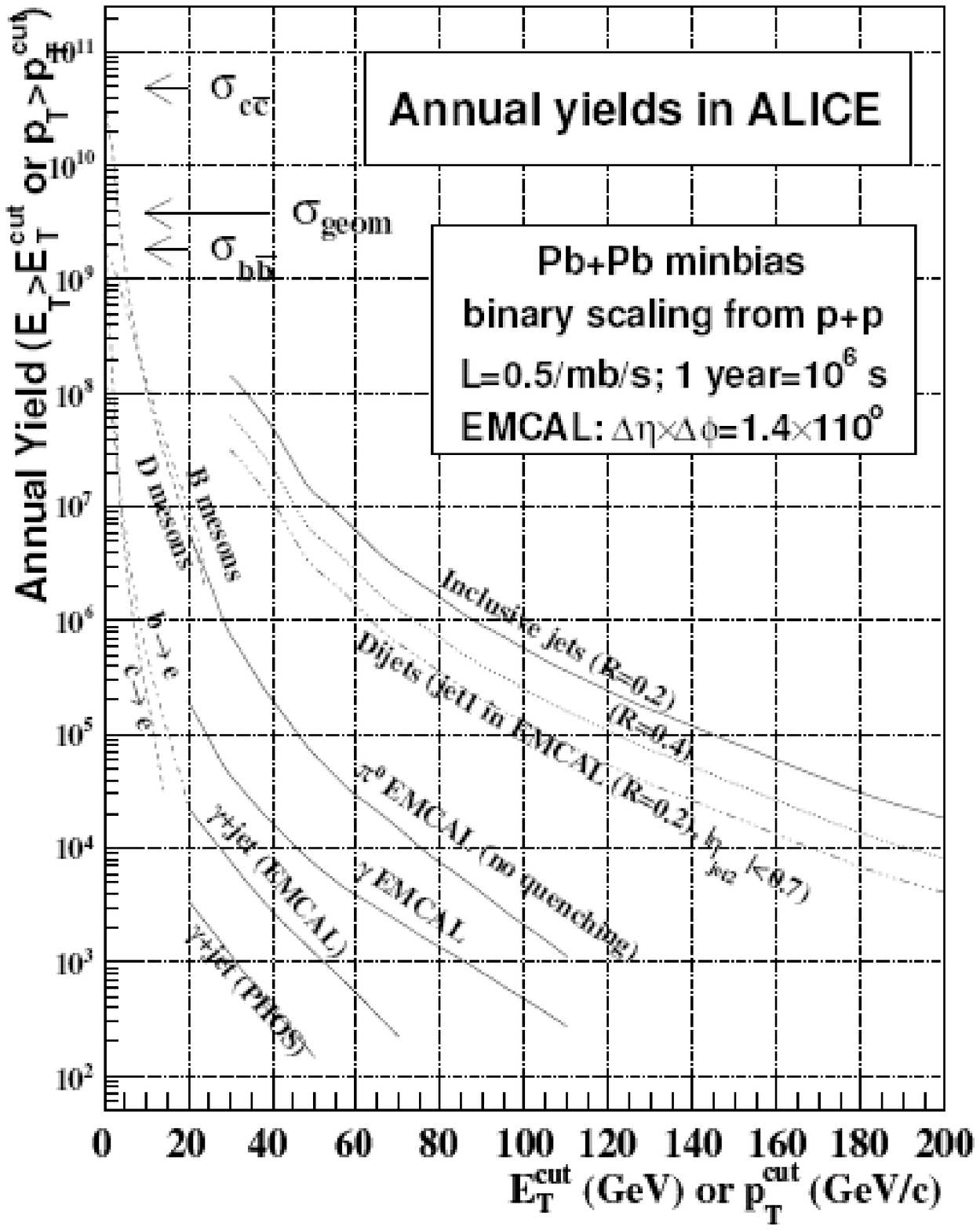}
\end{center}
\caption{Left: Cross sections for various hard processes ($\sigma_{\rm PbPb}^{hard}=A^2 \sigma_{\rm pp}^{hard}$) in PbPb minimum bias collisions in the range $\sqrt{s_{NN}} = 0.01\div 14$ TeV. Figure taken from $^{17}$. Right: 
Expected annual yields in the ALICE EMCal acceptance for various hard processes for minimum bias PbPb collisions at 5.5 TeV. Figure taken from $^{18}$.}
\label{fig1}
\end{figure}

\begin{figure}[htb]
\vskip -2cm
\begin{minipage}[h]{6.cm}
\begin{center}
\includegraphics[width=6cm]{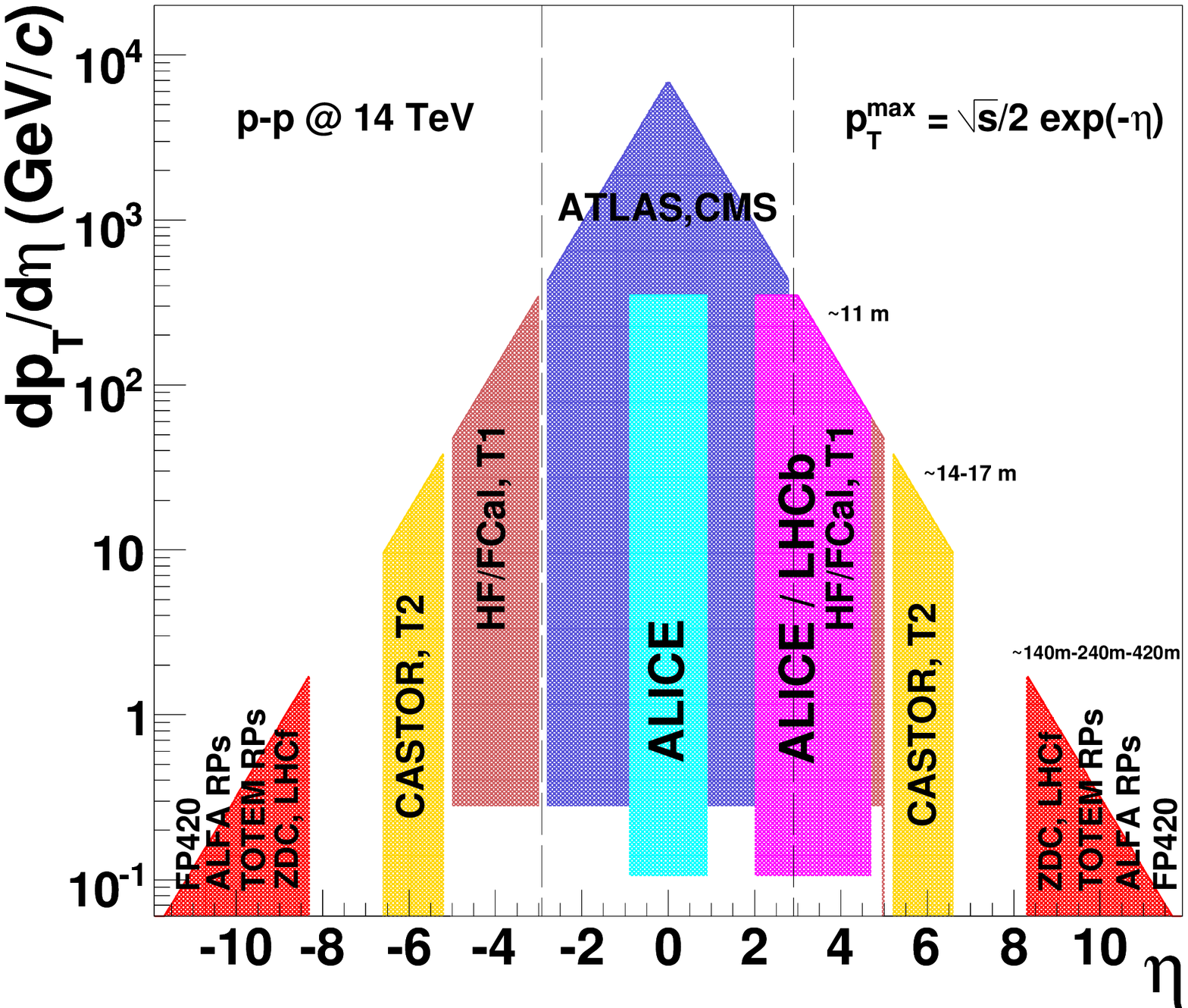}
\end{center}
\end{minipage}
\begin{minipage}[h]{6.cm}
\includegraphics[width=7.3cm]{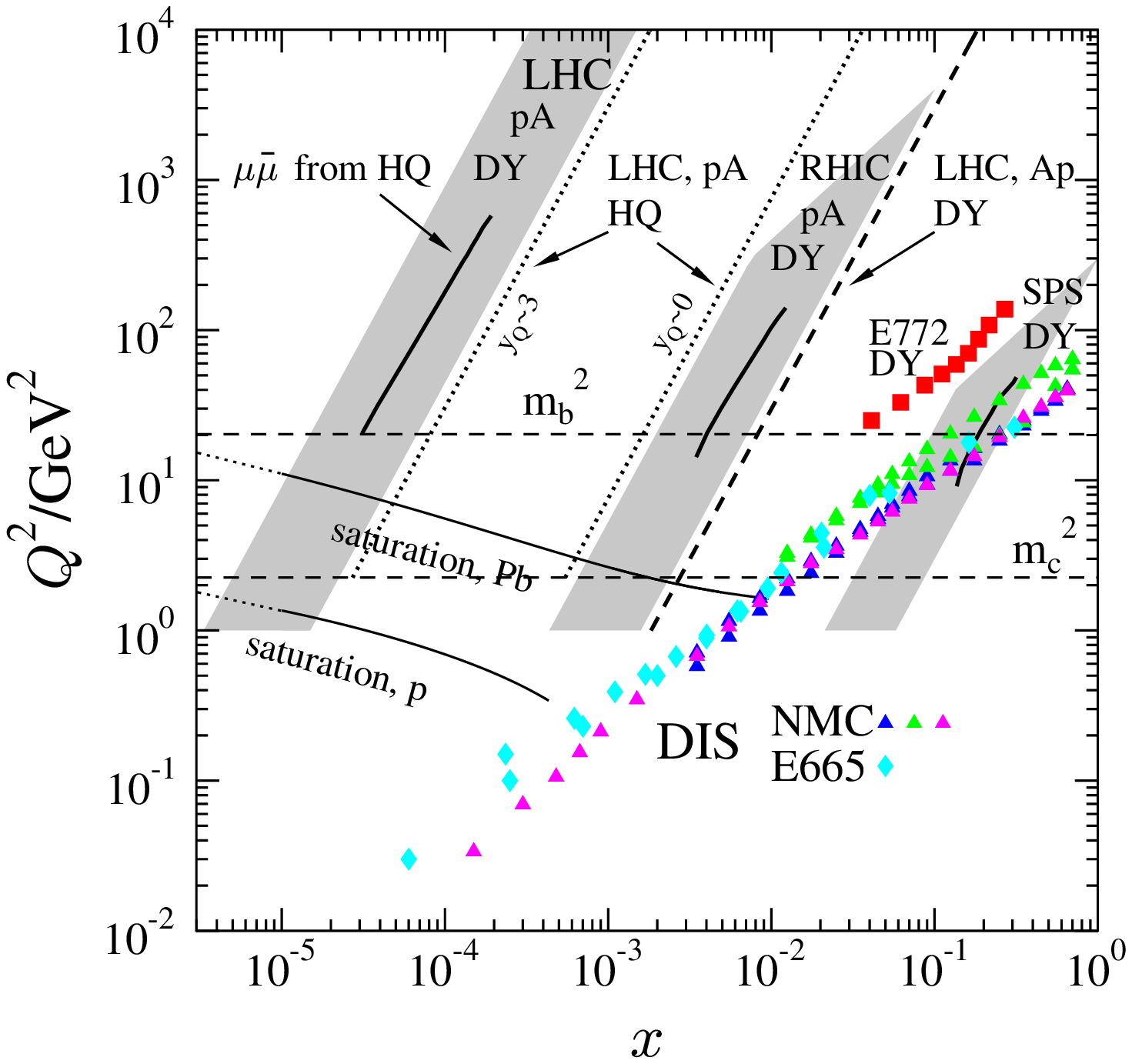}
\end{minipage}
\vspace*{-2cm}
\vspace*{8pt}
\caption{Left: Approximate $p_T-\eta$ coverage of current and proposed spectrometers and calorimeters
at the LHC. Figure taken from $^{21}$. Right: 
Resolution power ($Q^2$) $\times$ momentum fraction ($x$) coverage of the SPS, RHIC and LHC experiments for parton densities (grey bands and solid lines), compared with the regions covered by previous lepton-nucleus and proton-nucleus experiments (colored markers). Figure taken from $^9$.}
\label{fig2}
\end{figure}

The increase in center-of-mass energy of almost a factor 30 with respect to RHIC, together with the complementary detector capabilities, will offer new measurements with respect to those presently available. As evident examples:
\begin{itemize}
\item The yield of particles with large mass or transverse momentum - hard probes \cite{Accardi:2004be,Accardi:2004gp,Bedjidian:2004gd,Arleo:2004gn} - will be much more abundant, and some of them will be measured for the first time (with high statistics) in heavy-ion collisions, like $\Upsilon$ or $Z^0+{\rm jet}$ production (see Fig. \ref{fig1}, taken from \cite{d'Enterria:2008ge,Cortese:2008zza}).
\item Calorimeter capabilities of ATLAS and CMS \cite{D'Enterria:2007xr,Steinberg:2007nm} will allow for measuring jets both at central and non-central rapidities\footnote{Jets in heavy-ion collisions have been measured for the first time by the STAR Collaboration at RHIC \cite{jets,Putschke:2008wn}.}. ALICE will also be equipped with an electromagnetic calorimeter \cite{Cortese:2008zza}.
\item The kinematical coverage of the parton densities inside proton and nuclei will greatly exceed that available at the SPS and RHIC (see Fig. \ref{fig2}, taken from \cite{Accardi:2004be,d'Enterria:2008is}).
\end{itemize}

The aim of this review is to present a comprehensive compilation of the existing predictions for the heavy -ion programme at the LHC, not to discuss the current interpretation of available experimental data. Nevertheless, I will briefly indicate the main, 'standard' claims extracted from the experimental programmes at the SPS and RHIC. {\it Let me stress that none of these claims are devoid of alternative explanations, and that their presentation will doubtlessly contain some personal bias}. I will use them {\it only} to motivate the discussion of the new opportunities at the LHC and the discriminating power of the forthcoming measurements there.

The standard claims at RHIC are the following (the reader may find extensive discussions and references to the relevant experimental data in \cite{Heinz:2000bk,rhic,Back:2004je,Arsene:2004fa,Adams:2005dq,Gyulassy:2004zy,Jacobs:2004qv,Muller:2007rs,d'Enterria:2006su}):
\begin{itemize}
\item Multiplicities at RHIC are much lower than pre-RHIC expectations \cite{Bass:1999zq,Armesto:2000xh}. The standard interpretation is that particle production in the collisions shows a large degree of coherence due to initial state effects.
\item The elliptic flow measured in the collisions can be well reproduced by calculations within ideal hydrodynamics with a very early thermalization (or isotropization) time and small room for shear viscosity. This is currently interpreted in terms of the creation of some form of matter which (nearly) equilibrates very early and behaves like a quasi-ideal fluid.
\item The yield of high transverse momentum particles of different species measured at RHIC is strongly depleted in comparison with the expectations of an incoherent superposition of nucleon-nucleon collisions (as suggested by the collinear factorization theorems and confirmed by experimental data on weakly interacting perturbative probes). This fact, named jet quenching, together with the absence of such depletion in dAu collisions, is understood as the creation of a partonic medium, very opaque to energetic partons traversing it.
\end{itemize}

On the basis of these observations, it has been claimed that partonic matter, with an energy density larger than required by lattice QCD (see e.g. \cite{lattice,Cheng:2007jq,Fodor:2007ue}) for the phase transition from hadronic matter to the Quark-Gluon Plasma (QGP) to occur, has been formed. Such matter is extremely opaque to fast color charges traversing it, and its collective expansion closely resembles that of an ideal fluid. These two latter facts suggest that the produced matter is strongly coupled, which is in opposition to the naive picture of the QGP as an ideal parton gas and is not contradicted by lattice data which show some deviation from the Stefan-Boltzmann law and a finite value of the conformal anomaly up to temperatures larger than several times the deconfinement temperature.

Many questions remain open both in the experiment (e.g. suppression of heavy-flavor production or unbiased jet measurements, in nucleus-nucleus collisions) and on the theory sides (can the observed phenomena be explained within pQCD or do they require strong coupling?; what is the correct implementation and actual role of bulk and shear viscosity in hydrodynamical calculations?; how can such an early isotropization be achieved?; can the initial state - the nuclear wave function - be described by perturbative methods?; how can we compute particle production in such a dense environment?;$\dots$).

In this review of predictions for the heavy-ion programme at the LHC\footnote{Similar efforts done for RHIC can found in \cite{Bass:1999zq,Armesto:2000xh}.} I will classify them into different groups according to the following scheme: Those observables which characterize the produced medium itself, which I will call {\it bulk observables} (or soft probes, as they refer to particles with momentum scales of the order of the typical momentum scale of the medium - the temperature if thermalization were achieved); and those whose expectation in the absence of any medium can be calculated by perturbative methods in QCD (pQCD) (thus characterized by a momentum scale much larger than both $\Lambda_{QCD}$ and the 'temperature' of the medium), commonly referred to as {\it hard probes} \cite{Accardi:2004be,Accardi:2004gp,Bedjidian:2004gd,Arleo:2004gn}. Not being the subject of this review, I will provide few references to introduce the different subjects - I refer the reader to \cite{qgp1,qgp2,qgp3}.

I will start this review by some qualitative expectations for the LHC, based on simple arguments (in this respect see also \cite{Borghini:2007ub}). Through such discussion I aim to show how a single observable - charged multiplicity at mid-rapidity - strongly influences most other predictions. Then I will turn to detailed predictions on bulk observables. I will review those on multiplicities, collective flow, hadrochemistry at low transverse momentum, correlations and fluctuations. Next I will discuss hard and electromagnetic probes: particle production at large transverse momentum and jets, heavy quarks and quarkonia, and photons and dileptons\footnote{Concerning photons, their production at low momentum cannot not be described within pQCD but they have customarily  become part of the general item of hard and electromagnetic probes.}. Then I will review briefly pA collisions \cite{Accardi:2004be}. I will conclude with a summary and a discussion about the potentiality of the measurements at the LHC - particularly those made during the first run - to further substantiate or, on the contrary, disproof the picture of the medium that has arisen from the SPS and RHIC.

Most of the material that I will review is based on what was presented at the CERN Theory Institute on {\it Heavy Ion Collisions at the LHC - Last Call for Predictions}, held at CERN from May 14th to June 8th 2007, co-organized by Nicolas Borghini, Sangyong Jeon, Urs Achim Wiedemann and myself \cite{Abreu:2007kv,Armesto:2008fj}. I apologize in advance to those whose contributions I may unwillingly skip. I also apologize for not including any prediction for ultra-peripheral collisions (UPC) - see recent excellent reviews in \cite{upc,d'Enterria:2008sh}.

\section{Qualitative expectations}
\label{qualitative}

In principle, the reliability of the predictions for a given observable made within the framework of a given model is as good as the understanding of the existing experimental situation on that observable and related ones - provided the model contains the physical ingredients relevant for the extrapolation. It turns out that predictions for most observables, both for soft and hard probes, demand some parameter fixing which, in the most favorable case, can be related to a single measurable quantity. Such a quantity is usually the charged multiplicity at mid-rapidity or pseudorapidity which, in a more or less model-dependent way, can be related with energy densities, temperatures,$\dots$ of the medium at some given time.

In this Section, I will review some qualitative or semi-quantitative expectations for central PbPb collisions at the LHC.
The aim here is not to provide realistic or definite numbers (actually I will be most conservative in the estimates, so very probably the quantities for the LHC are underestimated in comparison to those at RHIC), but more or less stringent bounds, and to show explicitly how different predictions become affected or determined by a single observable, namely charged multiplicity at mid-rapidity. Let me note that a collection of data-driven predictions can be found in \cite{Borghini:2007ub}. While this latter collection, in its aim to being as model-independent as possible, is complementary to the one to be presented in the next Sections, it overlaps in spirit what will be presented here.

In Table \ref{armestotable1} I show the results within the Monte Carlo code \cite{Amelin:2001sk}  for the number of participants, of collisions and the charged multiplicity at mid-rapidity and pseudorapidity in central PbPb collisions at $\sqrt{s_{NN}}=5.5$ TeV. While these quantities are obtained in the framework of a given simulator, they will serve for the purpose of illustration in this Section. They will also be employed to better allow a comparison among different predictions for multiplicities at mid-(pseudo)rapidity in Subsection \ref{multi}.

\begin{table}[h]
\caption{Results in the Monte Carlo code in $^{39}$ for the mean impact parameter, number of participants and binary nucleon-nucleon collisions, and charged multiplicity at mid-(pseudo)rapidity, for different centrality classes defined by the number of participants, in central PbPb collisions at $\sqrt{s_{NN}}=5.5$ TeV.}
\label{armestotable1}
\begin{center}
\begin{tabular}{|c|c|c|c|c|c|}
\hline
\% & $ \langle b \rangle$ (fm) & $\langle N_{part} \rangle $ & $\langle N_{coll} \rangle$ & $dN_{ch}/dy|_{y=0}$ & $dN_{ch}/d\eta|_{\eta=0}$ \\ \hline
$0\div 3$ & 1.9 & 390 & 1584  & 3149 & 2633 \\ \hline
$0\div 5$ & 2.4 & 375 & 1490 & 2956 & 2472 \\ \hline
$0\div 6$ & 2.7 & 367 & 1447 & 2872 & 2402 \\ \hline
$0\div 7.5$ & 3.0 & 357 & 1390 & 2759 & 2306 \\ \hline
$0\div 8.5$ & 3.1 & 350 & 1354 & 2686 & 2245 \\ \hline
$0\div 9$ & 3.2 & 347 & 1336 & 2649 & 2214 \\ \hline
$0\div 10$ & 3.4  & 340 & 1303 & 2583 & 2159 \\ \hline
\end{tabular}
\end{center}
\end{table}

For the purpose of fixing one reference centrality class, I will define it by a number of participants $N_{part}=350$. In the following and unless otherwise stated, when referring to RHIC and the LHC I will be making reference to AuAu collisions at top RHIC energy, and PbPb collisions at the LHC, for a central centrality class defined by $N_{part}=350$. 

Let me start with multiplicities, as they are a key observable which will determine many other predictions. As stated previously, expectations for other observables from collective flow to jet quenching, depend on the scaling of certain quantities e.g. initial energy density, which are related in some more or less direct way with the final multiplicity measured in the event. Thus, many predictions are provided for some specific values of parameters which may be linked with a multiplicity.

Predictions for multiplicities can be discussed in the following way: A lower bound comes from the wounded nucleon model \cite{Bialas:1976ed} in which the multiplicity in nuclear collisions is expected to be proportional to the number of participant nucleons. This proportionality is also the limiting value expected by models which consider extremely strong shadowing effects. On the other hand, an upper limit can be set by the proportionality to the number of binary nucleon-nucleon collisions $N_{coll}$, as expected both in models of particle production which suppose a dominance of hard, perturbative processes (using the collinear factorization theorem \cite{Collins:1985gm,Collins:1985ue}, inclusive particle production is proportional to the product of the fluxes of partons in projectile and target which in the totally incoherent limit is proportional to the number of nucleon-nucleon collisions) and in soft models of particle production in absence of shadowing corrections (see e.g. \cite{Capella:1999kv}) through the cutting rules \cite{Abramovsky:1973fm}.

\begin{figure}[htb]
\begin{center}
\includegraphics[width=10cm]{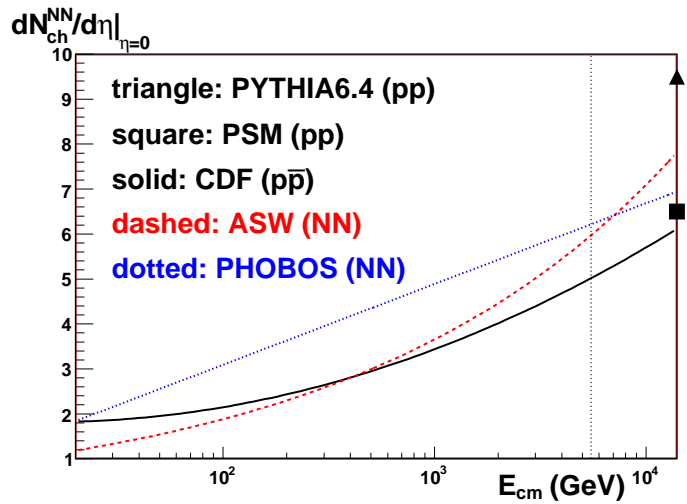}
\end{center}
\caption{Charged multiplicity at mid-pseudorapidity in nucleon-nucleon collisions versus center-of-mass energy, from different parametrizations (CDF for p$\bar{\rm p}$ collisions $^{46}$, ASW $^{47}$ and PHOBOS $^{35}$ for nucleon-nucleon collisions) and Monte Carlo simulators (PSM1.0 $^{39}$, and PYTHIA6.4 $^{48}$ as shown in $^{12}$, for pp collisions; these two points are included just for the purpose of illustration as they depend on the set of parameters used for the simulation).}
\label{fig3}
\end{figure}

On the basis of these considerations, the multiplicity can then be written in the following way (see also the discussions in \cite{Armesto:2000xh}):
\begin{equation}
\left.\frac{dN_{ch}^{AA}}{d\eta}\right|_{\eta=0}=\left.\frac{dN_{ch}^{NN}}{d\eta}\right|_{\eta=0}\left[\frac{1-x}{2}\,N_{part}+xN_{coll}\right],\ \ 0<x<1,
\label{eq1}
\end{equation}
with the superscript $NN$ referring to nucleon-nucleon collisions - an average of pp, pn and nn\footnote{At large energies and at central rapidities, particle production should be determined by partons with small momentum fraction (which can be estimated using $2\to 1$ kinematics as $x\sim m_T/\sqrt{s_{NN}}$, with $m_T=\sqrt{p_T^2+m^2}$ the transverse mass of the produced particle). At such small momentum fractions, isospin symmetry is expected to hold.}. Shadowing effects and energy-momentum constraints \cite{Capella:1999kv} tend to decrease $x$. As an example, values extracted from RHIC data at $\sqrt{s_{NN}}=19.6$ and 200 GeV \cite{Back:2004dy} are $x\simeq 0.13$. For nucleon-nucleon collisions, I will use the proton-(anti)proton data shown in Fig. \ref{fig3}. The three lines correspond to the parametrization of Sp$\bar{\rm p}$S and Tevatron data by CDF \cite{Abe:1989td}
\begin{equation}
\left.\frac{dN_{ch}^{NN}}{d\eta}\right|_{\eta=0}({\rm CDF})=2.5-0.25\ln s_{NN}+0.023\ln^2 s_{NN},
\label{cdf}
\end{equation}
to the parametrization in  \cite{Armesto:2004ud}
\begin{equation}
\left.\frac{dN_{ch}^{NN}}{d\eta}\right|_{\eta=0}({\rm ASW})=0.47 \,(s_{NN})^{0.144} (N_{part})^{0.089}=0.50 \,(s_{NN})^{0.144}
\label{asw}
\end{equation}
and to the PHOBOS parametrization in the contribution by Busza in \cite{Abreu:2007kv},
\begin{equation}
\left.\frac{dN_{ch}^{NN}}{d\eta}\right|_{\eta=0}({\rm PHOBOS})=-0.5+0.39\ln s_{NN}
\label{phobos}
\end{equation}
(note that this parametrization was obtained from fits to nucleus-nucleus data and thus it was not intended to describe nucleon-nucleon data).
I will also assume, as suggested by RHIC data \cite{Back:2004dy}, that the energy and centrality dependences of charged particle yields at mid-rapidity decouple. Considering all this, I show in Table \ref{table2} some naive predictions for the LHC\footnote{For the same centrality class defined by $N_{part}=350$, the corresponding charged multiplicity at $\eta=0$ at top RHIC energy from the PHOBOS parametrization \cite{Abreu:2007kv} is 635.}. The predictions from the wounded nucleon model ($x=0$) lie in the range $900\div 1100$, while those from a scaling with the number of collisions lie in the range $6800 \div 8400$. The latter agree with the expectations in 1995 as shown in the ALICE Technical Proposal \cite{alicetp}. The former roughly coincide with the expectations (1100, \cite{Borghini:2007ub}) from limiting fragmentation (extended longitudinal scaling) and a self-similar trapezoidal shape of the $\eta$-distribution between RHIC and LHC energies. Let us note that, as discussed in \cite{Borghini:2007ub}, charged multiplicities larger than $\sim 1650$ will be difficult to reconcile with limiting fragmentation.

\begin{table}[h]
\caption{Charged multiplicity at central pseudo-rapidity in PbPb collisions at LHC energy for $N_{part}=350$ ($N_{coll}=1354$) from Eq. (\ref{eq1}), for three different predictions of the corresponding multiplicity in pp collisions, see Fig. \ref{fig3}.}
\label{table2}
\begin{center}
\begin{tabular}{|c|c|c|c|}
\hline
pp extrapolation & $dN^{pp}_{ch}/d\eta|_{\eta=0}$ & $x$ & $dN^{PbPb}_{ch}/d\eta|_{\eta=0}$ \\ \hline
ASW & 5.97 & 0  &  1050 \\ \hline
ASW & 5.97 & 0.13  &  1950 \\ \hline
ASW & 5.97 & 1  &  8100 \\ \hline
CDF & 5.02 & 0 & 900 \\ \hline
CDF & 5.02 & 0.13 & 1650 \\ \hline
CDF & 5.02 & 1 & 6800 \\ \hline
PHOBOS & 6.22 & 0 & 1100 \\ \hline
PHOBOS & 6.22 & 0.13 & 2050 \\ \hline
PHOBOS & 6.22 & 1 & 8400 \\ \hline
\end{tabular}
\end{center}
\end{table}

Now one can try to estimate a lower bound for the energy density in this reference centrality class defined by $N_{part}=350$. For this and for the forthcoming discussions in this Section, I will consider three possibilities for multiplicities:
\begin{itemize}
\item Case I, the smallest one in Table \ref{table2}, 900;
\item Case II, the maximum multiplicity allowed by limiting fragmentation \cite{Borghini:2007ub}, 1650;
\item And Case III, a value of 2600 which is representative of the highest recent predictions for the LHC (see Subsection \ref{multi}).
\end{itemize}
I use the Bjorken estimate \cite{Bjorken:1982qr} and the arguments about the average formation (proper) time for particle production, $\langle \tau_{form}\rangle$, in \cite{rhic}:
\begin{eqnarray}
\langle\epsilon\rangle(\langle \tau_{form}\rangle)&\ge& \frac{1}{\langle \tau_{form}\rangle{\cal A}}\,\langle m_T^{measured}\rangle \left.\frac{dN^{measured}}{d\eta}\right|_{\eta=0}\nonumber \\
&\approx& \frac{\langle m_T^{measured}\rangle^2}{\pi R_A^2} \, \frac{3}{2}\,  \left.\frac{dN_{ch}^{measured}}{d\eta}\right|_{\eta=0}.
\label{bjorken}
\end{eqnarray}
In this equation $\langle \tau_{form}\rangle\approx \langle m_T^{measured}\rangle^{-1}$, ${\cal A}=\pi R_A^2$ is an upper bound for the overlapping area for central collisions with $R_A=1.12\, A^{1/3}$ fm the nuclear radius, $m_T$ is the transverse mass and the super-index $measured$ indicate that these are the final quantities measured in the detectors. For top RHIC energy, using $\langle m_T^{measured}\rangle=0.57$ GeV as given by PHENIX \cite{rhic} (this quantity is weakly dependent on centrality), and taking $dN_{ch}^{measured}/d\eta|_{\eta=0}=635$ as given by the PHOBOS parametrization in \cite{Abreu:2007kv}, I get $\langle\epsilon\rangle(\langle \tau_{form}\rangle=0.35 \  {\rm fm})\ge 12$ GeV/fm$^3$. For the LHC, one has to estimate the increase in $\langle m_T^{measured}\rangle$ with collision energy. For that, I use the parametrization for $\langle p_T\rangle\left(\sqrt{s}\right)$ by UA1 \cite{Albajar:1989an} and adjust the hadron mass to get the value given by PHENIX i.e.
\begin{equation}
\langle p_T\rangle\left(\sqrt{s}\right)=0.4-0.03\ln\left(\sqrt{s}/{\rm GeV}\right)+0.0053 \ln^2\left(\sqrt{s}/{\rm GeV}\right)\ \  [{\rm GeV}],
\label{ua1}
\end{equation}
and $m=0.42$ GeV. Then I get, for Case I, $\langle\epsilon\rangle(\langle \tau_{form}\rangle=0.29\  {\rm fm})\ge 22$ GeV/fm$^3$. Therefore, the most conservative estimates for the LHC indicate a multiplicity increase of a factor $900/635\simeq 1.4$ and an increase of a factor $\sim 2$ in energy density at formation time, with respect to top RHIC energy (or $\langle\epsilon\rangle(\langle \tau_{form}\rangle=0.29\  {\rm fm})\ge 42$ and 66 GeV/fm$^3$ for Cases II and III respectively).

Now, and for the purpose of illustrating some qualitative behaviors, I turn to the eventual equilibration and dynamical evolution of the created system. For that I will use generic arguments based on the Bjorken ideal hydrodynamical scenario in one spatial dimension \cite{Bjorken:1982qr}, see \cite{qgp3,Hirano:2008hy} for reviews of the hydrodynamical description of heavy-ion collisions. First, one needs the energy density at the time when hydrodynamical evolution is initialized, i.e. a thermalization or isotropization time. The estimates at RHIC lie in the range $0.17\div 1$ fm \cite{rhic,Back:2004je,Arsene:2004fa,Adams:2005dq,qgp3,Niemi:2008ta} in ideal hydro (0.17 fm is the crossing time of two Au nuclei at RHIC), and similar values for studies including viscosity \cite{viscous,Chaudhuri:2008je,Song:2008hj}\footnote{On viscous hydrodynamics, see the recent review \cite{Romatschke:2009im}.}. I will take an intermediate time $\tau_{therm}^{RHIC}\simeq 0.6$ fm as a reference value. To extrapolate to the LHC, it looks plausible that a system with larger density thermalizes faster. Using the ideas \cite{cgcini,Kharzeev:2000ph} in the Color Glass Condensate (CGC, see the review in \cite{qgp3})\footnote{In the CGC, the multiplicity is proportional to the saturation scale squared $Q_s^2$ and its energy and centrality dependences roughly factorize, while the thermalization time is expected to be inversely proportional to $Q_s$.}, I will assume that the thermalization time scales like the inverse square root of the multiplicity at $\eta=0$. Therefore one expects
\begin{equation}
\frac{\langle\tau_{therm}^{LHC}\rangle}{\langle\tau_{therm}^{RHIC}\rangle}\simeq 0.85,\  0.62, \ 0.49
\label{tauth}
\end{equation}
for Cases I, II, III respectively. So the thermalization time at the LHC is $\tau_{therm}\lesssim 0.5$ fm. Assuming free streaming ($\epsilon \propto 1/\tau$ in the one-dimensional case) from formation time to thermalization, the corresponding lower bound for the energy density is
\begin{equation}
\langle\epsilon\rangle(\langle \tau_{therm}^{LHC}\rangle\simeq0.5\  {\rm fm})\ge 12\ \  {\rm GeV/fm}^3,
\label{epstauth}
\end{equation}
again factor $\sim 2$ larger than the one obtained for RHIC. If one assumes that the thermalization time decreases with increasing particle density, then larger multiplicities at the LHC will favor smaller thermalization times and thus larger energy densities at thermalization. For example, in the model used for illustration, one-dimensional free streaming plus CGC, $\langle\epsilon\rangle(\langle \tau_{therm}\rangle \propto (dN_{ch}^{measured}/d\eta|_{\eta=0})^{3/2}$ (modulo logarithmic corrections).

\begin{figure}[htb]
\begin{center}
\includegraphics[width=10cm]{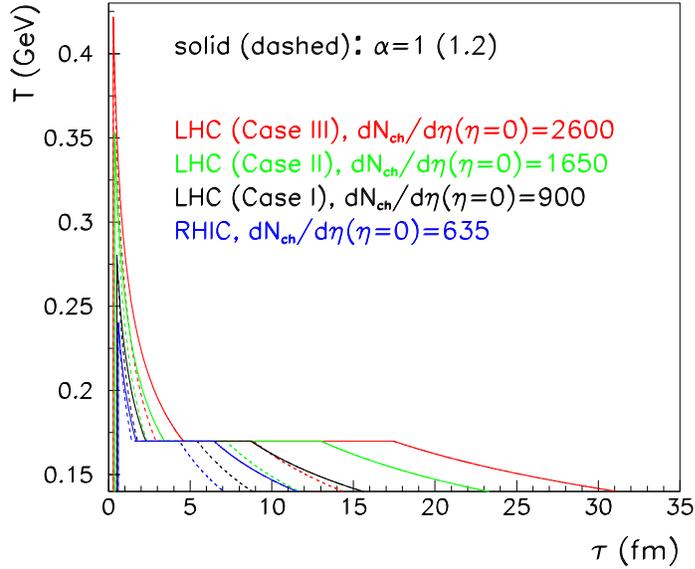}
\end{center}
\caption{Temperature versus proper time in the Bjorken model for the four scenarios in Table \ref{table3} and for two values of $\alpha$ in Eq. (\ref{bjhydro}): $\alpha=1$ and 1.2.}
\label{fig4}
\end{figure}

Now I will consider the evolution of the system, in order to illustrate the typical scales for the different phases of the system. To do so, I assume an ultra-relativistic ideal gas of 3 light quarks and gluons in the deconfined phase, and of 8 pseudo-scalar mesons in the confined phase. Using Bjorken estimate (\ref{bjorken}) and the Stefan-Boltzmann law, the values of the relevant quantities at $\langle\tau_{therm}\rangle$ are given in Table 3.
\begin{table}[h]
\caption{Values of the energy density, the temperature and the entropy density at $\langle\tau_{therm}\rangle$ for RHIC and the LHC.}
\label{table3}
\begin{center}
\begin{tabular}{|c|c|c|c|c|c|}
\hline
 & $dN^{AA}_{ch}/d\eta|_{\eta=0}$ & $\langle\tau_{therm}\rangle$ (fm) & $\langle\epsilon\rangle$  (GeV/fm$^3$) & $T_i$ (GeV) & $s_0$ (fm$^{-3}$) \\ \hline
RHIC & 635 & 0.6 & 6.8  & 0.241 &  38 \\ \hline
LHC & 900 (I) & 0.51 & 12.6  & 0.281 &  60 \\ \hline
LHC & 1650 (II) & 0.37 & 32.7  & 0.356 &  123 \\ \hline
LHC & 2600 (III) & 0.30 & 64.4 & 0.422 & 204 \\ \hline
\end{tabular}
\end{center}
\end{table}
Then I consider the evolution of the system using the Bjorken ideal hydrodynamical scenario in one spatial dimension \cite{Bjorken:1982qr} (see \cite{Beuf:2008vd} for recent developments in 1+1 ideal hydrodynamics)
 for both the confined and deconfined phases, but with a free parameter $\alpha$ to mimic a larger dilution rate due to transverse expansion:
\begin{equation}
\left(\frac{T}{T_0}\right)^3=\left(\frac{\tau_0}{\tau}\right)^\alpha,\ \ \frac{\epsilon}{\epsilon_0}=\left(\frac{\tau_0}{\tau}\right)^{4\alpha/3},\ \ \dots,
\label{bjhydro}
\end{equation}
with $\alpha=1$ corresponding to a pure longitudinal expansion. For a first-order phase transition, and assuming a deconfinement temperature of 170 MeV and a freeze-out temperature of 140 MeV, the evolution of the temperature is shown in Fig. \ref{fig4}. While the numbers shown in both the Figure and in Table \ref{table3} are most rough estimations, the plot illustrates some features common to more involved calculations: at the LHC the deconfined phase will last longer than at RHIC. The hadronic phase is not comparatively shorter than at RHIC (in this very schematic calculation using power-law evolutions of thermodynamical quantities), but its impact on some final observables (e.g. on photon or dilepton emission) could be expected to be smaller than at RHIC, due to the fact that the hadronic phase is restricted to the same range of temperatures but the partonic phase reaches higher $T$ at the LHC than at RHIC. On the other hand, it clearly shows that the larger the multiplicities, the longer-lived the deconfined phase will be.

\begin{figure}[htb]
\begin{center}
\includegraphics[width=6.3cm]{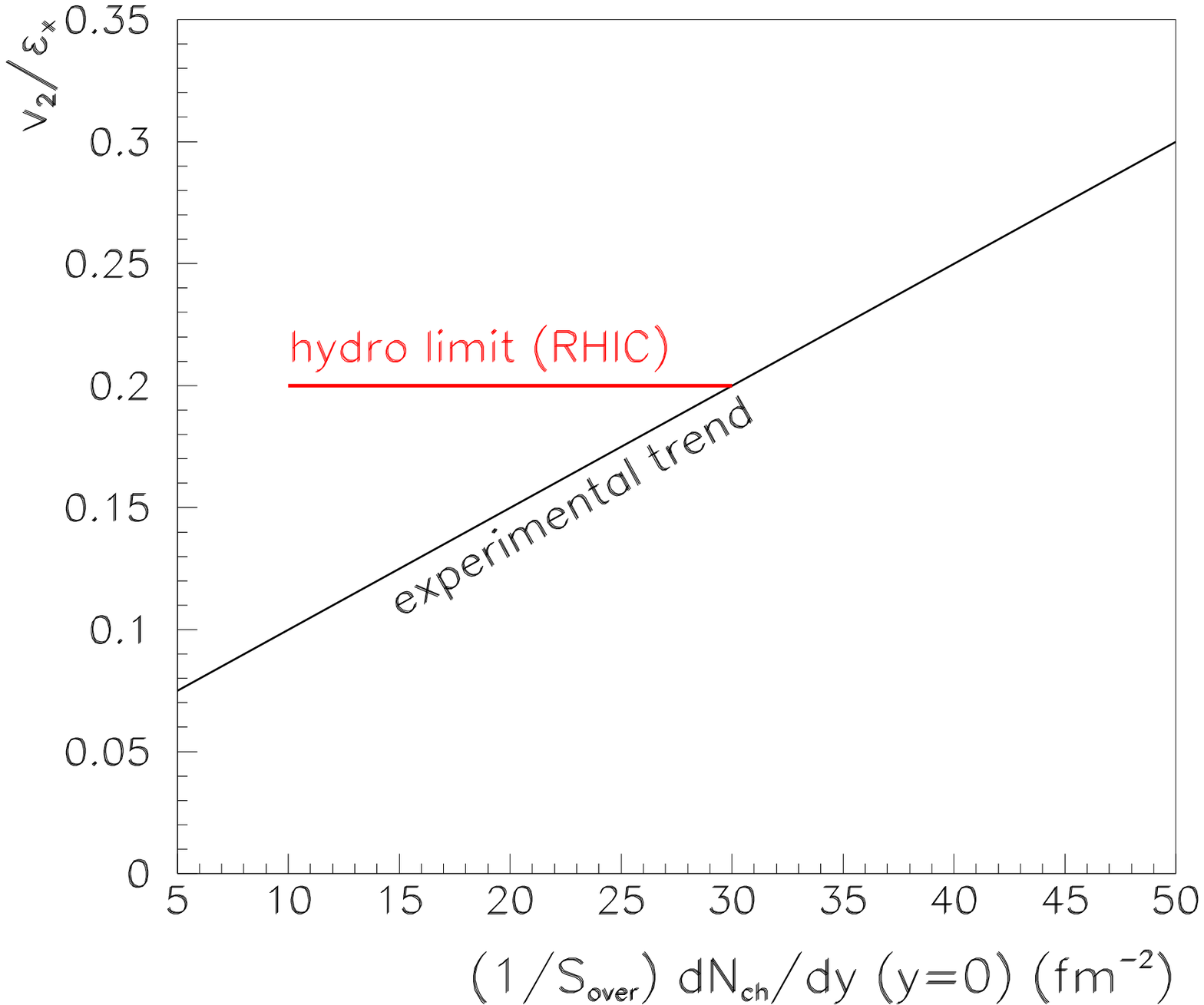}\hfill \includegraphics[width=6.3cm]{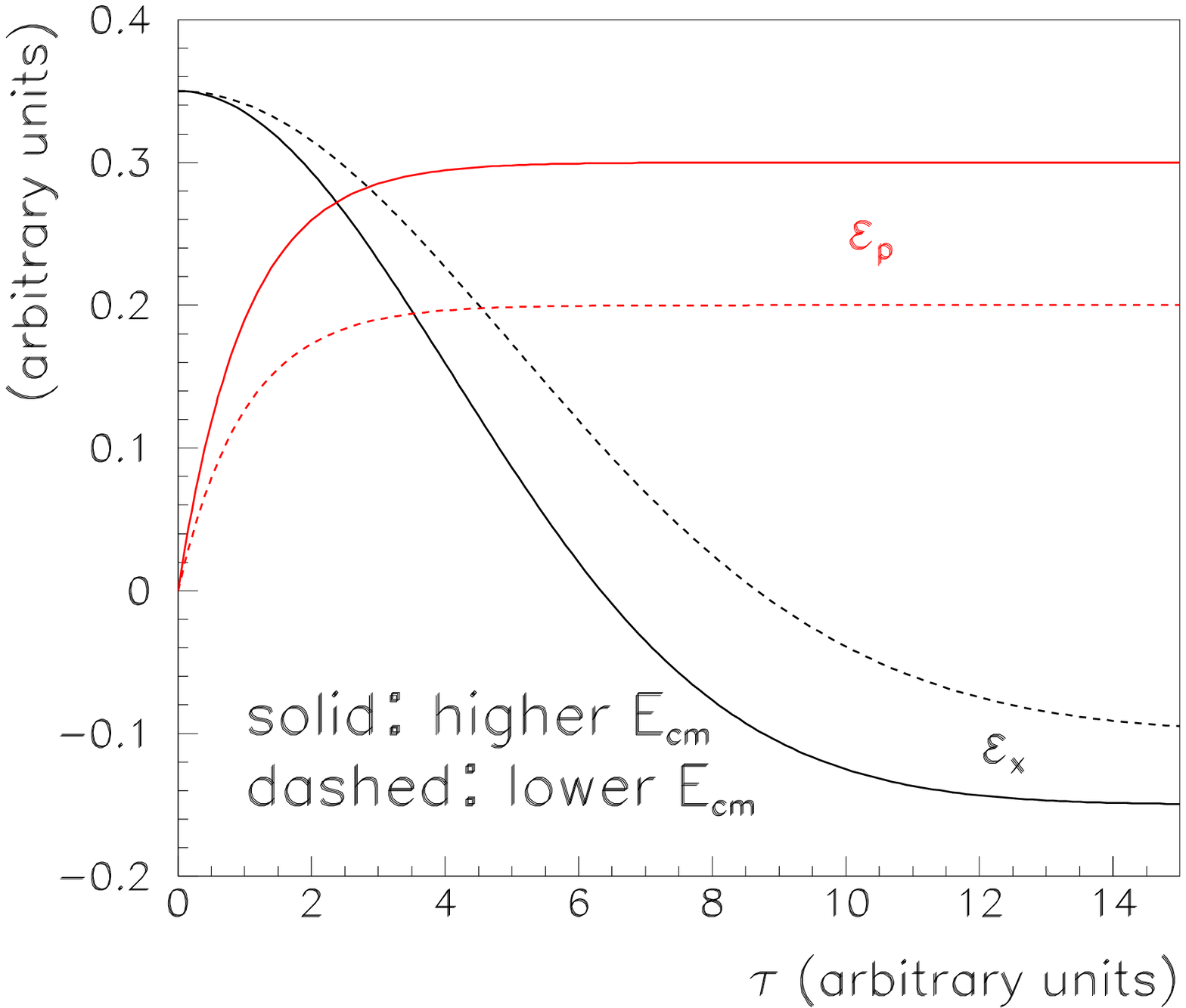}
\end{center}
\caption{Left: schematic plot showing the experimental trend (black) and the hydrodynamical limit (red line) of $v_2/\epsilon_x$ versus $(1/S_{over})dN_{ch}/dy|_{y=0}$. Right: schematic behavior of the spatial $\epsilon_x$ and momentum $\epsilon_p$ anisotropies versus proper time $\tau$ for lower (dashed) and higher (solid lines) energy densities at a fixed initial space anisotropy.}
\label{fig5}
\end{figure}

Now I will focus on the dependence on multiplicity of the elliptic flow $v_2$ integrated over transverse momenta\footnote{The discussion of the behavior of $v_2(p_T)$ requires a parallel discussion of the hadronization process which is far more involved.}. According to general arguments, see e.g. \cite{Voloshin:1999gs}, in the low-density limit the distortion of the azimuthal spectra with respect to the reaction plane $XZ$ (and thus the elliptic flow $v_2$) is proportional to the space anisotropy
\begin{equation}
\epsilon_x=\frac{\langle y^2-x^2\rangle}{\langle y^2+x^2\rangle}
\label{epsx}
\end{equation}
and to the density of scattering centers (or particle density) in the transverse plane $XY$,
\begin{equation}
\frac{v_2}{\epsilon_x}\propto \frac{1}{S_{over}}\left.\frac{dN_{ch}}{dy}\right|_{y=0}\, ,
\label{v2}
\end{equation}
with $S_{over}$ the overlap area for a given centrality class and the average in (\ref{epsx}) is done over the transverse energy density profile and, eventually, over the number of events. This relation is fulfilled by experimental data from lowest SPS to highest RHIC energies, see \cite{Alt:2003ab}, and is illustrated in Fig. \ref{fig5} left. It allows for a semi-quantitative relation between the multiplicity and the elliptic flow: I will assume that for AuAu or PbPb collisions at a given centrality class the spatial anisotropy, mainly determined by the geometry of the collision, and the overlap area are approximately the same and do not vary substantially with energy. Taking the slope of the experimental trend $\sim 0.005$ and for a point lying at (22,0.16)\footnote{These values are roughly those of the experimental data ($v_2=0.051$, $\epsilon_x=0.319$) \cite{Adler:2002pu} for a $20\div 30$ \% centrality class, which corresponds \cite{Niemi:2008ta} to an impact parameter $\sim 7.5$ fm and $N_{part}\sim 160$ both for RHIC and the LHC.}, increases in multiplicity by factors 1.5, 2.5, 4\footnote{These numbers are illustrative of the predictions for charged multiplicities at mid-pseudorapidity for $N_{part}=350$ at the LHC, 900, 1650 and 2600 - Cases I, II and III respectively -, compared with 635 at RHIC, and are applicable to other centralities provided the factorization between the centrality and energy dependences holds, see previous discussions.} translates into increases in $v_2/\epsilon_x$ of $\simeq 35$, 100, 205 \% respectively.

On the other hand, ideal hydrodynamics calculations 
\cite{Ollitrault:1992bk,Kolb:1999it} indicate a saturation or limiting value of $v_2/\epsilon$ versus $(1/S_{over})dN_{ch}/dy|_{y=0}$. The detailed value depends on the equation of state, on the details of initialization (see e.g. \cite{Lappi:2006xc} for a study of the influence of different initial conditions on the spatial anisotropy) and hadronization prescription, and on the treatment of the confined phase. The inclusion of viscous effects further reduces such limiting value \cite{viscous}. Besides, as illustrated in Fig. \ref{fig5} right, for a fixed initial spatial anisotropy, higher initial energy densities or temperatures imply larger density gradients which increase the final momentum anisotropy \cite{qgp3,Hirano:2008hy} defined as
\begin
{equation}
\epsilon_p=\frac{\langle T_{xx}-T_{yy}\rangle}{\langle T_{xx}+T_{yy}\rangle}\, ,
\label{ep}
\end{equation}
and thus increase $v_2/\epsilon_x$, as this momentum anisotropy is known \cite{Kolb:1999it} to be related with the observed $v_2\simeq \epsilon_p/2$. Numerical results \cite{Niemi:2008ta,Kestin:2008bh} within ideal hydrodynamics indicate increases in the transverse momentum integrated $v_2$ at $b\sim 7.5$ fm from RHIC to the LHC,  ranging from $\sim 15$ \% \cite{Kestin:2008bh} for charged multiplicities at mid-rapidity around 1200, to $\sim 40\div 60$ \% in \cite{Niemi:2008ta} for twice this multiplicity for central PbPb collisions. Results in viscous hydro \cite{viscous} yield increases $\sim 10$ \% for a charged multiplicity of 1800.

Let us finally discuss very briefly the influence of multiplicities on the standard observable for jet quenching, namely single inclusive particle suppression usually studied through the nuclear modification factor, defined for a given particle $k=h^\pm (ch),\pi^0,\dots$ as
\begin{equation}
R_{AA}(y,p_T)=\frac{\frac{dN^{AA}_k}{dydp_T}}{\langle N_{coll}\rangle \frac{dN_k^{NN}}{dydp_T}}\,,
\label{raadef}
\end{equation}
with $\langle N_{coll}\rangle$ the average number of binary nucleon-nucleon collisions in the considered centrality class.
A simple model\footnote{This simplistic model, whose sole aim is allowing for a discussion of the competing effects of density increase and different biases, is by no means quantitative.  For example, it does not consider in any detail geometrical biases like the surface bias, it does not take into account fragmentation and it assumes a pure power law behavior of the hadronic spectra which is true neither in data nor in pQCD. It is based on ideas developed in models of radiative energy loss in e.g. \cite{Baier:2001yt,Salgado:2003gb} but not restricted to these models - e.g. models with collisional energy loss also result in some probability of no energy loss.} to discuss this is the following: Let us assume {\it for a fixed geometry} (i.e. fixed length or eventual dynamical expansion) that partons can escape the medium without losing any energy with probability $p_0$, while they may lose some energy $\Delta E$ with probability $1-p_0$. Considering a spectrum $\propto 1/p_T^n$ ($n=8$ roughly describes the spectrum in $pp$ collisions at mid-rapidity at RHIC \cite{rhic}) I get
 \begin{equation}
R_{AA}(y,p_T)=p_0+\frac{1-p_0}{(1+\epsilon)^n}\,,\ \ \epsilon=\frac{\Delta E}{p_T}\, . 
\label{raamod}
\end{equation}
\begin{figure}[htb]
\begin{center}
\includegraphics[width=10cm]{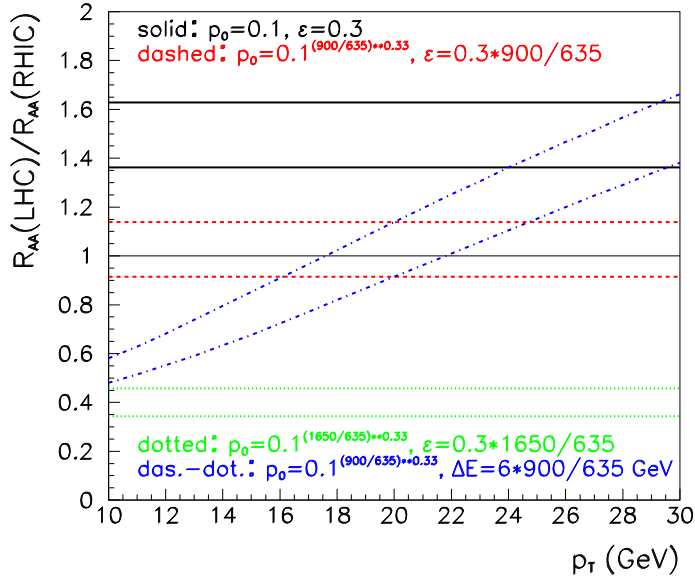}
\end{center}
\caption{Ratio of nuclear modification factors at the LHC and at RHIC from Eq. (\ref{raamod}). Different line styles refer to different parameters $p_0$ and $\epsilon$ or $\Delta E$, see the legends in the plot, while lower and upper lines of each style correspond to spectral power-law exponents $n=6$ and 5 respectively.}
\label{fig6}
\end{figure}

In Fig. \ref{fig6} I show the ratio of nuclear modification factors at the LHC and at RHIC. For RHIC, I have chosen $p_0=0.1$ and $\epsilon=0.3$ which produce a flat $R_{AA}(p_T)\simeq 0.21$ which qualitatively corresponds with that observed for $\pi^0$'s in central AuAu collisions at RHIC for $10$ GeV $<p_T<$ 20 GeV. To extrapolate to the LHC situation, I have chosen two values of $n=6,5$ and either the same values of $p_0$ and $\epsilon$, or these values modified by the expected ratio of multiplicities in Cases I and II (Case III is not illustrated for clarity of the plot), or a modification in which $\Delta E$, and not $\epsilon$,  scales with multiplicity at mid-rapidity. Different options produce evidently different results (e.g. the flatter the spectrum, $n=5$ compared to $n=6$, the larger the ratio; the larger the multiplicity, the smaller the ratio), a fact which stresses the need of a control of the reference spectrum and of the geometry or dynamical behavior of the medium in order to extract quantitative conclusions about the medium properties from measurements of the nuclear modification factor.

\section{Bulk observables}
\label{bulk}

Now I turn to the predictions for observables which directly characterize the medium produced in the collisions. These bulk observables correspond to particles with momentum scales of the order of the typical scales of the medium - the temperature if thermalization is achieved -, thus the name of soft probes that has been used to designate them.

In the following, the use of names of authors will correspond usually to those predictions contained  in the compilation \cite{Abreu:2007kv,Armesto:2008fj}, while those predictions not contained there will be referenced in the standard way. I refer the reader to the compilation \cite{Abreu:2007kv} for further information and model description of the former - a given contribution in  \cite{Abreu:2007kv} can be found by looking for the name of the authors in the Section devoted to the corresponding observable.

In this Section I will review consecutively: multiplicities, collective flow, hadrochemistry at low transverse momentum, correlations and fluctuations.

\subsection{Multiplicities}
\label{multi}

\begin{figure}[htb]
\begin{center}
\includegraphics[width=13.5cm,height=13.5cm]{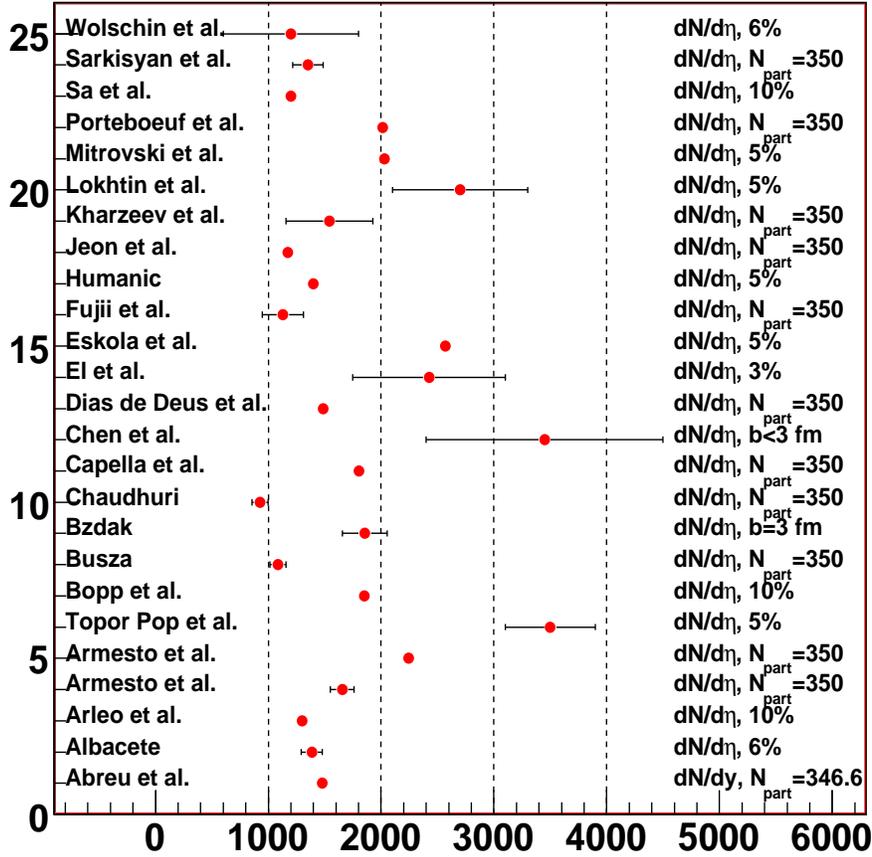}
\end{center}
\caption{Predictions for multiplicities in central Pb-Pb collisions at the LHC. On the left the name of the authors can be found. On the right, the observable and centrality definition is shown. The error bar in the points reflects the uncertainty in the prediction. See the text for explanations.}
\label{fig7}
\end{figure}

Charged particle multiplicity at mid-(pseudo)rapidity is a first-day observable at the LHC. Many groups have produced such predictions, see a compilation in Fig. \ref{fig7} where 25 predictions are shown\footnote{A compilation containing a smaller number of predictions can be found in \cite{Armesto:2008fj}.}.

\begin{figure}[htb]
\begin{center}
\includegraphics[width=14.cm,height=14.cm]{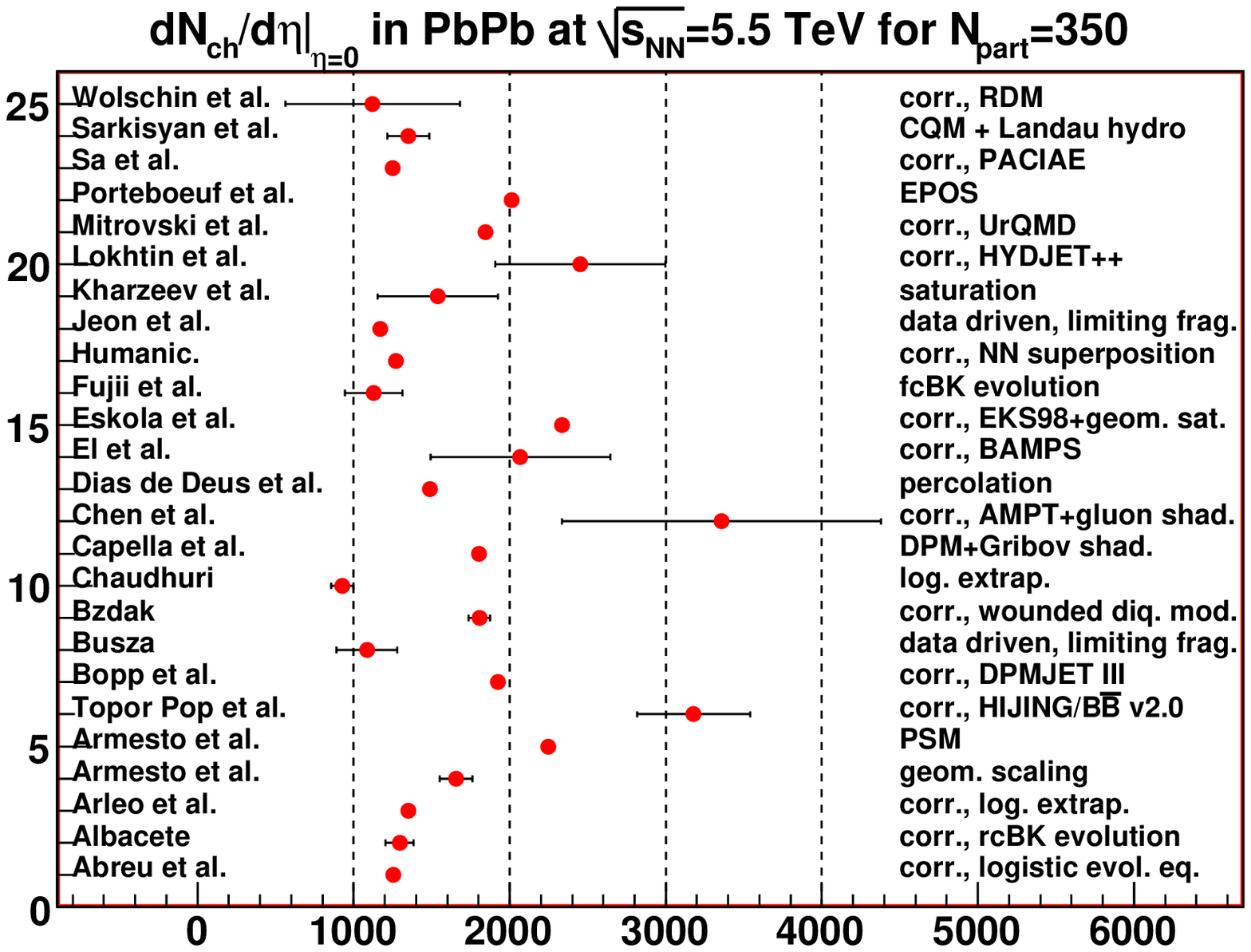}
\end{center}
\caption{Predictions for multiplicities in central Pb-Pb collisions at the LHC. On the left the name of the authors can be found. On the right, I indicate whether a correction has been applied or not, and provide a brief indication of some key ingredients in the model. The error bar in the points reflects the uncertainty in the prediction. See the text for explanations}
\label{fig8}
\end{figure}

Different groups provide predictions for different centrality classes. For a more accurate comparison, I re-scale them to a common observable ($dN_{ch}/d\eta|_{\eta=0}$) and centrality class ($\langle N_{part}\rangle =350$) using the model \cite{Amelin:2001sk}. The re-scaling factors can be read off Table \ref{armestotable1} and the corrected results found in Fig. \ref{fig8}. The re-scaling being made using a given model, its accuracy cannot be taken as very high, but it should reduce the uncertainties in the comparison to a 10 \% level.

Let me start by describing briefly the different predictions presented in the plots. A rough classification, for mere organizational purposes, can be made into the following items:

\begin{enumerate}

\item Monte Carlo simulators of nuclear collisions. These models include many different physical ingredients to be combined in a consistent manner. They all take into account energy-momentum and quantum number conservation in a detailed way. While sometimes the physical ingredients are similar between different models, the details of the implementation lead to different results.
\begin{itemize}

\item The PSM model \cite{Amelin:2001sk} contains a soft component with contributions from both the number of collisions and of participants which lead to the creation of color strings, satisfying roughly Eq. (\ref{eq1}) with $x$ calculable within the model and tending to 1 as energy constraints becomes less and less important with increasing energy. It also contains a hard component using standard pQCD, in which nuclear parton densities (npdf's) are used. Finally, string fusion is introduced as a collective mechanism in the soft component.

\item The HIJING/B$\bar{\rm B}$ \cite{ToporPop:2007hb,Abreu:2007kv} model contains a soft component proportional to the number of participants, and a hard component proportional to the number of collisions which also considers npdf's. It includes mechanisms for baryon number transport from the fragmentation to the central rapidity regions (string junctions), and introduces collectivity through an enhanced string tension - color ropes. The different predictions reflect the uncertainties in the increased string tension.

\item The DPMJET model \cite{Bopp:2004xn,Abreu:2007kv} is similar to the PSM, but it includes string junction transport, percolation of strings as a collective mechanism and the strong shadowing proposed for the soft sector in \cite{Capella:1999kv}.

\item The AMPT model \cite{Lin:2004en,Abreu:2007kv} considers a parton cascade initialized by HIJING \cite{Wang:1991hta} with subsequent hadronization via strings and a hadron transport. The different predictions correspond to the different npdf's used.

\item The HYDJET++ model \cite{Lokhtin:2009be} contains a soft, thermalized component which is treated hydrodynamically, and a hard component treated through PYTHIA (and PYQUEN, see Subsection \ref{highpt}). The error bar corresponds to a variation of the minimum transverse momentum for the hard component from 7 GeV (larger multiplicity) to 10 GeV (smaller multiplicity).

\item The UrQMD model \cite{Mitrovski:2008hb} contains a soft component, and a hard component through PYTHIA \cite{Sjostrand:2006za}, with a detailed space-time evolution of the pre-hadronic and hadronic degrees of freedom.

\item The EPOS model \cite{Drescher:2000ha,Abreu:2007kv} contains similar ideas to those of PSM and DPMJET but aims to account for energy-momentum conservation at the level of the cross sections (usually the cross sections are computed ignoring energy-momentum constraints which are applied a posteriori on the mechanism of particle production), and contains a detailed model for the soft-hard transition, for the treatment of the hadronic remnants and a separation between a dense core, eventually treated via hydrodynamical evolution, and a dilute corona which hadronizes via strings.

\item The PACIAE model \cite{Sa:2008fw} contains a parton cascade initialized by PYTHIA with hadronization via string formation and decay and a hadron transport. Collective effects are introduced through the increase of the string tension which, in this case, produces both a harder spectrum in transverse momentum and higher masses - as in previous approaches which consider increased string tensions -, and an enhancement of particle production with large longitudinal momentum.

\end{itemize}

\item Models based on saturation ideas.

\begin{itemize}

\item Abreu et al. \cite{DiasdeDeus:2007wb,Abreu:2007kv} is based on a non-linear, logistic evolution equation which resembles the Balitsky-Kovchegov (BK) equation in high-density QCD (see the review in \cite{qgp3}), for fixed-size dipoles. It admits an analytic solution and shows limiting fragmentation for some restricted parameter space.

\item Albacete \cite{Albacete:2007sm,Abreu:2007kv} is a prediction based on the running-coupling BK equation, with multiplicities computed through the use of $k_T$-factorization and local parton-hadron duality (LPHD). The error bar reflects the uncertainties in the extrapolation coming from the freedom to fix the parameters at RHIC.

\item Armesto et al. \cite{Armesto:2004ud,Abreu:2007kv} is a prediction based on the extension of the geometric scaling observed in lepton-proton collisions to proton-nucleus and nucleus collisions, and on LPHD. It provides a pocket formula for multiplicities, Eq. (\ref{asw}), in which the energy and centrality dependences explicitly factorize. The error bar reflects the uncertainties in the nuclear size dependence of the saturation scale extracted from lepton-nucleus data.

\item Eskola et al. \cite{Abreu:2007kv} use a pQCD approach supplemented with a geometric saturation ansatz. The obtained multiplicities and energy densities are used as input for an ideal hydrodynamical calculation.

\item Fujii et al. \cite{Abreu:2007kv} use the fixed-coupling BK equation plus limiting fragmentation together with $k_T$-factorization and LPHD. The error bar corresponds to the different initial conditions for evolution.

\item Kharzeev et al. \cite{Kharzeev:2004if,Abreu:2007kv} use the saturation ideas together with $k_T$-factorization and LPHD. The error bar corresponds to the different options in which the saturation scale grows with energy or saturates.

\end{itemize}

\item Data-driven predictions.

\begin{itemize}

\item Arleo et al. \cite{Abreu:2007kv} is a logarithmic extrapolation of multiplicities at RHIC which is used as input for ideal hydrodynamical calculations at low transverse momentum coupled to pQCD at large transverse momentum.

\item Busza \cite{Abreu:2007kv} is a data-driven extrapolation based on the logarithmic increase of particle densities from SPS to RHIC and on the factorization of energy and geometry dependences.

\item Chaudhuri \cite{Chaudhuri:2008je} is a data-driven extrapolation based on the logarithmic increase of multiplicities at RHIC which is used as input for a viscous hydrodynamical calculation.

\item Jeon et al. \cite{Abreu:2007kv} is a data-driven extrapolation based on limiting fragmentation (considering not only the slope of the pseudorapidity distribution near beam rapidity but also the curvature) and the logarithmic increase of particle densities from SPS to RHIC. A small, not visible error bar is due to the different choice of parameters in the fits to existing data.

\end{itemize}

\item Others.

\begin{itemize}

\item Bzdak \cite{Bzdak:2008gw} uses a variant of the wounded nucleon model \cite{Bialas:1976ed} in which the relevant degrees of freedom are not nucleons but quarks and diquarks. In order to obtain predictions for multiplicity, the results in the model for the number of wounded quarks and diquarks have been supplemented by the multiplicities for nucleon-nucleon collisions in Table \ref{table2}, with the error bar reflecting the uncertainty in the latter.

\item Capella et al. \cite{Abreu:2007kv} is a soft model in which the multiplicity gets contributions from both the number of collisions and of participants, supplemented with a very strong shadowing\footnote{This very strong shadowing corresponds to ideas very close to those of saturation but formulated in a soft domain in which pQCD techniques are not applicable and  phenomenological models are required.} related with diffraction in lepton-proton collisions  \cite{Capella:1999kv}.

\item Dias de Deus et al. \cite{Abreu:2007kv} use a model in which the multiplicity, proportional to the number of collisions, is decreased by a geometric factor, given by two-dimensional continuum percolation and related with the fraction of transverse area occupied by the overlapping sources of particles (strings).

\item El et al. \cite{El:2007vg,Abreu:2007kv} is a parton cascade initialized by CGC conditions. The parton cascade includes both $2\leftrightarrow 2$ and $2\leftrightarrow 3$ processes and uses LPHD to relate the output of the cascade with the final multiplicities. The error bar reflects the uncertainty in the extrapolation of the saturation scale in the CGC initial conditions from RHIC to LHC energies.

\item Humanic \cite{Humanic:2008nt} is a superposition model based on a geometrical ansatz to determine the number of pp collisions, which are modeled through PYTHIA. A space-time picture of hadronization is also included which allows a link to a hadron cascade.

\item Sarkisyan et al. \cite{Sarkisyan:2005rt} is a model based on the constituent quark model which leads to a participant-like picture similar to that in the model of Bzdak, with the energy deposition in the collision considered within Landau hydrodynamics (see e.g. \cite{Wong:2008ta} for a recent review).

\item Wolschin et al. \cite{Kuiper:2006si,Abreu:2007kv} is a relativistic diffusion equation in rapidity and time of the Fokker-Planck type. The error bar reflects the uncertainties in the extrapolation of the diffusion parameters from RHIC to the LHC.

\end{itemize}

\end{enumerate}

From the plots one can conclude that most predictions lie in the range $1000\div 2000$. It should be noted than a value lower than 1000 could be, depending on the corresponding value for pp collisions\footnote{pp collisions at the same energy as PbPb may not occur before several year of successful pp data-taking. Therefore, for the first runs an interpolation between p$\bar{\rm p}$ collisions at Tevatron and pp collisions at LHC energies could be required.}, in conflict with participant scaling. On the other hand, a value larger than 2000 will be a challenge for saturation physics. Monte Carlo simulators, due to their complexity, do not include yet many recent theoretical developments, e.g. none implements saturation effects. This might be the reason why they tend to give the largest values.

Finally, multiplicities show a decreasing tendency with time from 1995 \cite{alicetp}, through pre-RHIC predictions \cite{Armesto:2000xh}, until now. This is due to the inclusion of collective effects which imply a large degree of coherence in particle production like saturation, strong color fields, percolation, or strong gluon shadowing. This strong coherence can be understood as a decrease in the number of sources which contribute independently to multiparticle production. Proposals to find evidence of these ideas in correlations will be discussed in Subsection \ref{correl}. They would also leave an imprint in multiplicity distributions, see Bopp et al. in \cite{Abreu:2007kv}. Also the pseudorapidity distributions are informative: for example, the model by Abreu et al. \cite{DiasdeDeus:2007wb,Abreu:2007kv} shows a extremely wide plateau in rapidity (along $\sim 8$ units).

Now I review the predictions for baryon transport. Since this is an important observable from the point of view of the hadrochemistry, it could be included in Subsection \ref{hadroch}. But it is also a global characteristic of the collision which goes beyond the single number, $dN_{ch}/d\eta|_{\eta=0}$, mainly discussed so far.

\begin{figure}[htb]
\begin{center}
\includegraphics[width=8.cm]{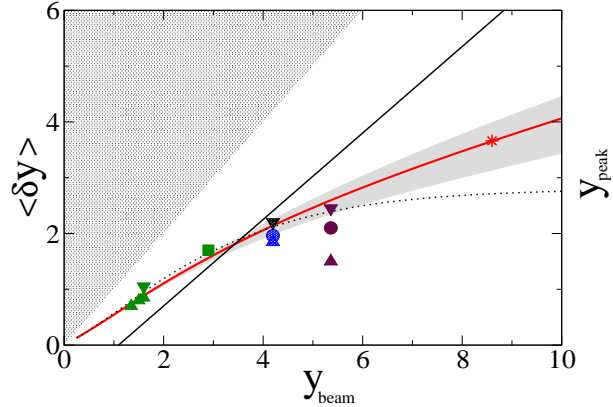}
\end{center}
\caption{Mean rapidity shift of net baryons as a function of
beam rapidity $y_{beam}$ in the model in
$^{88}$. Solid and dashed lines correspond to different options for the behavior of the saturation scale. The solid straight
line shows the prediction for the position of the fragmentation
peaks. The star at $y_{beam}\simeq 8.5$ is the prediction for central PbPb collisions at the LHC.
 Experimental data can be found in
$^{89}$. Figure taken from
$^{88}$.}
\label{fig8b}
\end{figure}

The general prediction for the net proton number (p $- \bar{\rm p}$) at mid-rapidity is smaller than 4 for central PbPb collisions at the LHC, to be compared with the value $5\div 8$ in central AuAu at top RHIC energy \cite{Bearden:2003hx}. This is so in models of different kinds, ranging from approaches with the baryon junction mechanism or other baryon transport effects, as HIJING/B$\bar{\rm B}$, DPMJET or the EPOS model, hydrodynamical models like Eskola et al., the diffusion equation of Wolschin et al., the statistical model of Rafelski et al., see Subsection \ref{hadroch}, the saturation model in \cite{MehtarTani:2008qg} (see Fig. \ref{fig8b} for the energy evolution of the mean rapidity shift $\langle \delta y\rangle=|\langle y _{net\ baryon}\rangle-y_{beam}|$  in this approach), or the model \cite{AlvarezMuniz:2009kh}
based on string formation with momentum fractions taken from parton distribution functions and string fragmentation through the Schwinger mechanism.

Note that I have focused on predictions for charged multiplicities at mid-rapidity well covered by all heavy-ion detectors at the LHC and, thus, a true first-day observable. Predictions for the total charged multiplicity in all phase space also exist, see e.g. Busza in \cite{Abreu:2007kv} (or even 
\cite{Gubser:2009sx} for predictions in strongly coupled super-symmetric Yang-Mills theories computed through the use of the AdS/CFT correspondence).

\subsection{Collective flow}
\label{flow}

Here I turn to collective flow - another first-day observable. Specifically, I will discuss elliptic flow at mid-rapidity, both integrated over and as a function of transverse momentum. In this Subsection I will concentrate on elliptic flow for hadrons, while that for photons will be discussed in the corresponding Subsection \ref{photons}.

Concerning the $p_T$-integrated $v_2$, the expectation both from data-driven estimations and from more involved, model-dependent calculations, is for it to increase when going from RHIC to the LHC. Such increase, as discussed in Section \ref{qualitative} and in \cite{Borghini:2007ub}, looks stronger in data-driven extrapolations than in hydrodynamical models. Numerical results \cite{Niemi:2008ta,Kestin:2008bh} within ideal hydrodynamics indicate increases in the transverse momentum integrated $v_2$ at $b\sim 7.5$ fm from RHIC to the LHC,  ranging from $\sim 15$ \% \cite{Kestin:2008bh} for charged multiplicities at mid-rapidity around 1200, to $\sim 40\div 60$ \% in \cite{Niemi:2008ta} for twice this multiplicity in central PbPb. Results in viscous hydro \cite{viscous} yield increases $\sim 10$ \% for a charged multiplicity of 1800. Let me note that by viscous hydro I mean calculations considering shear viscosity but neglecting bulk viscosity. Studies on the impact of the latter are at the very beginning \cite{Torrieri:2009ms,Denicol:2009am,Monnai:2009ad}.

\begin{figure}[htb]
\begin{center}
\includegraphics[width=8.cm]{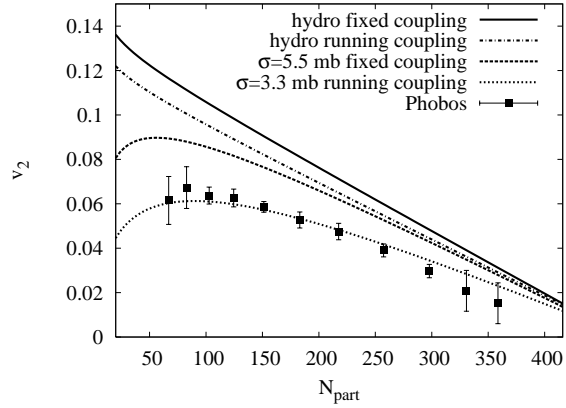}
\end{center}
\caption{$v_2$ versus $N_{part}$ at RHIC (lower line) and at the LHC (upper lines), for different values of the parameters in Eq. (\ref{knudsen}). The normalization is not determined in the model. Experimental data are from PHOBOS
$^{96}$. Figure taken from $^{35}$.}
\label{fig9}
\end{figure}

Concerning the difference between ideal hydrodynamics and non-ideal scenarios, a consequence of a larger density of the medium is that the ideal hydrodynamical behavior will be better fulfilled at the LHC than at RHIC \cite{Drescher:2007cd}, as a higher density implies a smaller mean free path and a faster thermalization. Thus the behavior of the medium at the LHC is expected to be closer to that of an ideal fluid than at RHIC, if one assumes that the medium at RHIC shows only partial thermalization i.e. that the mean free path is not yet much smaller than the system size. This is illustrated in Fig. \ref{fig9}, where
\begin{equation}
v_2\propto \frac{\epsilon_x}{1+K/0.7},\,\ \ K^{-1}=\frac{\sigma}{S_{over}}\frac{dN_{tot}}{dy}\frac{1}{\sqrt{3}}\,,
\label{knudsen}
\end{equation}
with $K$ the Knudsen number, $S_{over}$ the overlap area, $\sigma$ the typical cross section between constituents of the medium and $1/\sqrt{3}$ comes from the speed of sound of an ideal ultra-relativistic gas. The ideal hydrodynamical limit is reached for $K\to 0$, and $S_{over}$, $\epsilon_x$ and $dN_{ch}/dy$ are provided through initial conditions, see Drescher et al. in \cite{Abreu:2007kv} for details.

\begin{figure}[htb]
\begin{center}
\includegraphics[width=6cm,angle=270]{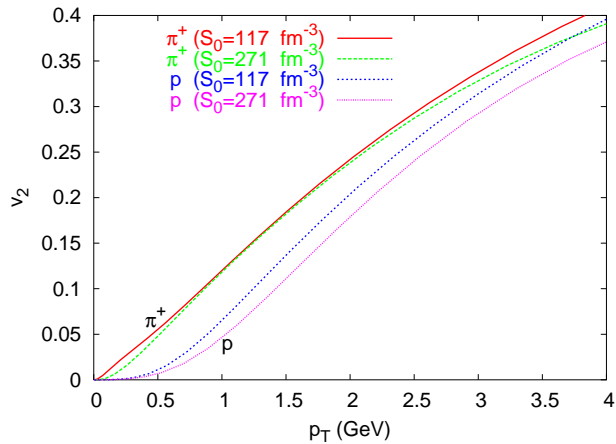}
\end{center}
\caption{$v_2$ versus $p_T$ from ideal hydrodynamical calculations for different entropy densities corresponding to RHIC and LHC situations, for positive pions and protons, for $b=7$ fm. Figure from 
$^{67}$.}
\label{fig10}
\end{figure}

Now I turn to $v_2(p_T)$. First I will discuss the expectations within the framework of hydrodynamical models. From the matching of pQCD with hydrodynamical spectra, see e.g. Eskola et al. in \cite{Abreu:2007kv,Niemi:2008ta}, hydrodynamical calculations are expected to be valid up to larger transverse momentum, $p_T<3\div 4$ GeV, at the LHC than at RHIC. In ideal hydrodynamical calculations, a very similar $v_2(p_T)$ at RHIC and at the LHC is expected for pions at $p_T<2$ GeV, while the $v_2(p_T)$ for protons is expected smaller (as illustrated in Fig. \ref{fig10}), 
see Bluhm et al., Kestin et al., Eskola et al. \cite{Abreu:2007kv} and \cite{Niemi:2008ta,Kestin:2008bh,Chojnacki:2007rq} for computations corresponding to initial charged multiplicities which range, in central PbPb, from $\sim 1200$ in Kestin et al. to $\sim 2300$ in Eskola et al. Note that even a decrease of $v_2(p_T)$ does not necessarily imply a decrease in the $p_T$-integrated $v_2$ - actually all models show the opposite behavior -, as it has to be convoluted which a $p_T$-spectrum which is harder at the LHC than at RHIC. If fact the small increase of $v_2(p_T)$ shown in \cite{Niemi:2008ta}, less than 10 \%, translates  into a much larger increase of $p_T$-integrated $v_2$ than other predictions, as discussed above. On the other hand, the available calculation within viscous hydrodynamics \cite{Chaudhuri:2008je} shows a decrease.

The difference between different predictions comes not only from the different initial conditions\footnote{For example, the importance of the initial conditions for the application of hydrodynamical calculations  has been recently discussed in \cite{Petersen:2009vx}.} (as illustrated in \cite{Niemi:2008ta}), but  also from details of the calculations (made in either two \cite{Niemi:2008ta} or three (all others) spatial dimensions), the equation of state in both the confined and deconfined phases and its matching (see e.g. Bluhm et al. for a study of the influence of the equation of state), the treatment of the hadronic phase, the hadronization procedure (e.g. a statistical method in \cite{Chojnacki:2007rq}),$\dots$

Now I focus on other approaches. The Monte Carlo simulators AMPT \cite{Lin:2004en} and EPOS \cite{Drescher:2000ha} give results \cite{Abreu:2007kv} at the LHC which are very close to those at RHIC for pions, while the former shows a decrease of $v_2(p_T)$ for protons. The simulator in \cite{Humanic:2008nt} gives sizably smaller $v_2(p_T)$ at the LHC than at RHIC in spite of the fact that this model contains hadronic rescattering which, naively thinking, should increase $v_2(p_T)$ due to the larger densities at the LHC.

\begin{figure}[htb]
\begin{center}
\includegraphics[width=8cm]{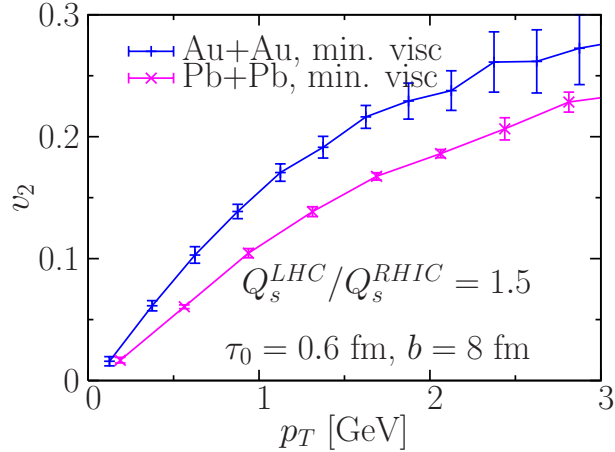}\end{center}
\caption{$v_2$ versus $p_T$ from the MPC parton cascade of Molnar, for RHIC and LHC situations for $b=8$ fm.
Figure taken from
$^{35}$.}
\label{fig11}
\end{figure}

The parton cascade MPC by Molnar \cite{Molnar:2001ux,Abreu:2007kv}, which considers $2\leftrightarrow 2$ partonic collisions, provides interesting information on the relation between viscous hydrodynamical calculations and transport results. The author chooses the parameters in the transport equation so as the shear viscosity is fixed to be $\eta/s\leq (4 \pi)^{-1}$ (the equality corresponds to the so-called minimal viscosity bound \cite{Policastro:2001yc}). The results show a decrease in $v_2(p_T)$ when going from RHIC to the LHC, see Fig. \ref{fig11},
with all dependence on multiplicity encoded in the relation of the saturation scale with multiplicity.

Finally, the absorption model of Capella et al. \cite{Abreu:2007kv} considers the absorption of the produced particles moving along paths in the medium, with increasing absorption with increasing length of traversed matter. Such model predicts a strong increase when going from RHIC to the LHC, due to the increasing medium density (as indicated in the previous Subsection, this model predicts a charged multiplicity at mid-pseudorapidity $\sim 1800$ for $N_{part}=350$).

Therefore, data on both $p_T$-integrated $v_2$ and on $v_2(p_T)$, together with the measurements of the multiplicity, may help to verify whether the origin of the elliptic flow is a collective expansion (and thus thermalization or isotropization has been achieved and hydrodynamical models are applicable) or thermalization has been achieved only partially. In the latter case, a sizable increase is expected in $v_2(p_T)$ for $p_T< 2$ GeV, while in the former a decrease or a mild increase generically results. A sizable decrease would favor some viscosity effects, though the issue of the dependence on the initial conditions for hydrodynamical evolution should be settled for firm conclusions to be extracted. Note that, while finite temperature pQCD calculations (valid for $T\gg T_{dec}$, see e.g. \cite{Arnold:2000dr} and references therein) indicate that the shear viscosity to entropy ratio should increase with temperature,
\begin{equation}
\frac{\eta}{s}\propto \frac{1}{\alpha^2_s(T)\ln\alpha_s^{-1}(T)}\,, 
\label{etas}
\end{equation}
with $\alpha_s(T)$ decreasing with increasing temperature, the behavior of this and other transport coefficients (like e.g. the bulk viscosity) for realistic temperatures close to the deconfinement temperature $T_{dec}$ is not clear yet.

\subsection{Hadrochemistry at low transverse momentum}
\label{hadroch}

\begin{figure}[htb]
\begin{center}
\includegraphics[width=7.5cm]{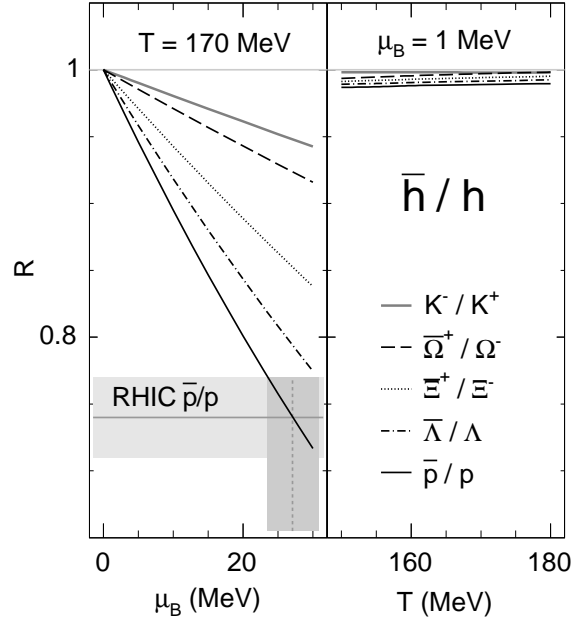}
\end{center}
\caption{Antiparticle/particle ratios $R$ as a function of $\mu_B$ for $T = 170$ MeV (left) and as a
function of $T$ for $\mu_B = 1$ MeV (right). The horizontal line at 1 is meant to guide the eye.
The $\bar{\rm p}$/p ratio (averaged over the data of the 4 RHIC experiments at $\sqrt{s_{NN}} = 200$ GeV) is
displayed (gray horizontal line) together with its statistical error (gray band). As illustrated,
$\mu_B \approx 27$ MeV (dashed line) can be read off the Figure directly within the given accuracy
(vertical gray band).
Figure taken from
$^{102}$.}
\label{fig12}
\end{figure}

Hadrochemistry is a key observable to disentangle the mechanism of particle production. Statistical models constitute the most popular framework to discuss it. Within statistical models, predictions are done normally in the grand-canonical ensemble valid for large systems\footnote{In the grand-canonical ensemble it is possible to predict the particle ratios without any reference to the total multiplicity i.e. to the total volume of the system. This case is one of the very few in which predictions can be done in absence of such information.}. The relevant parameters, fireball temperature and baryochemical potential $\mu_B$\footnote{The strangeness suppression factor used at lower energies within the grand-canonical ensemble is 1 at RHIC energies and this value is assumed for the LHC. } are extrapolated from the results extracted at lower energies. The results obtained by different groups (Andronic et al. and Kraus et al. in \cite{Abreu:2007kv} and  \cite{Kraus:2009xa})  are shown in Fig. \ref{fig12}  and Table \ref{table4}. It can be observed that even the $\bar{\rm p}$/p ratio takes values very close to 1 in the expected range of $T\simeq 160\div 175$ MeV and $\mu_B\simeq 0\div 6$ MeV. Concerning these extrapolations, $\bar{\rm p}$/p is particularly sensitive to the value of $\mu_B$, while the ratios of multi-strange baryons to non-strange particles are particularly sensitive to the temperature, see e.g. \cite{Kraus:2009xa}.

\begin{table}[hbt]
\caption{Predictions of the thermal model for hadron ratios in central
Pb+Pb collisions at LHC, for $\mu_B=0.8$ MeV and $T=161$ MeV. The numbers in parentheses represent the error
in the last digit(s) of the calculated ratios. Table taken from Andronic et al. in $^{35}$.}
\label{table4}
\begin{center}
\begin{tabular}{|c|c|c|c|c|c|}\hline
$\pi^-/\pi^+$ & $K^-/K^+$ & $\bar{p}/p$ & $\bar{\Lambda}/\Lambda$ &
$\bar{\Xi}/\Xi$ & $\bar{\Omega}/\Omega$ \\ \hline
1.001(0) & 0.993(4) & 0.948$^{-0.013}_{+0.008}$ & 0.997$^{-0.011}_{+0.004}$ &
1.005$^{-0.007}_{+0.001}$ & 1.013(4) \\ \hline
$p/\pi^+$ & $K^+/\pi^+$ & $K^-/\pi^-$ & $\Lambda/\pi^-$ &
$\Xi^-/\pi^-$ & $\Omega^-/\pi^-$ \\ \hline
0.074(6) & 0.180(0) & 0.179(1) & 0.040(4) & 0.0058(6) & 0.00101(15) \\ \hline
\end{tabular}
\end{center}
\end{table}

\begin{figure}
\begin{center}
\includegraphics[width=7cm]{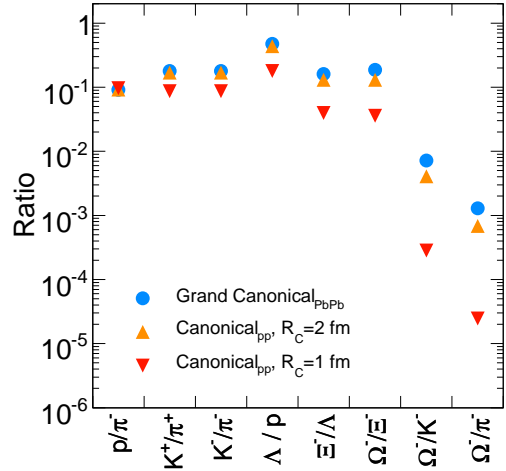}
\end{center}
\caption{Predictions for various particle ratios using different values for the cluster
size $R_C$. Figure taken from
$^{102}$.}
\label{fig13}
\end{figure}

For smaller systems, e.g. smaller nuclear sizes or peripheral collisions, the grand-canonical ensemble is not expected to provide a good description of particle production. While the traditional way of addressing the question for strangeness production is the use of a strangeness suppression factor - thus assuming a chemically non-equilibrated system, the proposal in \cite{Kraus:2009xa,Kraus:2007hf} is to keep strangeness conservation in smaller volumes, called clusters. The effects on particle ratios of the consideration of clusters of different sizes can be seen in Fig. \ref{fig13} and in \cite{Kraus:2007hf}.

Another proposal within statistical models is the non-equilibrium scenario of Rafelski et al. in \cite{Rafelski:2005jc}. This scenario shows a sensitivity to the total multiplicity in the central region and predicts, with respect to the chemically equilibrated one, an enhancement of multi-strange and single strange resonance yields, and a decrease of non-strange resonances (the prediction for net-baryon yields has been commented in the Subsection \ref{multi}). Results can be seen in Table \ref{table5} \cite{Abreu:2007kv}.

\begin{table}
\caption{Predictions for particle yields at the LHC for different scenarios by Rafelski et al.: chemically equilibrated (second column), chemically non-equilibrated but with the same freeze-out temperature as the previous one (third column), and chemically non-equilibrated with a different temperature but for the same multiplicity as the previous non-equilibrated case. The '*' refers to input values, while the subindex $vis$ refers to values observable in the ALICE TPC ($|\eta|<0.9$), $S$ denotes the entropy, $V$ the volume and $b$ the baryon number. The slashes are used to give the particle yields with/without weak decays. Table taken from Rafelski et al. in $^{35}$.}
\label{table5}
\begin{center}
\begin{tabular}{|c| c | c | c |  }
\hline
$T$ [MeV]        &$140^*$& $140^*$& $162^*$    \\
$dV/dy$[ fm$^3$]        &2036  &4187  &6200$^*$    \\
$dS/dy$          &7517  & 15262  &  18021   \\
$dN_{\rm ch}/dy|_{y=0}$ &$1150^*$    &$2351 $      &$2430 $     \\
$dN_{\rm ch}^{\rm vis}/dy$&$1351 $  &$2797^* $      &$ 2797$    \\
\hline
$p$   &$25/45 $  &$49/95$    &$66/104 $     \\
$ b-\bar b $  &2.6&5.3&6.1 \\
$(b+\bar b)/h^-$  &0.335&0.345&0.363 \\
$0.1\cdot\pi^\pm$            &49/67  &99/126  &103/126    \\
${\rm K}^\pm$        &94   &207   &175       \\
$\phi $            &14  &33    &23       \\                                
$\Lambda$          &19/28 &41/62    &37/50         \\
$\Xi^-$             &4  &9.5   &5.8   \\                     
$\Omega^-$         &0.82 &2.08   &0.98     \\    
\hline
$\Delta^{0},\,\Delta^{++} $
                 &4.7   &9.3   &13.7        \\
$K^*_0(892)$ & 22&48&52 \\
$\eta$ &62&136&127 \\
$\eta'$ & 5.2 &11.8&11.5\\
$\rho$ &36&73 &113\\
$\omega$ &32& 64& 104\\
$f_0$ & 2.7&5.5 & 9.7\\
\hline
K$^+ /\pi^+_{\rm vis}$   & 0.165 &0.176  &  0.148 \\
$\Xi^-/\Lambda_{\rm vis}$ &0.145 & 0.153& 0.116\\
$\Lambda(1520)/\Lambda_{\rm vis}$ &0.043 &0.042 &0.060 \\
$\Xi(1530)^0/\Xi^-$ &0.33 &0.33 &0.36 \\
$\phi/{\rm K}^+$ &0.15 & 0.16&0.13 \\
$K^*_0(892)/K^-$ & 0.236&0.234 &0.301 \\
\hline
\end{tabular}
\end{center}
\end{table}

The different scenarios for statistical production lead to marked differences in particle yields in heavy-flavor production, which will be commented on in Subsection \ref{heavy}.

\begin{figure}[htb]
\begin{center}
\includegraphics[width=6.5cm,height=5cm]{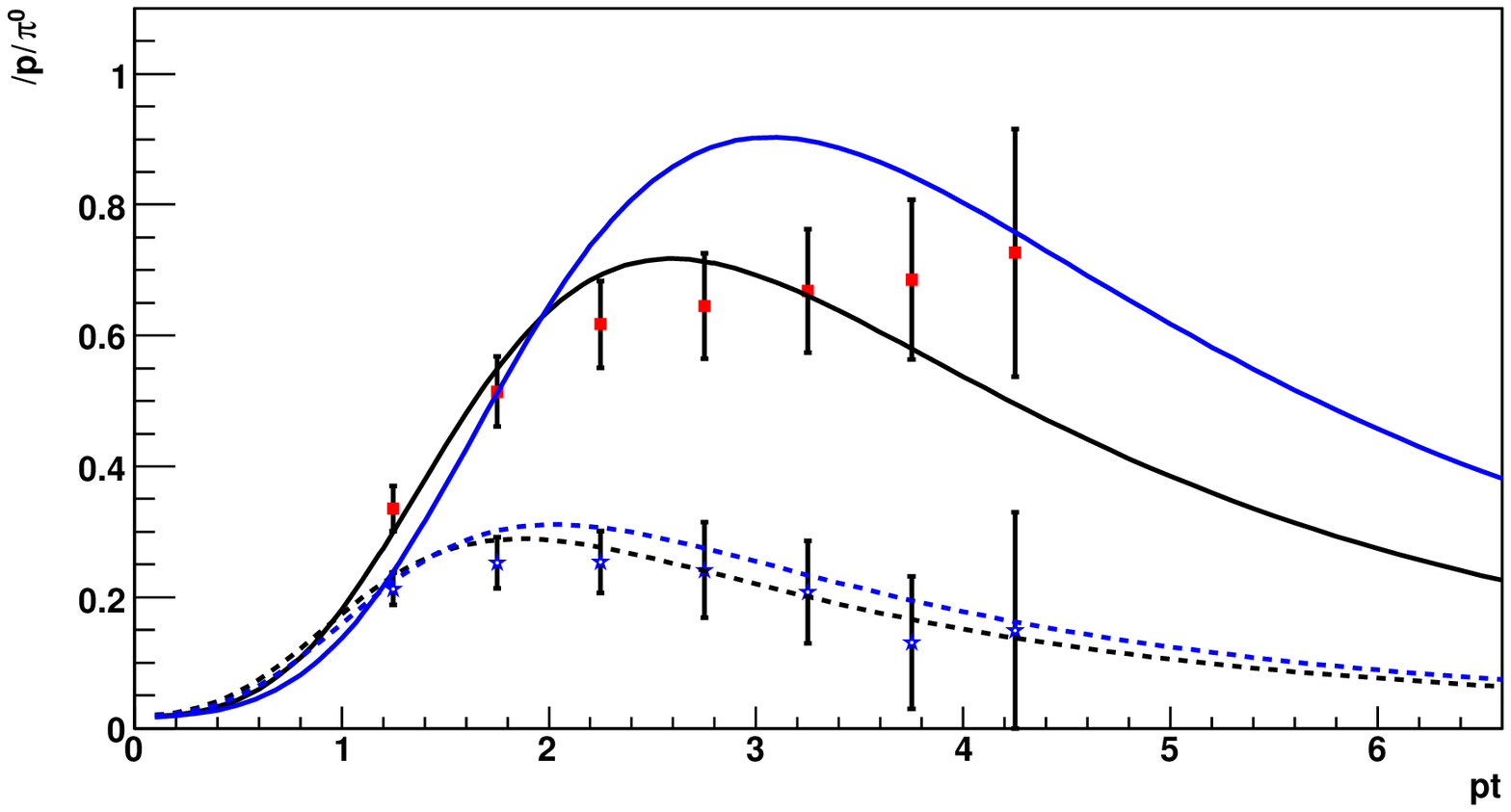}\hfill \includegraphics[width=6cm]{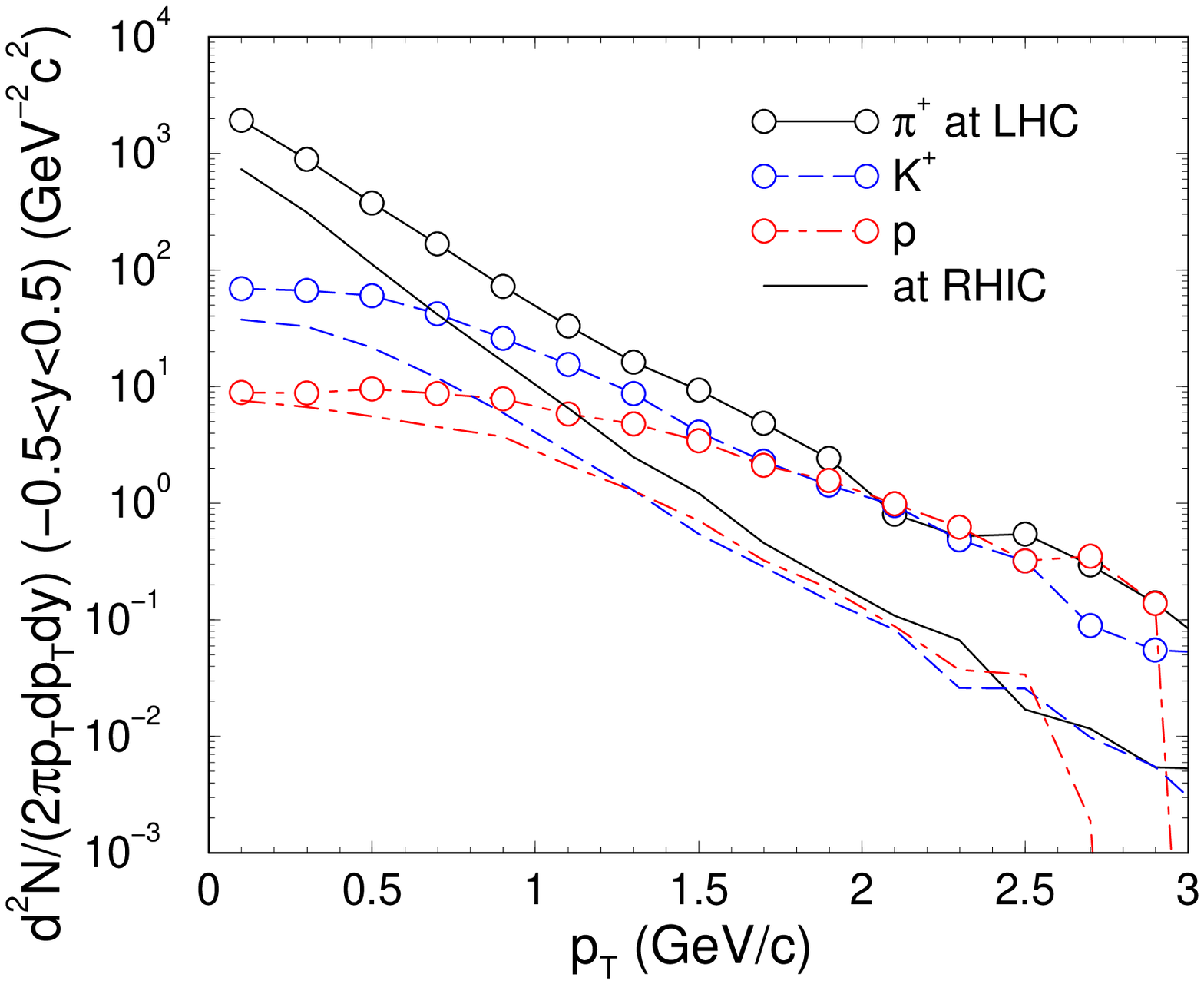}
\end{center}
\caption{Left: Ratio $\bar{\rm p}/\pi^0$ at RHIC (black)  and at the LHC (blue) from the percolation model of Cunqueiro et al., for central (upper lines) and peripheral (lower lines) collisions. 
Right: Transverse momentum spectra for various particle species in AMPT, Chen et al., at RHIC (lines) and at the LHC (symbols joined by lines), for $b< 3$ fm. Figures taken from
$^{35}$.}
\label{fig14}
\end{figure}

Now I turn to non-statistical models. These models mainly focus on the baryon-to-meson ratios, whose large values measured at intermediate $p_T\sim 3$ GeV at RHIC\cite{rhic,Back:2004je,Arsene:2004fa,Adams:2005dq}, much larger than those measured in nucleon-nucleon collisions, constitute the (anti)-baryon anomaly which has triggered many new ideas.

\begin{figure}[htb]
\begin{center}
\includegraphics[width=8cm]{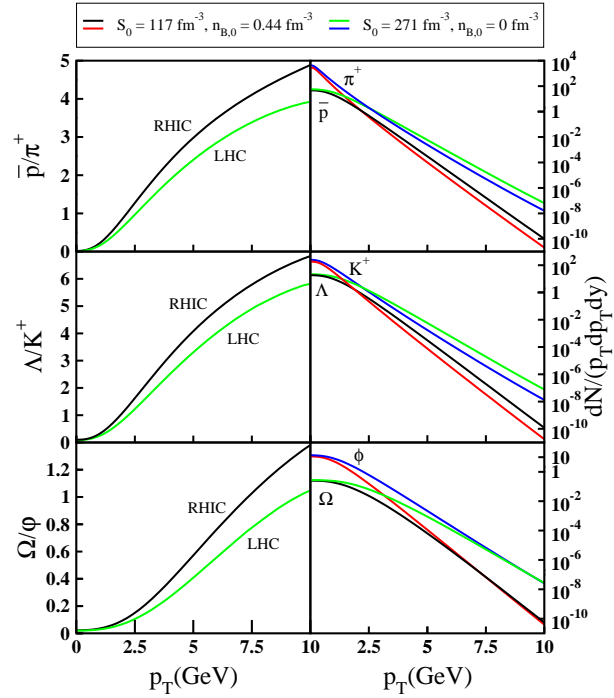}
\end{center}
\caption{Antiparticle/particle ratios (left) and transverse momentum spectra (right) for different particle species, in ideal hydrodynamics, for two scenarios corresponding to RHIC and the LHC.
Figure taken from
$^{67}$.}
\label{fig15}
\end{figure}

There are several available predictions: by ideal hydrodynamical models (Kestin et al. \cite{Kestin:2008bh,Abreu:2007kv}), by recombination models as implemented in AMPT (Chen et al. \cite{Abreu:2007kv}) and by models which consider a higher string tension like percolation models (Cunqueiro et al. \cite{Abreu:2007kv}) and HIJING/B$\bar{\rm B}$ by Topor Pop et al. \cite{Abreu:2007kv}, see Figs. \ref{fig14} and \ref{fig15}. In general, hydrodynamical and recombination models predict larger baryon-to-meson ratios than models which consider an increased string tension in nucleus-nucleus collisions with respect to nucleon-nucleon.

Let me comment that the percolation model of Cunqueiro et al.  \cite{Abreu:2007kv} predicts a Cronin effect - a nuclear modification factor above one - for protons at mid-rapidity in central PbPb collisions. This is at variance with most extrapolations or theoretical expectations which predicts a disappearance of the Cronin effect with increasing collision energy, see e.g. \cite{Albacete:2003iq,JalilianMarian:2005jf}.

While every non-statistical prediction is linked to a multiplicity scenario, it is not so easy to see the effect of a variation of multiplicity on the results for hadrochemistry of different models. In principle, hydrodynamical and recombination models would benefit from a larger multiplicity due respectively  to the larger applicability of hydrodynamics (larger duration of the hydrodynamical phase) and due the larger density in recombination models, with less restrictions due to finite density and volume. For models which consider a higher string tension (in nucleus-nucleus than in pp collisions), an increase in the string tension implies a reduction of multiplicity and an increase in baryon/strangeness production. Therefore, in these models an increase of multiplicity originating from a smaller string tension would imply a reduction of the effects characteristic of the enhanced string tension scenario. Obviously these most crude expectations can only be substantiated by further calculations for different multiplicity scenarios within the models.

Finally, let me mention that the possibility of producing charmed exotic states in heavy-ion collisions at the LHC has also been addressed, see Lee et al. in \cite{Abreu:2007kv}.

\subsection{Correlations}
\label{correl}

Now I turn to correlations. First, I will indicate the predictions for the Hanbury-Brown-Twiss (HBT) interferometry (see the recent review \cite{Lisa:2008gf}).

\begin{figure}[htb]
\begin{center}
\includegraphics[width=6cm]{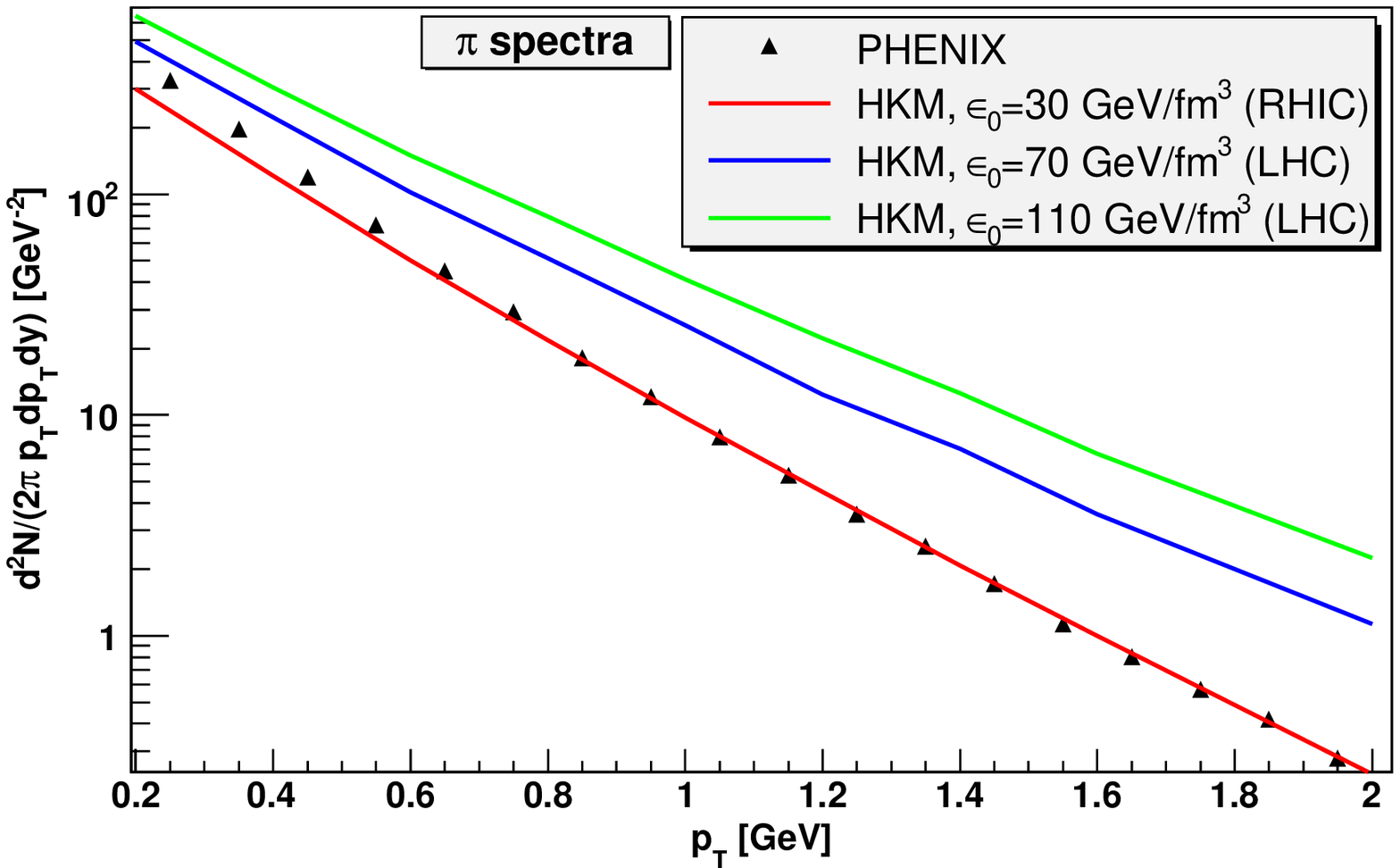}\hfill \includegraphics[width=6cm]{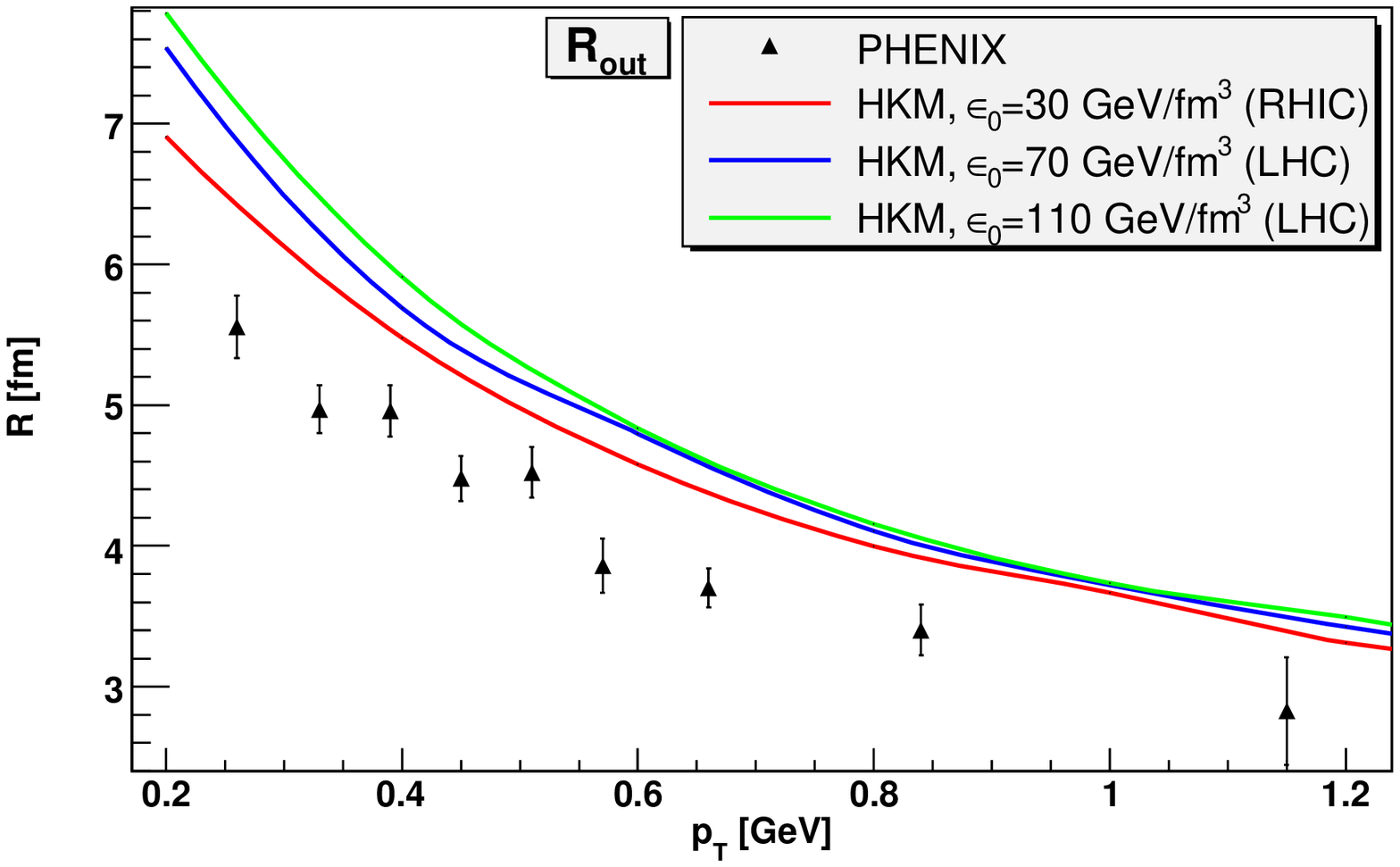}
\includegraphics[width=6cm]{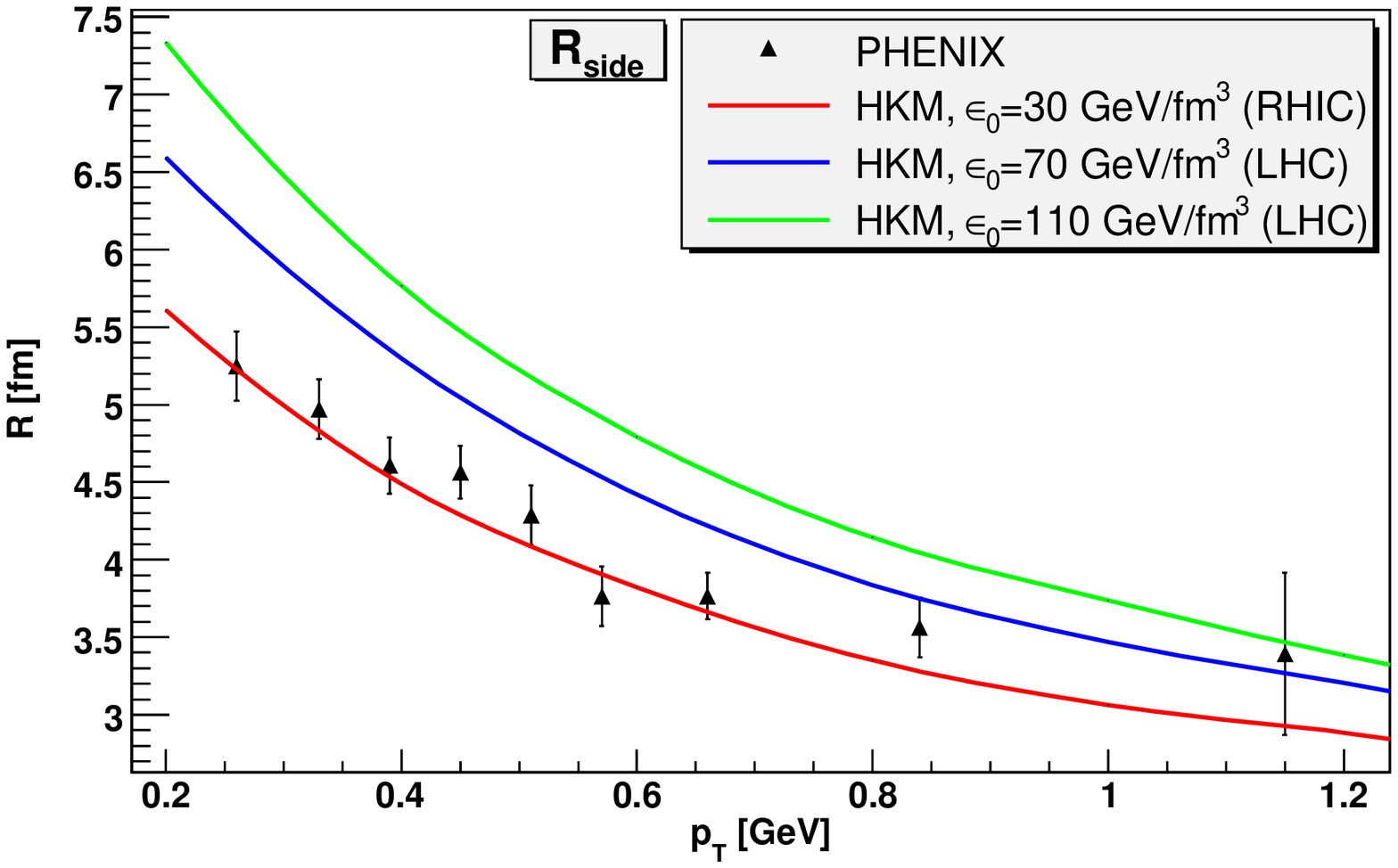}\hfill \includegraphics[width=6cm]{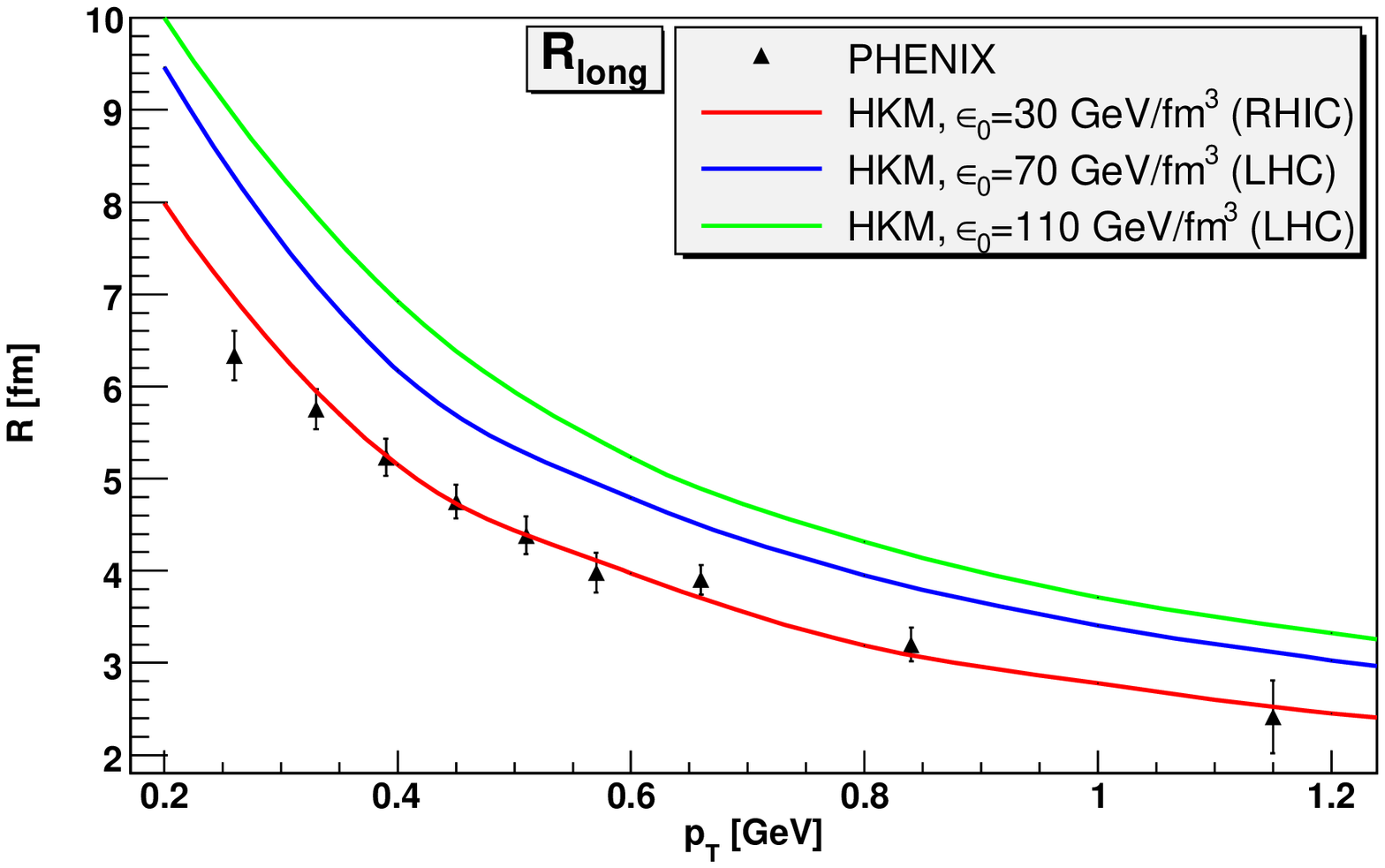}
\end{center}
\caption{Transverse momentum spectrum of pions and behavior of the pion HBT radii from Sinyukov et al., for different scenarios of initial energy densities.
Figure taken from
$^{35}$.}
\label{fig16}
\end{figure}

The generic expectation \cite{Borghini:2007ub} is that all HBT radii $R_{out}$, $R_{side}$ and $R_{long}$ will increase when going from RHIC to the LHC. This is substantiated by several calculations using ideal hydrodynamics, like Frodermann et al. (with the transition out-of-plane to in-plane shape clearly reflecting in the radii), or in the hydro-kinetic approach of Karpenko et al. and Sinyukov et al. \cite{Abreu:2007kv} (see Fig. \ref{fig16}). Both these calculations consider ideal hydrodynamics but different hadronization procedures, the latter intending to consider some out-of-equilibrium features. This is also the case in the calculations in the AMPT model by Chen et al. in \cite{Abreu:2007kv} and the hydro+statistical model \cite{Chojnacki:2007rq,Kisiel:2008ws} which combines a hydrodynamical behavior with hadronization through the statistical method. The corresponding results can be seen in Table \ref{table6}.

\begin{table}[hbt]
\caption{Predictions for the HBT radii at RHIC/LHC from two different models: Chen et al. in
$^{35}$ for $b=0$ and $0.3< k_T < 1.5$ GeV, and the hydrodynamical plus statistical model in
$^{97}$, for two pion multiplicity scenarios at the LHC, 558 and 1193.
}
\label{table6}
\begin{center}
\begin{tabular}{|c|c|c|}\hline
RHIC/LHC &  Chen et al. & $^{97}$
 \\ \hline
$R_{out}$ &  3.60/4.23 & 5.4/6.0$\div$6.5 \\ \hline
$R_{side}$ & 3.52/4.70 & 4.3/5.3$\div$6.3 \\ \hline
$R_{long}$ & 3.23/4.86 & 6.1/7.6$\div$8.6\\ \hline
\end{tabular}
\end{center}
\end{table}

Within ideal hydrodynamics, the features of the HBT radii which are not in agreement with RHIC data - too large $R_{long}$ and $R_{out}/R_{side}$, and the behavior of $R_{out}$ and $R_{side}$ with the relative momentum of the pair - will also  be present at the LHC. A piece of knowledge still missing in this context is the effect of viscosity on the HBT radii \cite{Muronga:2004sf} and the effects of pre-thermalization dynamics \cite{Pratt:2008qv}.

On the other hand, partial thermalization, which implies a departure of the ideal hydrodynamical behavior \cite{Drescher:2007cd}, may also help to reduce the ratio $R_{out}/R_{side}$ \cite{Gombeaud:2009fk} in agreement with RHIC data. If this is the case, then the expectation that the situation at the LHC will be closer to ideal hydrodynamics will reflect in an increase of this ratio when going from RHIC to the LHC. Besides, the inclusion of minijets modifies the behavior of the HBT radii and of the chaoticity parameter with respect to pure hydrodynamical predictions \cite{Lokhtin:2009be}.

\begin{figure}[htb]
\begin{center}
\centerline{}
\vskip 1.5cm
\includegraphics[width=8cm]{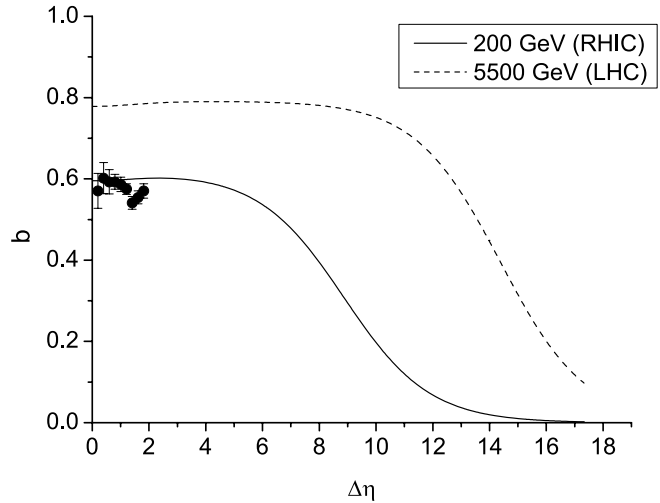}
\end{center}
\caption{Forward-backward correlation correlation strength for different values of the rapidity gap  $\Delta \eta=\eta_F-\eta_B$between the forward and backward windows at RHIC and at the LHC, from a two-step scenario. Preliminary data are from 
$^{114}$.
Figure courtesy of the authors of
$^{113}$.}
\label{fig17}
\end{figure}

Correlations can also be useful to clarify the mechanism of particle production. Correlations in rapidity were proposed long ago, see e.g. \cite{Capella:1978rg}, as a measurement sensitive to the distribution of particle sources. More specifically, defining two rapidity intervals denoted by $F$ and $B$ with multiplicities $n_F$ and $n_B$ respectively, the correlation strength $b$ (sometimes denoted as $\sigma^2_{FB}$) is defined as
\begin{equation}
\langle n_F\rangle (n_B)=a+bn_B,\ \ b=\frac{D^2_{FB}}{D^2_{BB}}=\frac{\langle n_Fn_B\rangle-\langle n_F\rangle\langle n_B\rangle}{\langle n_B^2\rangle-\langle n_B\rangle^2}\,.
\label{bparam}
\end{equation}
Predictions exist \cite{Brogueira:2007ub} for such quantity at the LHC, see Fig. \ref{fig17} and Dias de Deus et al. \cite{Abreu:2007kv}, in the framework of a two-step scenario which considers first the formation and interaction of particle emitters (coherent along large rapidity regions) which subsequently decay into the observed particles (see also \cite{Brogueira:2009nj}).

Many explanations try to address the existence of such long range correlations, see e.g. \cite{Armesto:2006bv} and references therein. It has been linked to the so-called ridge phenomenon measured at RHIC (see e.g. \cite{vanLeeuwen:2008pn}): the existence of a two-particle correlation narrow in azimuth but extended along several units of pseudorapidity in AuAu collisions. While a quantitative description is missing, present qualitative explanations are based (e.g. \cite{Dumitru:2008wn}) on the coupling of particle production correlated along a long rapidity range to the collective flow. An extended correlation as predicted by e.g. the two-step scenario mentioned above, together with the fact that the collective flow is expected to last longer at the LHC than at RHIC, should make this phenomenon more prominent at the LHC.

\subsection{Fluctuations}
\label{fluctu}

\begin{figure}[htb]
\begin{center}
\includegraphics[width=10cm]{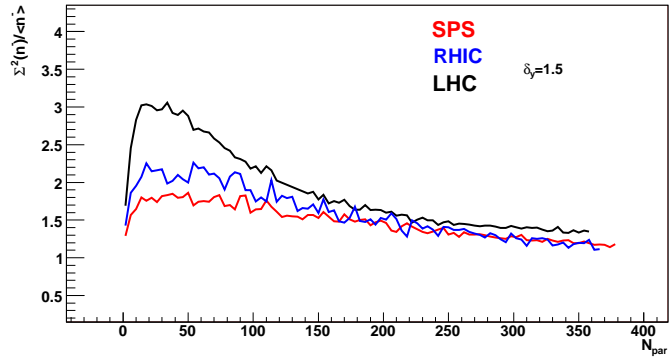}
\end{center}
\caption{Scaled variance of negative particles versus the number of participants in PbPb at top SPS, AuAu at top RHIC, and PbPb at LHC energies, from bottom to top, in the percolation model of Cunqueiro et al.
Figure taken from
$^{35}$.}
\label{fig18}
\end{figure}

Many types of fluctuations have been proposed and analyzed as possible signatures of a phase transition in ultra-relativistic heavy-ion collisions: in multiplicity, charge, baryon number, transverse momentum,... The results at SPS and RHIC energies are not clear - the evidence of a non-statistical or non-trivial origin of fluctuations at SPS and RHIC is still under debate -, which has prevented predictions for the LHC. Available predictions are for the multiplicity fluctuations (Cunqueiro et al. in \cite{Abreu:2007kv}) quantified through the scaled variance of negative particles,
\begin{equation}
\frac{\Sigma^2(n^-)}{\langle n^-\rangle}=\frac{\langle (n^-)^2\rangle-\langle n^-\rangle^2}{\langle n^-\rangle}
\label{scvar}
\end{equation}
measured in a given rapidity interval $\delta y$. The predictions, shown in Fig. \ref{fig18}, indicate a non-monotonic behavior at some given number of participants (a change of slope at some $N_{part}$ smaller with increasing energy) which is, in the framework of this model, indicative of the existence of a percolation phase transition. Note that in this model, as in others, multiplicity fluctuations are linked to those in transverse momentum.

On the other hand, Torrieri in \cite{Abreu:2007kv} proposes the use of fluctuations of particle ratios e.g. of kaons and pions,  as measurements sensitive to the mechanism of particle dynamics: the fully equilibrated scenario of the grand-canonical ensemble should show a different behavior from the other ensembles or non-statistical scenarios.

\section{Hard and electromagnetic probes}
\label{hard}

In this Section I review the available predictions for those probes of the medium whose yields can be, in the absence of a medium,  computed through perturbative techniques - hard probes. These are high transverse momentum particle production, heavy-quark and quarkonium production, and photon and dilepton production at large momentum or mass\footnote{Concerning the total charm and charmonium cross sections and their production at low momentum, doubts exist on whether they can be reliably computed in pQCD - either at fixed order or via resummation techniques - or not, see e.g. \cite{Vogt:2007aw} for a discussion on the uncertainties for charm.}. For photons and dileptons, i.e. electromagnetic probes, I will also consider their production at low momentum or mass, though their calculation lies, in principle, beyond the reach of perturbative techniques. Extensive studies on hard probes at the LHC can be found in \cite{Accardi:2004be} (pA collisions and benchmark studies), \cite{Accardi:2004gp} (particle production at high transverse momentum and jets), \cite{Bedjidian:2004gd} (heavy quarks and quarkonia) and \cite{Arleo:2004gn} (photons and dileptons).

\subsection{Particle production at large transverse momentum and jets}
\label{highpt}

The suppression of the yield of hadrons at large transverse momentum measured at RHIC \cite{rhic,Back:2004je,Arsene:2004fa,Adams:2005dq} - the jet quenching phenomenon - is one of the most important subjects of current research and debate in the field. It is most commonly quantified through the nuclear modification factor (\ref{raadef}) and usually attributed to the energy loss of the leading parton which fragments onto the measured hadron, see e.g. the standard reviews in \cite{qgp3,rel} and the more recent one \cite{d'Enterria:2009am} (more specific information about radiative energy loss which is the reference explanation can be found in \cite{CasalderreySolana:2007zz,Majumder:2007iu}, and about studies of the energy loss in strongly coupled super-symmetric Yang-Mills plasmas through the AdS/CFT correspondence in \cite{Gubser:2009sn}).

One comment on the definition of the region that I call of large transverse momentum is in order. At RHIC, such region - usually taken at $p_T>7\div 10$ GeV -, is determined by that in which the characteristics of fragmentation become those in absence of any medium, i.e. in pp, and where fragmentation or hadronization is expected to be described by standard pQCD techniques so no collectivity in hadronization (e.g. recombination) seems to be required. More specifically, the baryon-to-meson anomaly disappears, the nuclear modification factor for different species becomes similar, etc. Note that this definition is not free from ambiguities as new effects included in the models (for example, in the transition from recombination to perturbative fragmentation) may shift it. At the LHC, due to the larger densities and larger expected collectivity, such region may start at larger $p_T$ than at RHIC, a question which only data will answer.

\begin{figure}[htb]
\begin{center}
\includegraphics[width=13.5cm,height=13.5cm]{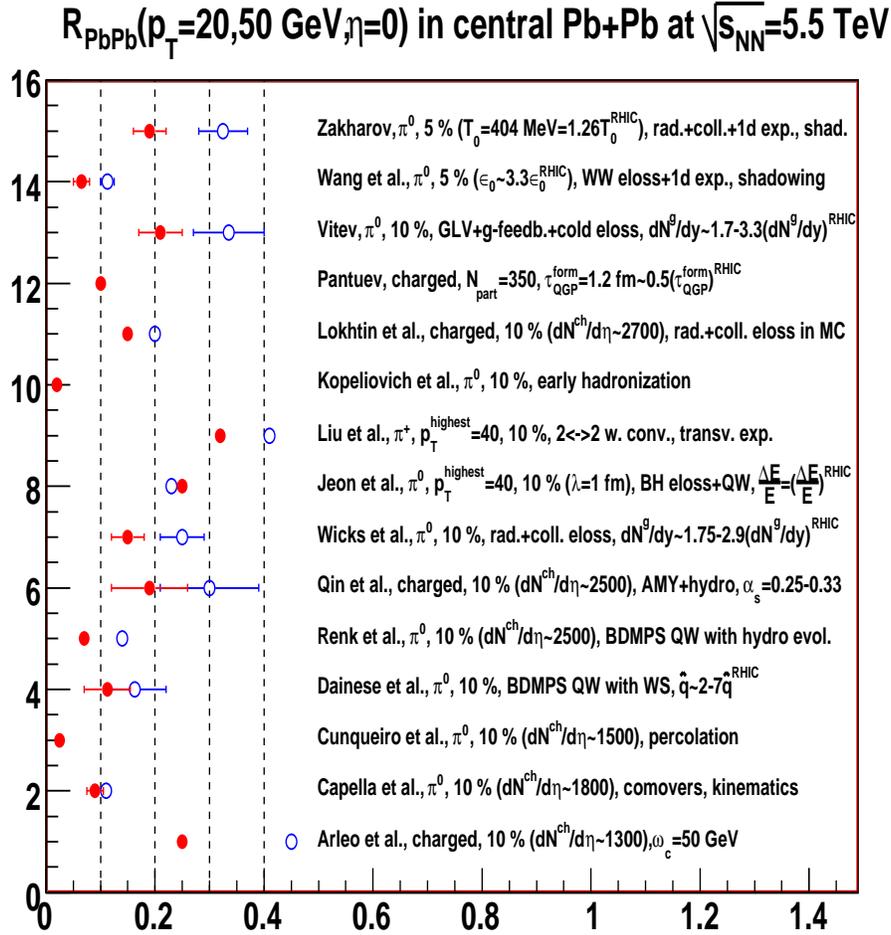}
\end{center}
\caption{Predictions for the nuclear modification factor in central Pb-Pb collisions at the LHC, for $p_T=20$ (red filled symbols) and 50 (when available, blue open symbols) GeV. On the right, the name of the authors, the particle, centrality definition and some model explanation is shown. The error bar in the points reflects the uncertainty in the prediction. See the text for explanations.}
\label{fig19}
\end{figure}

I start by reviewing the predictions for the nuclear modification factor in central PbPb collisions at the LHC. In Fig. \ref{fig19} I show 15 predictions for $R_{AA}$ at $p_T=20, 50$ GeV from different models. Differently from the case of multiplicities, where some easy re-scaling to a common centrality class was feasible, here such re-scaling is not possible as there is no simple relation between a change of density/multiplicity and the resulting energy loss and $R_{AA}$. Therefore, I simply indicate in the figure the centrality definition or the multiplicity or energy density (with respect to that at RHIC) for which the predictions were computed.

Different models use different parameters related with the medium density and the scattering strength of the parton with the medium. The most common one is the transport coefficient $\hat{q}$ which can be related locally to the energy density through \cite{Baier:2002tc}
\begin{equation}
\hat{q}(x,y,z,\tau)=c\cdot \epsilon^{3/4}(x,y,z,\tau),
\label{qhat}
\end{equation}
with $c$ some constant which in pQCD is expected to be of order 1\footnote{This proportionality is expected in pQCD \cite{Baier:2002tc} and also at strong coupling through the AdS/CFT correspondence, see Hong Liu in \cite{Abreu:2007kv} and \cite{Liu:2006ug}. Besides, the transport coefficient may acquire some energy dependence, see Casalderrey-Solana et al. in \cite{Abreu:2007kv} and \cite{CasalderreySolana:2007sw}. Attempts have been essayed to compute it from first principles in QCD, see Antonov et al. in \cite{Abreu:2007kv} and \cite{Antonov:2007sh}.} . Other models use the gluon density, the energy density at thermalization time, the value of $\alpha_s$, etc.

For descriptive purposes, the predictions can be classified into the following groups:
\begin{enumerate}

\item Models which consider only radiative energy loss (see \cite{Majumder:2007iu} for a comparison among the theoretical basis of the different models).
\begin{itemize}

\item Arleo et al. \cite{Abreu:2007kv,Arleo:2007bg} use fragmentation functions modified through their convolution with quenching weights - the probability of a given amount of energy loss - which are evaluated using a simplified radiation spectrum. The employed characteristic gluon frequency is $\omega_c=50$ GeV.

\item Dainese et al. \cite{Abreu:2007kv,Dainese:2004te}, the PQM model, use quenching weights calculated from the full radiation spectrum in the multiple soft scattering approximation and a static medium modeled by the initial overlap geometry. The different predictions correspond to different extrapolations of the transport coefficient from RHIC to LHC energies.

\item Renk et al. \cite{Abreu:2007kv,Renk:2006pk} use, as the previous model, quenching weights calculated from the full radiation spectrum in the multiple soft scattering approximation, but with a hydrodynamical modeling of the medium and the relation (\ref{qhat}).

\item Jeon et al. \cite{Abreu:2007kv,Jeon:2002dv} use a schematic model for the quenching weights which considers only an average energy loss.

\item Vitev \cite{Abreu:2007kv,Vitev:2005he} uses the GLV model with quenching weights with gluon feedback and one-dimensional Bjorken expansion, and higher-twist shadowing of parton densities. The different predictions correspond to different extrapolations of the gluon density from RHIC to the LHC.

\item Wang et al. \cite{Abreu:2007kv,Zhang:2007ja} use a model for medium-modified fragmentation functions and compute the yields at next-to-leading order. A Bjorken expansion is considered. The error bars correspond to the different parametrization of nuclear shadowing employed in the pQCD calculations.

\end{itemize}

\item Models which consider radiative and elastic energy loss.
\begin{itemize}

\item Qin et al. \cite{Abreu:2007kv,Qin:2007zz} use the AMY model with radiative and collisional energy loss in a medium which is modeled through ideal hydrodynamics. The error bars correspond to different values of $\alpha_s$.

\item Wicks et al.  \cite{Abreu:2007kv,Wicks:2007am} use the GLV model for radiative energy loss whose quenching weights are convoluted with those from elastic energy loss. The error bars correspond to different extrapolations of the gluon density from RHIC to the LHC.

\item Lokhtin et al.  \cite{Abreu:2007kv,Lokhtin:2005px}, the PYQUEN model, is a implementation within PYTHIA of radiative energy loss which considers a mean radiative energy loss distributed among some gluons, which are then allowed to do vacuum final state radiation until the branching process stops, after which they scatter elastically.

\item Zakharov \cite{Zakharov:2008kt} uses a model which consider quenching weights based on a single radiation spectrum in the multiple soft scattering approximation, plus elastic scattering, nuclear shadowing and Bjorken expansion of the medium. The error bars correspond to the different values at which the running coupling is frozen in the infra-red, and to considering a purely gluonic or a chemically equilibrated plasma.

\end{itemize}

\item Models with elastic energy loss plus parton conversions. Liu et al.  \cite{Abreu:2007kv,Liu:2006sf} consider production in pQCD with the possibility of elastic scattering in which conversion channels e.g. $q g \to gq$ or $gg \to q\bar{q}$, are included. This inclusion turns out to be of importance for the hadrochemistry at large transverse momentum, see below. For this model, the highest available $p_T$ for the predictions is 40, not 50 GeV.

\item Others.
\begin{itemize}

\item Capella et al.  \cite{Abreu:2007kv,Capella:2006fw} use a comover absorption scenario in which energy gain and loss terms are implemented in one-dimensional rate equations, plus strong shadowing. The error bands correspond to the different kinematics (considering $2\to 1$ or $2\to 2$ processes) to evaluate the shadowing.

\item Cunqueiro et al.  \cite{Abreu:2007kv,Cunqueiro:2007fn} consider a scenario in which percolation induces an increase in the string tension and a strong modification of the distribution of particle sources.

\item Kopeliovich et al.  \cite{Abreu:2007kv,Kopeliovich:2007yv} consider a sudden hadronization scenario in which hadrons are created very soon and interact strongly with the produced medium.

\item Pantuev  \cite{Abreu:2007kv,Pantuev:2008zz} consider the medium as composed by a thin transparent corona and a totally opaque core, which can be alternatively interpreted in terms of a formation time for the QGP. Estimations of the variation of this time when going from RHIC to the LHC allow for the predictions.

\end{itemize}
\end{enumerate}

While no simple quantitative conclusion can be extracted from this variety of models, it can be claimed that those which implement radiative or collisional energy loss generically predict a nuclear modification factor between $0.15\div 0.25$ at $p_T=20$ GeV and increasing with increasing $p_T$. Larger densities lead to larger suppressions, but the concrete value and the quantitative behavior with increasing $p_T$ are different for different models, a fact which is not only related with the theoretical model used for energy loss but also with the 'embedding' of such model in the medium.

\begin{figure}
\begin{center}
\includegraphics[width=5.8cm]{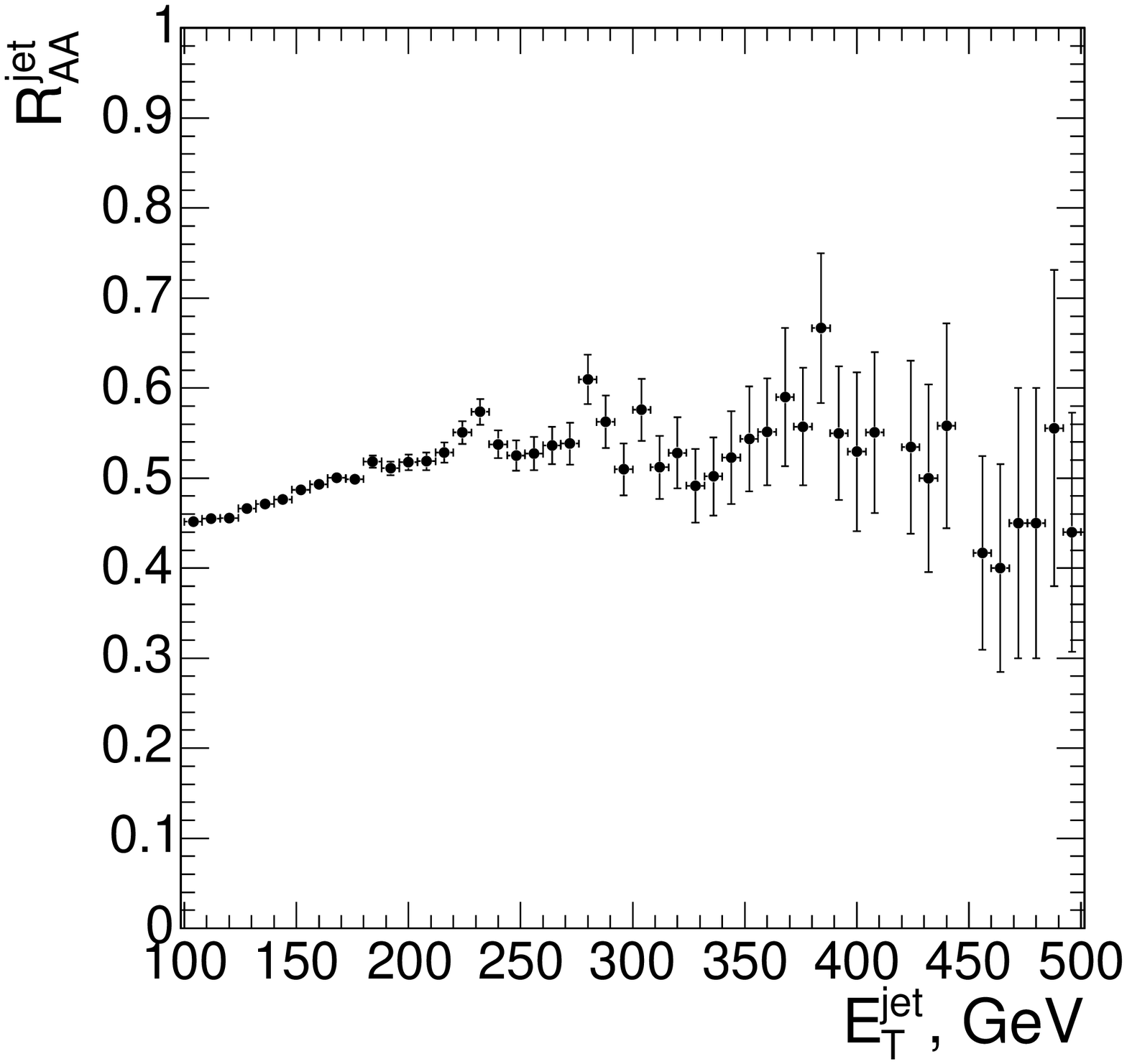}\hfill \includegraphics[width=5.8cm]{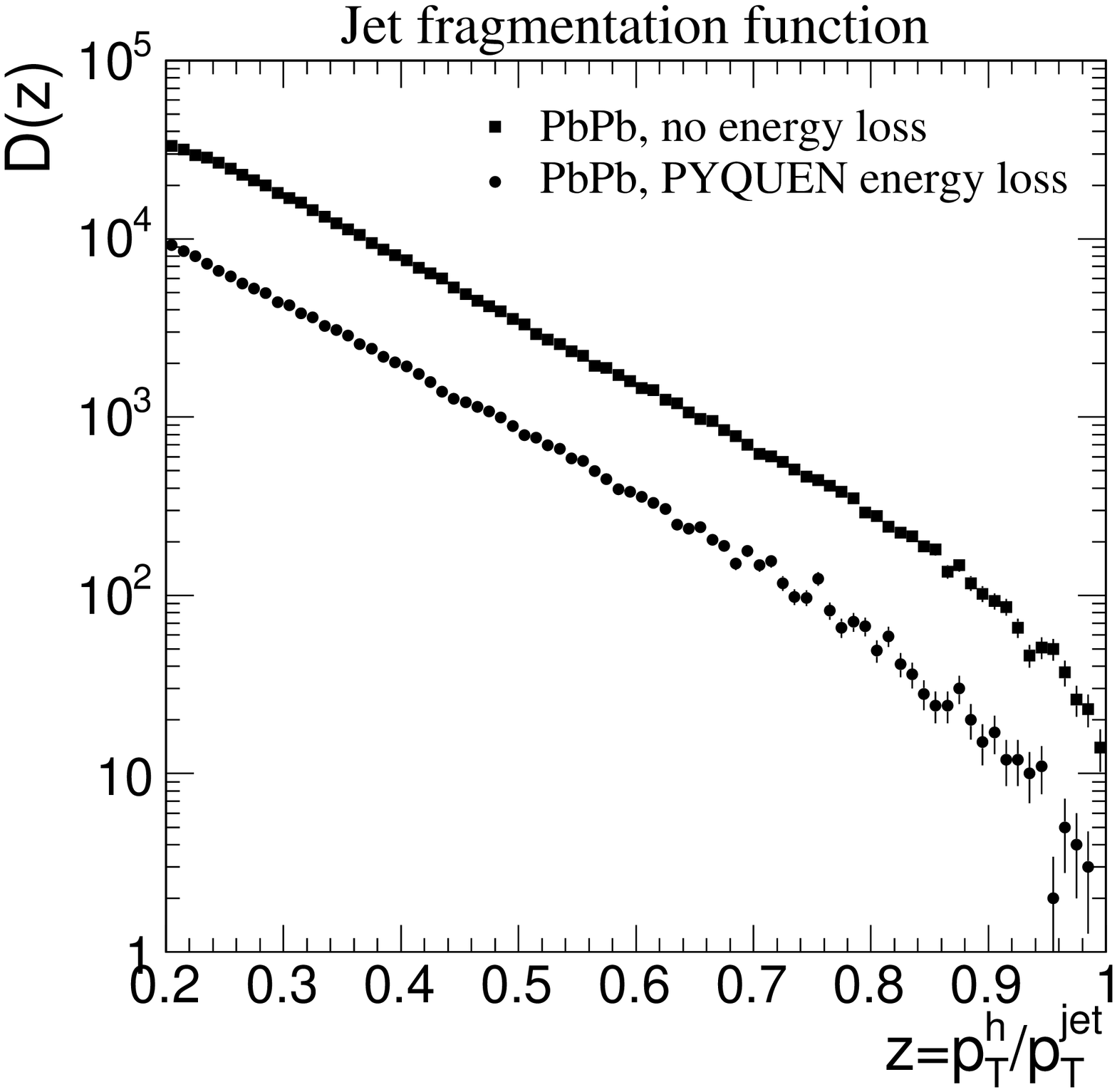}
\end{center}
\caption{Left: Nuclear modification factor for jets with $|\eta|<3$ (defined through a cone algorithm with $R=0.5$) versus the $E_T$ of the jet, for an integrated luminosity of 0.5 nb$^{-1}$, in the PYQUEN model.
Right: Fragmentation functions defined with respect to the $p_T$ of the jet, in the PYQUEN model. Figures taken from
$^{35}$.}
\label{fig20}
\end{figure}

On the other hand, jets will be very abundantly produced at the LHC, see e.g. \cite{Accardi:2004gp,d'Enterria:2008ge,Cortese:2008zza} and Fig. \ref{fig1} right. Provided the issues \cite{d'Enterria:2009am} of jet reconstruction through some algorithm, background subtraction (see e.g. \cite{Cacciari:2007fd}) and jet energy calibration are successfully addressed, they offer huge possibilities to verify the physical mechanism underlying the jet quenching phenomenon, both through the measurement of the $R_{AA}$ of jets (see Fig. \ref{fig20} left) as well as more differential observables such as jet fragmentation functions (see Fig. \ref{fig20} right), jet shapes,$\dots$

For such studies, and for the study of particle correlations, new theoretical tools have to be developed\footnote{E.g. the developments in the modified leading-logarithmic approximation (MLLA) approximation \cite{Borghini:2005em,Borghini:2009eq,Dremin:2006da,Armesto:2008qe,Ramos:2008qb}, the modifications of Dokshitzer-Gribov-Lipatov-Altarelli-Parisi (DGLAP) evolution \cite{Armesto:2007dt,Majumder:2009zu} or the inclusion of elastic energy loss in the parton cascade \cite{Domdey:2008gp}.}
and implemented in Monte Carlo simulators (semi-quantitative ideas were pioneered in \cite{Salgado:2003rv}, see e.g. \cite{Vitev:2008rz} for a recent study). This is an ongoing effort with several groups involved and several Monte Carlo generators becoming gradually available: PYQUEN \cite{Lokhtin:2005px}, Q-PYTHIA \cite{Armesto:2008qh}, JEWEL \cite{Zapp:2008gi}, YaJem \cite{Renk:2008pp}, etc.

\begin{figure}
\begin{center}
\includegraphics[width=12cm]{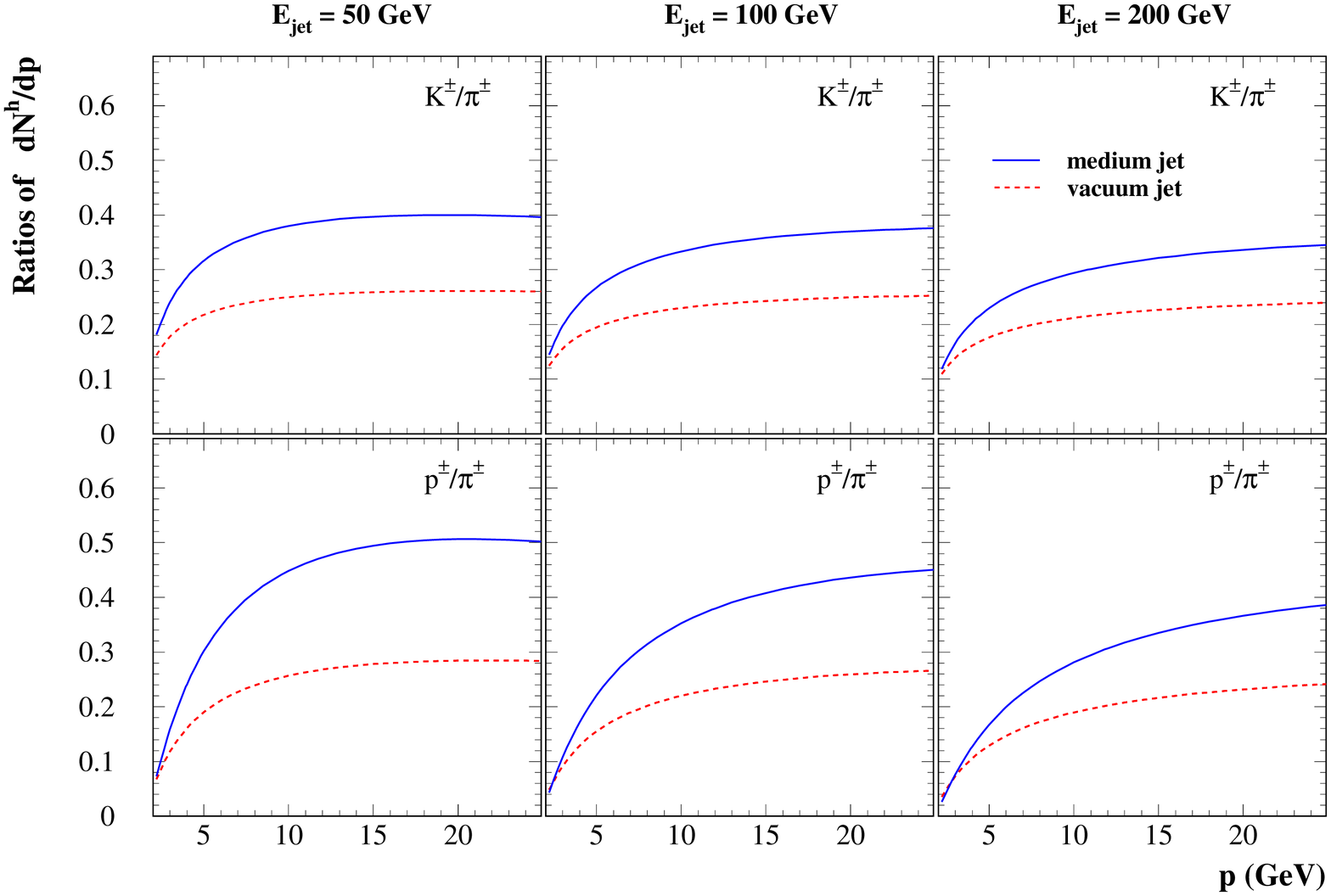}
\end{center}
\caption{Medium  modification of particle ratios within a jet  versus the momentum of the particles, in the MLLA approximation by Sapeta et al.
Figure taken from
$^{35}$.}
\label{fig21}
\end{figure}

Another aspect of great importance both to verify the origin of jet quenching and to understand the interplay between the energetic particles and the soft medium is hadrochemistry at large $p_T$. First, within the MLLA approximation and modeling the medium-modification of the final state radiation pattern through a multiplicative constant in the collinear parts of the splitting functions \cite{Borghini:2005em}, an enhancement of the ratio of baryons and strange mesons over pions due to medium effects is found within the fragmentation of a energetic parton
\cite{Sapeta:2007ad}, see Fig. \ref{fig21}.

\begin{figure}
\begin{center}
\includegraphics[width=5.8cm, height=6cm]{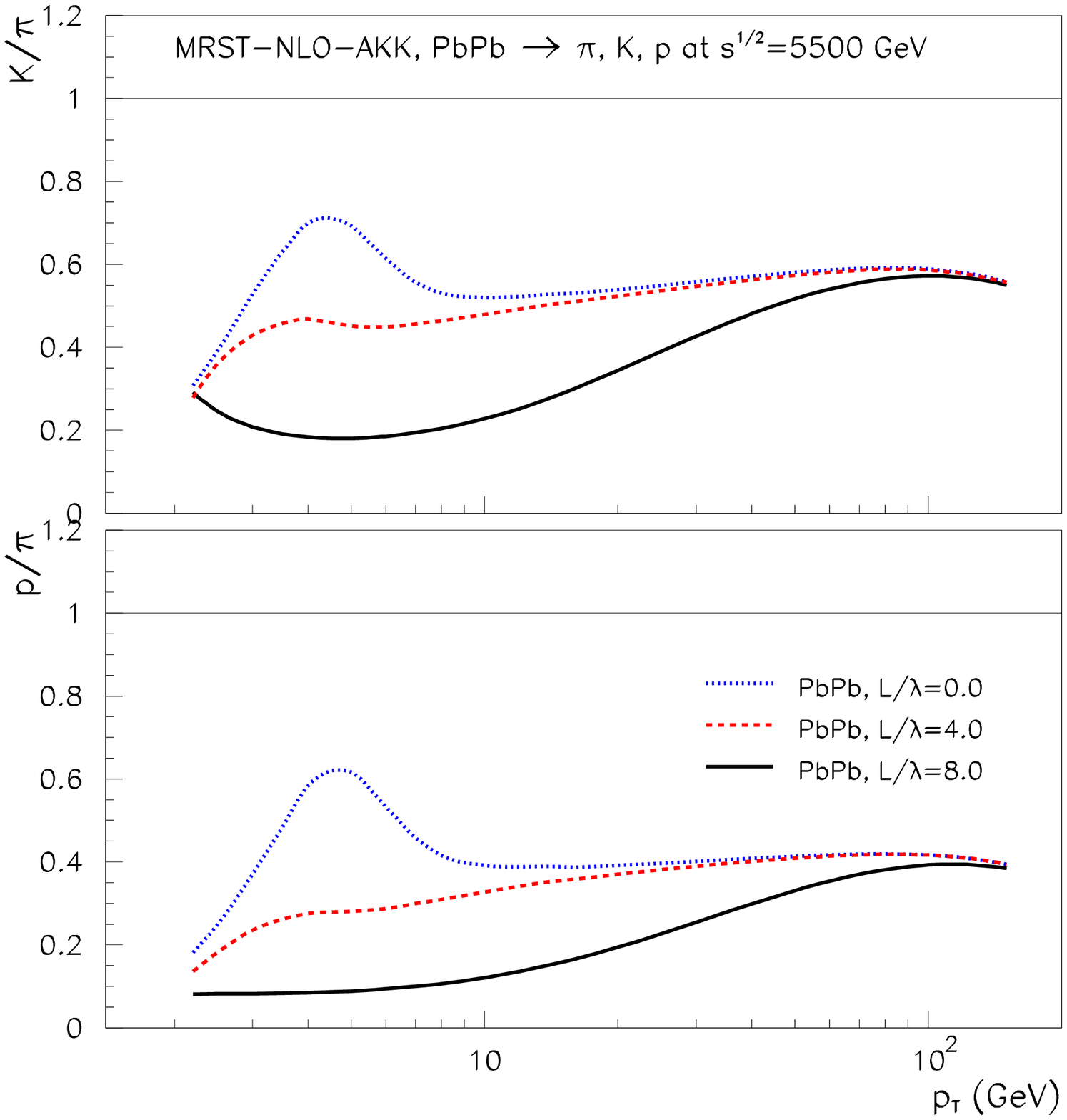}\hfill \includegraphics[width=5.8cm, height=6cm]{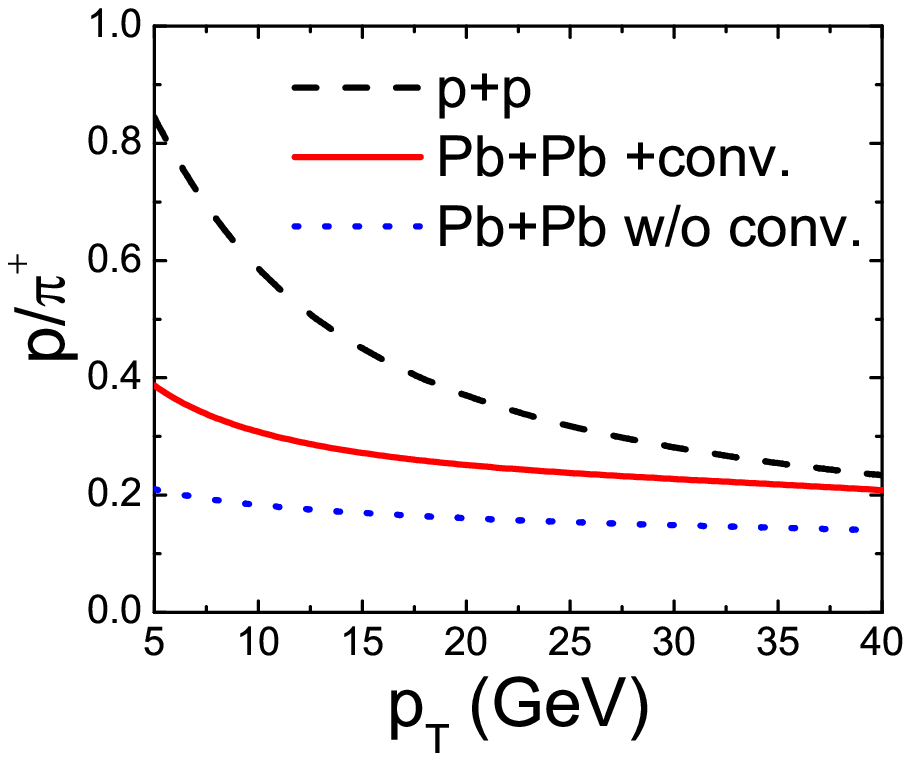}
\end{center}
\caption{Left: Particle ratios versus transverse momentum in pQCD with GLV energy loss for different opacities, from Barnafoldi et al.
Right: p/$\pi^+$ ratio at the LHC in pp, and in PbPb collisions with elastic energy loss and conversions, from Liu et al. Figures taken from
$^{35}$.}
\label{fig22}
\end{figure}

Second, the non-Abelian nature of radiative energy loss implies that gluons lose more energy than quarks due to their larger color charge - i.e. the value of the quadratic Casimir of the adjoint (3) and fundamental (4/3) representations in QCD, respectively. Therefore, hadrochemistry is affected, as different particles receive different contributions from the fragmentation of quarks and gluons and this relative contribution varies with particle momentum\footnote{In this respect, it should be noted that some fragmentation functions e.g. those of protons are badly constrained from available experimental data in absence of any medium e.g. in $e^+e^-$ or $pp$. Therefore medium-modification studies are subject to an uncertainty that can only be resolved with a better knowledge of the vacuum fragmentation functions, see e.g. \cite{:2008afa}.}. This can be seen in Fig. \ref{fig22} left where Barnafoldi et al. \cite{Abreu:2007kv} show the results for particle ratios of the GLV energy loss with different opacities $L/\lambda$, with $L$ the medium length and $\lambda$ the mean free path of partons in the medium.

Finally, conversions as discussed above in the model by Liu et al. \cite{Abreu:2007kv,Liu:2006sf}  also modify the hadrochemistry at large $p_T$, see Fig. \ref{fig22} right, and there is the possibility of recombination of partons from adjacent jets \cite{Hwa:2006zq}
which may also increase the baryon-to-meson ratio at large transverse momentum.

\begin{figure}
\begin{center}
\includegraphics[width=7cm]{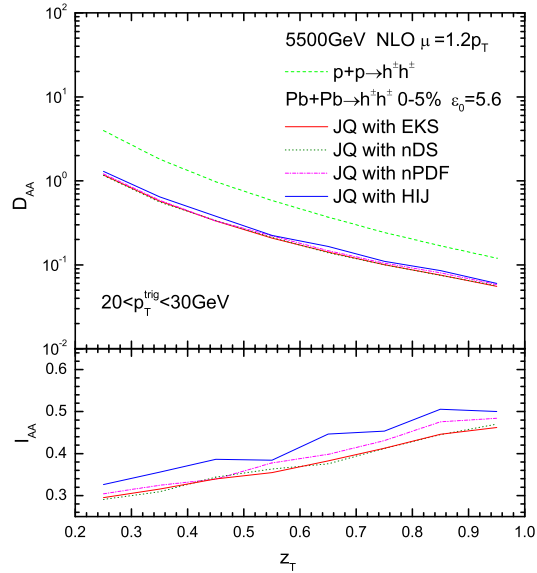}
\end{center}
\caption{Hadron-triggered fragmentation functions $D_{AA}(z_{T})$ and the corresponding medium modification factors
$I_{AA}(z_{T})$ in NLO pQCD for central Pb+Pb collisions at $\sqrt{s_{NN}} = 5.5$ TeV, by Wang et al. Different lines refer to pp and to PbPb with different nuclear parton densities.
Figure taken from
$^{35}$.}
\label{fig23}
\end{figure}

Now I focus on correlations at large transverse momentum and the disappearance of the backward azimuthal peak observed at RHIC. It is usually quantified through the hadron-triggered fragmentation functions for two hadrons $h_1,h_2$, a trigger particle $h_1$ which defines the near side hemisphere (azimuthal angle 0) with $p_T^{trig}$ and an associated particle $h_2$ in the away side hemisphere (around azimuthal angle $\pi$) with $p_T^{asso}$, reading
\begin{equation}
D_{AA}(z_T=p_T^{asso}/p_T^{trig},p_T^{trig}) = p_T^{trig}\frac{d\sigma^{h_1h_2}/dy^{trig}dp_T^{trig}dy^{asso}dp_T^{asso}}{d\sigma^{h_1}/dy^{trig}dp_T^{trig}}\,.
\label{daa}
\end{equation}
This quantity provides information on the conditional yield of particles in the backward hemisphere for a given trigger, and offers additional constraints (other than those coming from $R_{AA}(p_T)$) on the parameters characterizing the medium in models of energy loss \cite{Zhang:2007ja}. Predictions for the LHC exist, see \cite{Renk:2007rn} and Wang et al. in  \cite{Abreu:2007kv,Zhang:2007ja}, Fig. \ref{fig23}.

To conclude this Subsection, I will comment on one aspect for which no prediction is yet available, namely the wide structure observed in the backward azimuthal region when the $p_T$ of the associated particles is lowered to that of the particles in the bulk. The standard qualitative explanations, see \cite{d'Enterria:2009am} and references therein, go from deflection of jets in strong fields, to medium-induced radiation, Mach cones, Cherenkov radiation,$\dots$ Recent experimental analysis \cite{:2008nd} seem to disfavor the deflection of jets. But the nature of such structure has not yet been unambiguously established, mainly because it is not yet clear how the jet energy is transferred to the expanding medium. Much effort is currently devoted to it, see e.g. Bauchle et al. and Betz et al. for Mach cones in a hydrodynamical medium, Dremin for Cherenkov radiation and Mannarelli et al. for the energy evolution of the angular structure of the energy deposition of jets, in \cite{Abreu:2007kv}.

\subsection{Heavy quarks and quarkonia}
\label{heavy}

Heavy quark and quarkonium production and suppression are other standard hard probes, see the recent review \cite{Rapp:2009my}. Beginning with heavy-quark production, it offers the possibility of testing the expected hierarchy of radiative energy loss \cite{Dokshitzer:2001zm,Armesto:2005iq}:
$$\Delta E({\rm gluons})>\Delta E({\rm light}\ {\rm quarks})>\Delta E({\rm heavy}\ {\rm quarks}),$$
with the first inequality coming from the different color factors (as discussed in the previous Subsection), and the second from the suppression of radiation due to the mass of the parent parton.
Besides, collisional energy loss is expected to be more important for heavy quarks  \cite{Mustafa:2004dr} than for light partons, and the details of medium modeling (as a collection of static scattering centers, as a dynamical medium e.g.  in Djordjevic et al. in \cite{Abreu:2007kv} or in \cite{Djordjevic:2008iz},$\dots$). The measurement by PHENIX and STAR  \cite{Adare:2006nq,Abelev:2006db} of a nuclear modification factor much smaller than 1 for 'non-photonic' electrons (expected to come from the semi-leptonic decays of heavy flavors) has triggered a lot of activity. The LHC, with the new possibilities for heavy-flavor identification of beauty (and eventually of charm) \cite{lhc,Alessandro:2006yt,D'Enterria:2007xr} and for the measurement of non-photonic electrons (with the possibility of separating charm and beauty via correlations \cite{Mischke:2008qj,Mischke:2008af,Adare:2009ic}), together with the extended transverse momentum reach, offers an ideal testing ground for these ideas.

\begin{figure}
\begin{center}
\includegraphics[width=5.8cm, height=5.8cm]{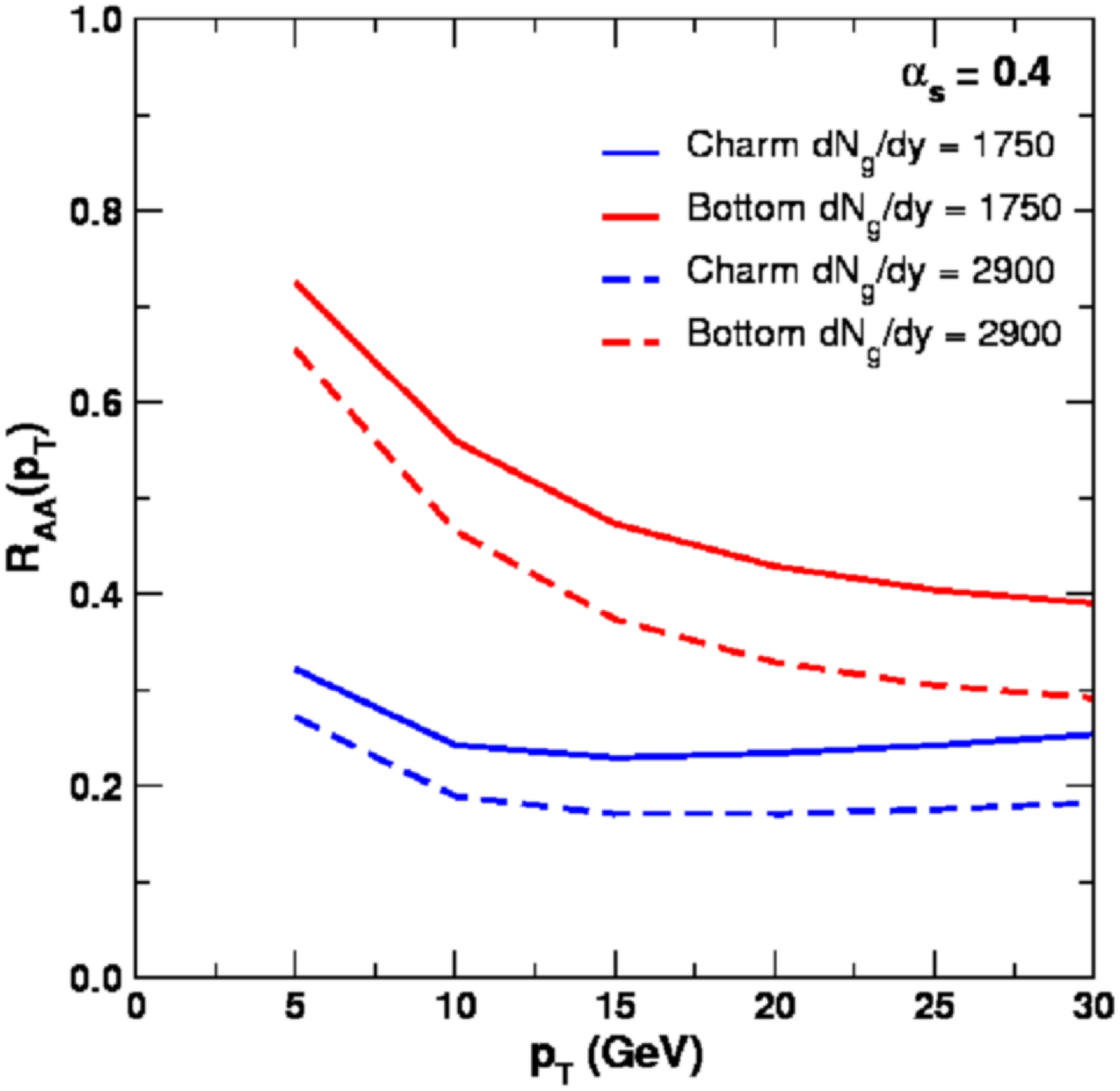}\hfill \includegraphics[width=5.8cm, height=6cm]{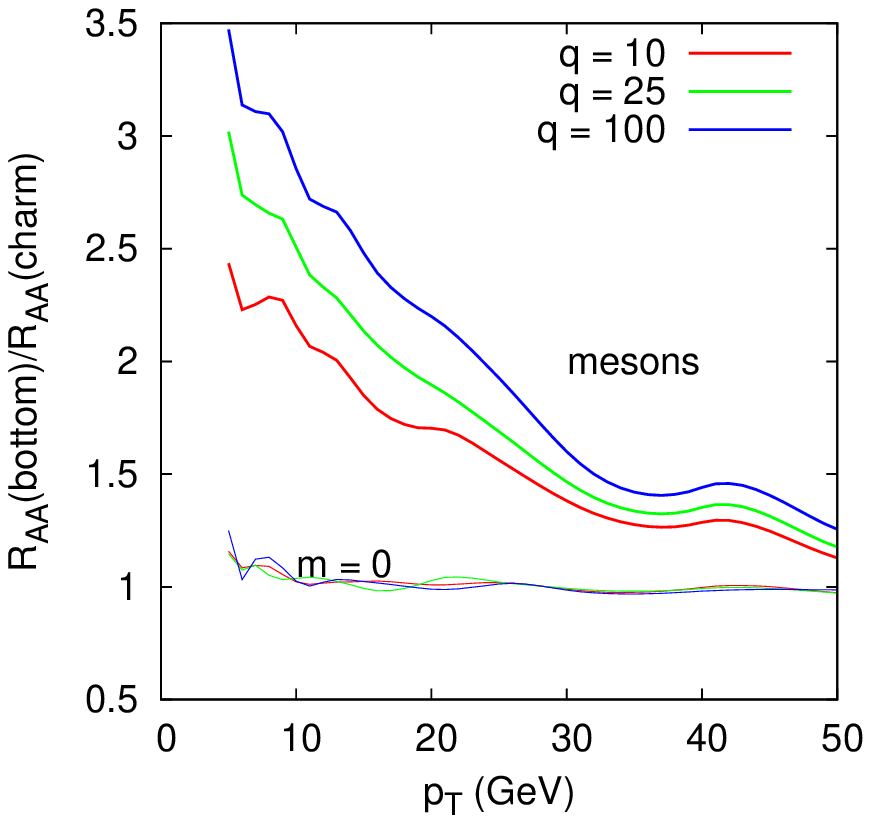}
\end{center}
\caption{Left: Nuclear modification factor for D and B mesons in the WHDG model for different gluon densities and $\alpha_s=0.4$, from Wicks et al.
Right: Ratio of nuclear modification factors for B over D mesons with radiative energy loss, for different transport coefficients and for the case where the mass effect on radiative energy loss is switched off, from Armesto et al. Figures taken from
$^{35}$.}
\label{fig24}
\end{figure}

\begin{figure}
\begin{center}
\includegraphics[width=6cm,height=5.cm]{ivanheavy.eps}\hfill \includegraphics[width=5.3cm, height=5.3cm]{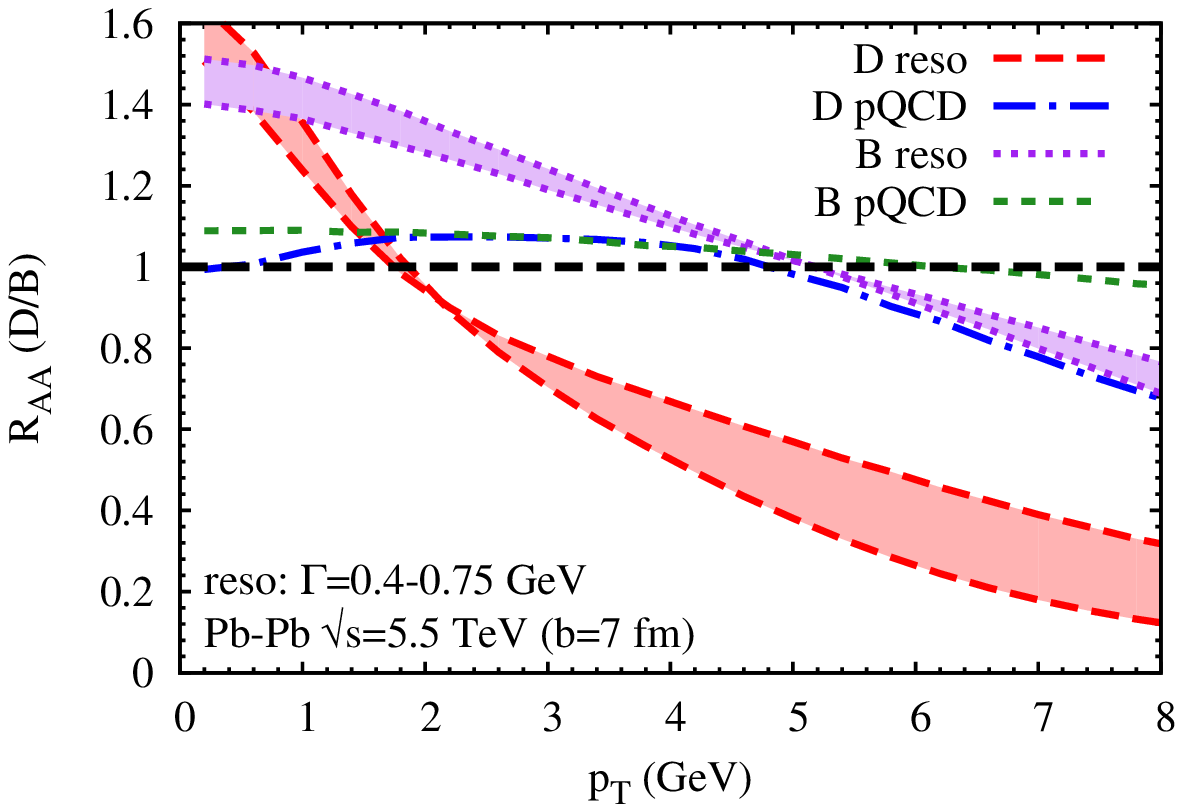}
\end{center}
\caption{Left: Nuclear modification factor of charm and beauty mesons and baryons for CuCu and AuAu collisions at RHIC, and for PbPb collisions at the LHC for two different gluon densities, from Vitev.
Right: Nuclear modification factors of D and B mesons in PbPb collisions for $b=7$ fm at the LHC, from pQCD and from scattering with resonances in the QGP, from van Hees et al. Figures taken from
$^{35}$.}
\label{fig25}
\end{figure}

\begin{figure}
\begin{center}
\includegraphics[width=6cm,height=5cm]{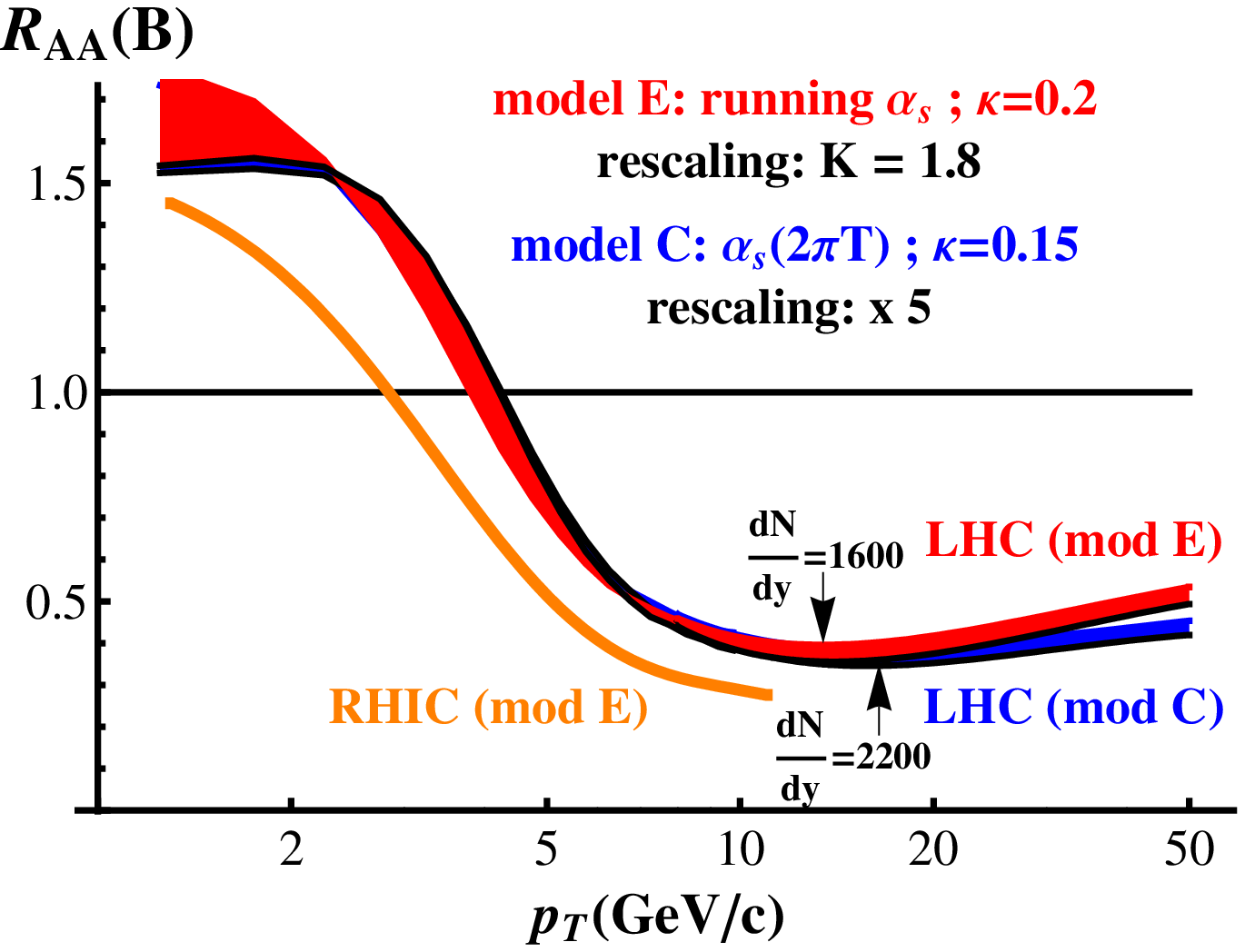}\hfill \includegraphics[width=6cm]{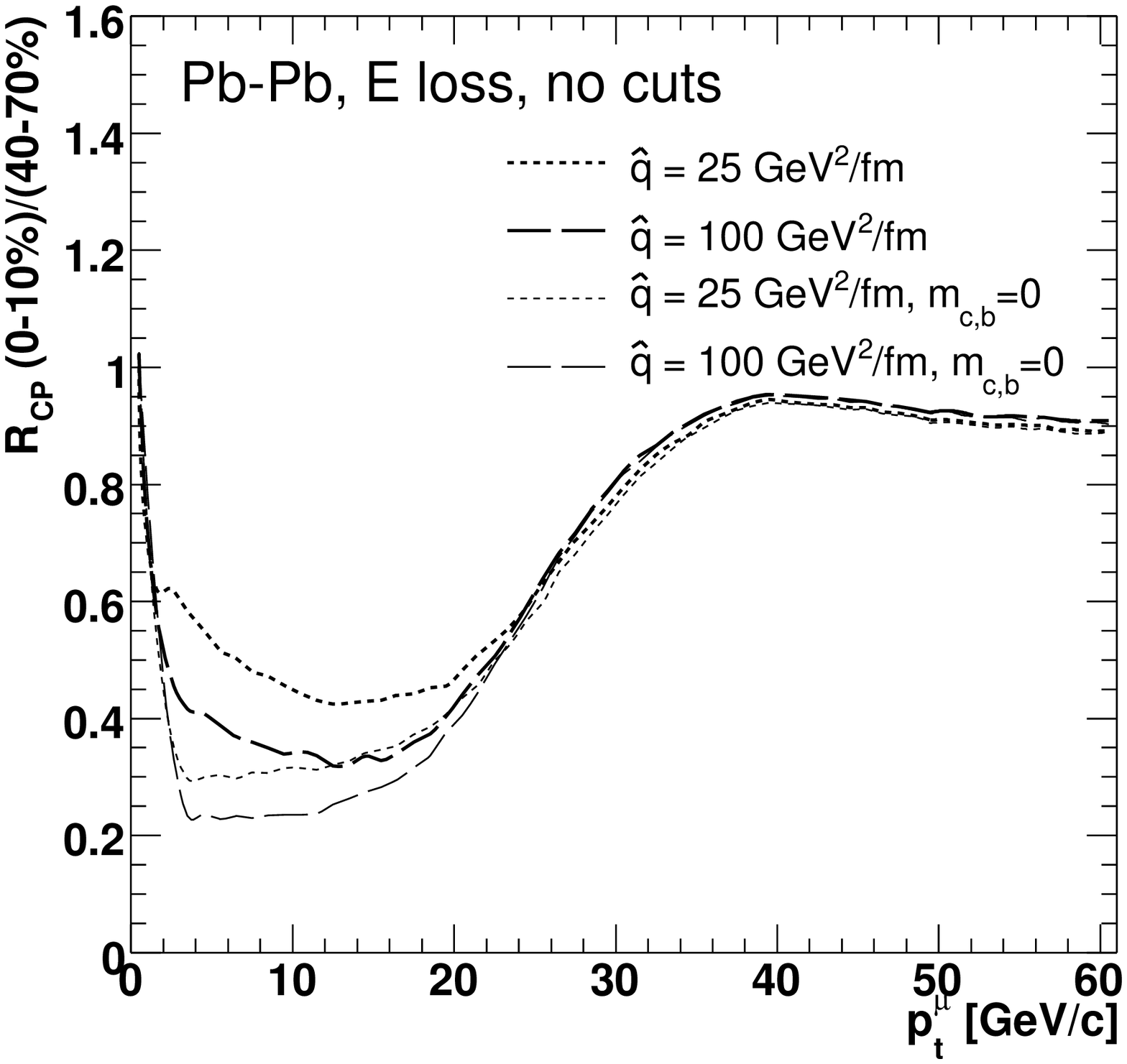}
\end{center}
\caption{Left: Nuclear modification factor for B mesons in central PbPb collisions at the LHC versus transverse momentum, in a model with collisional energy loss with fixed (model C) and running (model E) coupling constant. Bands are defined by different charged multiplicities at mid-rapidity. Figure taken from
$^{177}$.
Right: Nuclear modification factor $R_{CP} (p_T)$ for muons at the LHC, for different transport coefficients and for the case where the mass effect on radiative energy loss is switched off. Figure taken from
$^{178}$.}
\label{fig26}
\end{figure}

In Figs. \ref{fig24}, \ref{fig25} and \ref{fig26} I show available predictions for PbPb collisions at the LHC. Specifically, in Fig \ref{fig24} left I show the results of the collisional plus radiative energy loss model of Wicks et al. \cite{Wicks:2007am} which uses the DGLV model for radiative energy loss of heavy quarks, whose corresponding quenching weights are convoluted with those from elastic energy loss. The different lines correspond to different extrapolations of the gluon density from RHIC to the LHC. In Fig \ref{fig24} right I show the results from a purely radiative model by Armesto et al. \cite{Armesto:2005mz} which uses the quenching weights for heavy quarks computed in the multiple soft scattering approximation. The geometry in this model is considered as in the PQM model, see the previous Subsection. In this case, the results shown are not for the nuclear modification factor but for the ratio of nuclear modification factors of bottom and charm mesons, which clearly shows the mass effect on the energy loss in an accessible region of $p_T$.

In Fig. \ref{fig25} left the results of the model of Vitev \cite{Adil:2006ra} are shown (see also \cite{Sharma:2009hn}). This model considers a very fast hadronization for heavy-flavored mesons, which then dissociate through collisions in the QGP. In Fig. \ref{fig25} right van Hees et al. \cite{Rapp:2005at} consider both radiative losses of the heavy quarks and their strong scattering with resonances in the plasma through a diffusion equation.

In Fig. \ref{fig26} left the results of the collisional model of \cite{Gossiaux:2009mk} are provided. This model considers elastic energy loss for fixed and running coupling constants, with the bands defined by different extrapolations of multiplicities. Finally, in Fig. \ref{fig26} right I show the results from \cite{ConesadelValle:2007sw} for the $R_{CP}$ (the nuclear modification factor defined not with respect to nucleon-nucleon collisions but with respect to a peripheral class of events) for muons coming both from semi-leptonic decays of heavy flavors and from decays of electro-weak bosons. For the former an energy loss model equivalent to that in  \cite{Armesto:2005mz} is used. The latter are expected to show no medium (hot matter) effect, thus this measurement contains its own self-calibration with respect to cold nuclear matter effects.

All results presented here show a large suppression at $p_T\sim 10\div 20$ GeV (they have been computed at mid-rapidity except those in \cite{ConesadelValle:2007sw} which has been done for the ALICE muon arm covering $2.5<\eta<4$), and a gradual increase of the nuclear modification factor with $p_T$. 

On the other hand, in a strongly coupled super-symmetric Yang-Mills plasma, the dominant energy loss mechanism for a heavy quark, computed through the use of the AdS/CFT correspondence, is a drag force (valid for small velocities of the heavy quark), see  \cite{d'Enterria:2009am,CasalderreySolana:2007zz,Gubser:2009sn}. Calculations show \cite{Gossiaux:2009mk,Horowitz:2008zz} that this drag force results in a nuclear modification factor much flatter with $p_T$ than in pQCD-based models with elastic or radiative energy losses.

\begin{figure}
\begin{center}
\includegraphics[width=7cm]{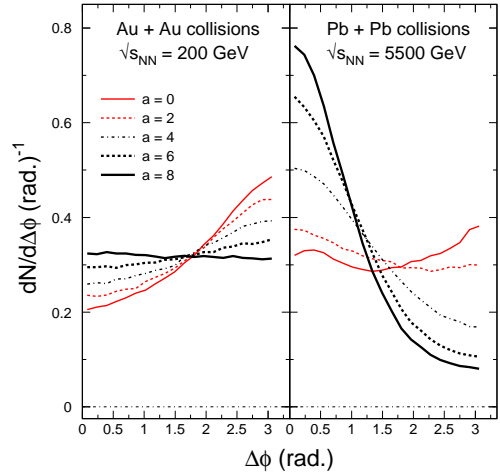}
\end{center}
\caption{D$\overline{\rm D}$
correlation as a function of relative azimuth angle $\Delta \phi$
with different drag parameter $a$ for central Au+Au collisions at
RHIC (left) and Pb+Pb collisions at LHC (right). The unit of the
drag coefficient $a$ is [10$^{-6}$(fm)$^{-1}$MeV$^{-2}$]. No $p_T$ cut has been 
applied. The initial temperature $T_0$ and thermalization time
$\tau_0$ are, respectively, 340 MeV and 0.6 fm at RHIC and 610
MeV and 0.3 fm at LHC. Figure taken from
$^{181}$.}
\label{fig26b}
\end{figure}

Besides, the effects of different shadowing mechanisms (Kopeliovich et al. \cite{Abreu:2007kv}), of light-to-heavy conversions \cite{Liu:2008bw} and of thermal production (see Chen et al. in \cite{Abreu:2007kv}) have been considered. Also, the possible characterization of the plasma through the de-correlation of D mesons coming from back-to-back charm-anticharm pairs  \cite{Zhu:2007ne,Tsiledakis:2009da} is under investigation, see Fig. \ref{fig26b}. Further, soft effects can modify the charm cross section with respect to usual expectations: an enhanced string tension as introduced in the 
HIJING/B$\bar{\rm B}$ \cite{ToporPop:2007hb} model leads to an enhancement of the charm cross section in heavy-ion collisions \cite{Pop:2009sd} which amounts to a $60\div 70$ \%  at RHIC energies and to an order of magnitude at the LHC.

Now I turn to the suppression of quarkonium - one of the canonical signatures of QGP formation since its proposal in 1986 \cite{Matsui:1986dk}. Unfortunately, this signal is plagued with uncertainties from cold nuclear matter effects, both from the lack of knowledge of the nuclear parton densities (see Section \ref{pa}) and the lack of understanding of nuclear absorption in cold, normal nuclear matter. The most recent phenomenological analysis \cite{Arleo:2006qk,Lourenco:2009sk} indicate that the absorption cross section of $J/\psi$ in nuclear matter is either constant or decreasing with increasing energy. On the contrary, theoretical models previous to RHIC data \cite{Braun:1997qw,Kopeliovich:2001ee} pointed to an increase of absorption with increasing energy. Although some progress  \cite{Capella:2006mb} has been made in understanding such unexpected feature, predictions for the LHC are still subject to large uncertainties.

\begin{figure}
\begin{center}
\includegraphics[width=12.5cm]{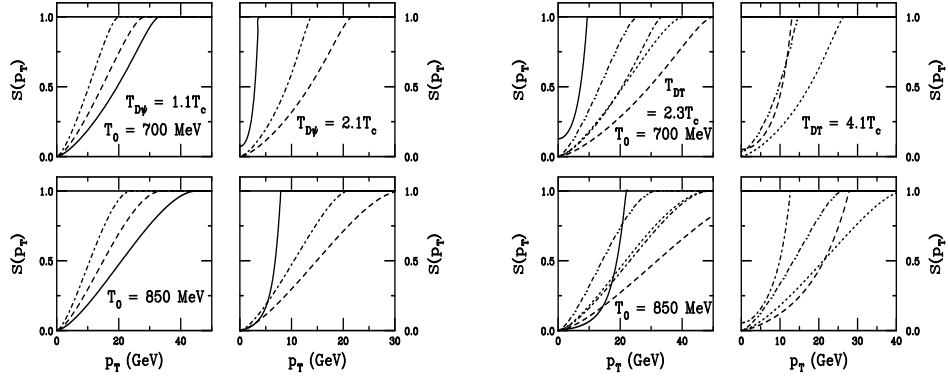}
\end{center}
\caption{The survival probabilities as a function of $p_T$ for the charmonium
(left-hand side) and bottomonium (right-hand side) states for initial 
conditions at the LHC.
The charmonium survival probabilities are $J/\psi$ (solid), $\chi_c$ 
(dot-dashed) and $\psi'$ (dashed) respectively.  The bottomonium survival
probabilities are given for $\Upsilon$ (solid), $\chi_{1b}$ (dot-dashed), 
$\Upsilon'$ (dashed), $\chi_{2b}$ (dot-dot-dash-dashed) and $\Upsilon''$ 
(dotted) respectively.  The top plots are for $T_0 = 700$ MeV while the bottom
are for $T_0 = 850$ MeV.  The left-hand sides of the plots for each state
are for the lower dissociation temperatures, $1.1T_c$ for the $J/\psi$
and $2.3T_c$ for the $\Upsilon$ while the right-hand sides show the results for
the higher dissociation temperatures, $2.1T_c$ for the $J/\psi$ and $4.1T_c$
for the $\Upsilon$. Figure taken from
$^{35}$.}
\label{fig27}
\end{figure}

On the other hand, the behavior of quarkonia in a QGP is not clear either. Lattice data \cite{Kaczmarek:2004gv} support a suppression pattern in which $\psi^\prime$ and $\chi_c$ melt just above the deconfinement temperature, while the $J/\psi$ survives up to temperatures close to $2T_c$ and the $\Upsilon$ up to sizably larger $T$. Contrariwise, potential models, see \cite{Mocsy:2008eg} for a discussion, suggest that the $J/\psi$ melts much closer to $T_c$ than indicated by lattice results. The actual suppression pattern could be tested by the transverse momentum dependence of the suppression of different quarkonium states, as illustrated by Vogt in \cite{Abreu:2007kv}, see Fig. \ref{fig27}. There, clear differences can be seen between the suppression sequence of the different states e.g. $\chi_c$ suppression disappears at smaller $p_T$ than the $J/\psi$ one if the dissociation temperature of $J/\psi$ is close to $T_c$, while the opposite happens if the $J/\psi$ dissociation temperature is much higher\footnote{While the naive expectation is that the suppression disappears with increasing $p_T$ due to the smaller time that the bound state stays in the plasma, there are proposals  \cite{Liu:2006nn} that the suppression should reappear at larger $p_T$ due to the larger 'effective' $T$ seen by the bound state when moving fast with respect to the medium.}.

\begin{figure}
\begin{center}
\includegraphics[width=6.5cm]{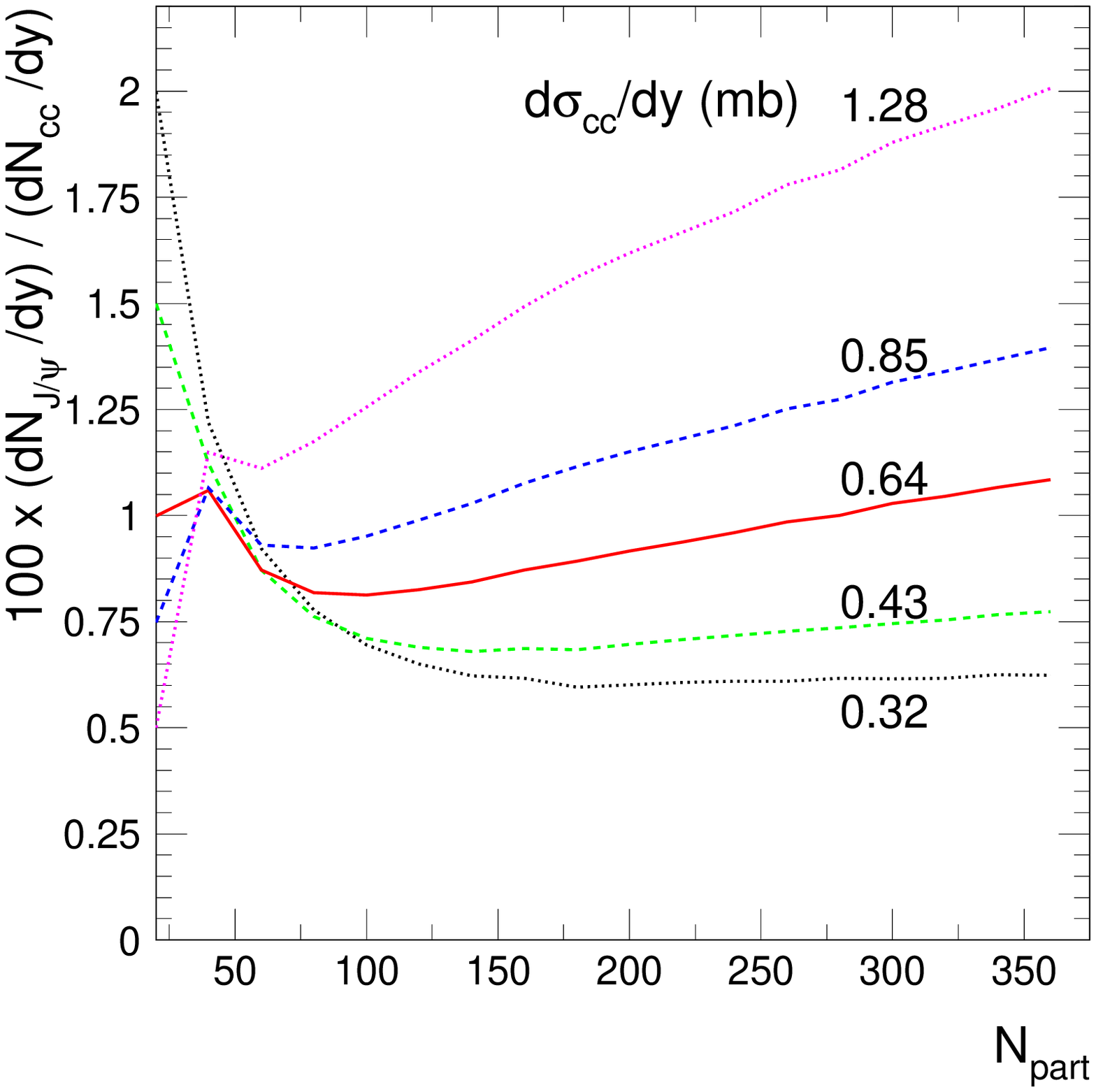}\hfill \includegraphics[width=5.5cm,height=5.5cm]{figthews.eps}
\end{center}
\caption{Left: Predictions for the centrality dependence of the $J/\psi$ yield versus the number of participants, relative to the charm production yield for different
values of the charm cross section indicated on the curves, from Andronic et al.
Right: $\langle p_T^2\rangle$ of the $J/\psi$ versus the number of binary nucleon-nucleon-collisions for different
  nuclear smearing parameters, for initial and in-medium production, from Thews et al. Figures taken from
$^{35}$.}
\label{fig28}
\end{figure}

Another signature of the existence of a deconfined state of quarks would be an enhancement of the quarkonium yield due to a recombination process in which quarks and anti-quarks from the plasma form bound states. Apart from having being proposed as a justification for the baryon-to-meson anomaly and the scaling of $v_2$ normalized to the quark number versus the quark kinetic energy observed at RHIC (see \cite{rhic,Back:2004je,Arsene:2004fa,Adams:2005dq}), such mechanism has been suggested to explain the apparent larger suppression at forward than at central rapidities measured at RHIC \cite{Adare:2006ns}, an effect which goes in opposition to a density-driven suppression - the system is expected to be more dilute far from mid-rapidity. For the purpose of predictions, the largest uncertainty in the recombination mechanism comes from both the charm and the bulk multiplicities at LHC energies, see Fig. \ref{fig28} left
\cite{Andronic:2006ky} for the predictions for a fixed total multiplicity and different charm cross sections. The recombination mechanism could be tested by the different dependence of the quarkonium average transverse momentum on centrality with and without recombination, see Fig. \ref{fig28} right
\cite{Thews:2005vj}. In this latter plot initial production correspond to the initial yield of $J/\psi$ coming from the hard collisions (in which $J/\psi$ is formed from $c\bar{c}$ pairs from the same nucleon-nucleon collision), while in-medium formation corresponds to recombination. The parameter $\lambda$ accounts for the difference of the transverse momentum of the $J/\psi$ in pp and in pA, assuming for the latter a proportionality with the  number of binary nucleon-nucleon collisions $N_{coll}$,
\begin{equation}
\langle p_T^2\rangle_{pA}=\langle p_T^2\rangle_{pp}+\lambda^2(N_{coll}-1).
\label{smea}
\end{equation}

\begin{figure}
\begin{center}
\includegraphics[width=6cm]{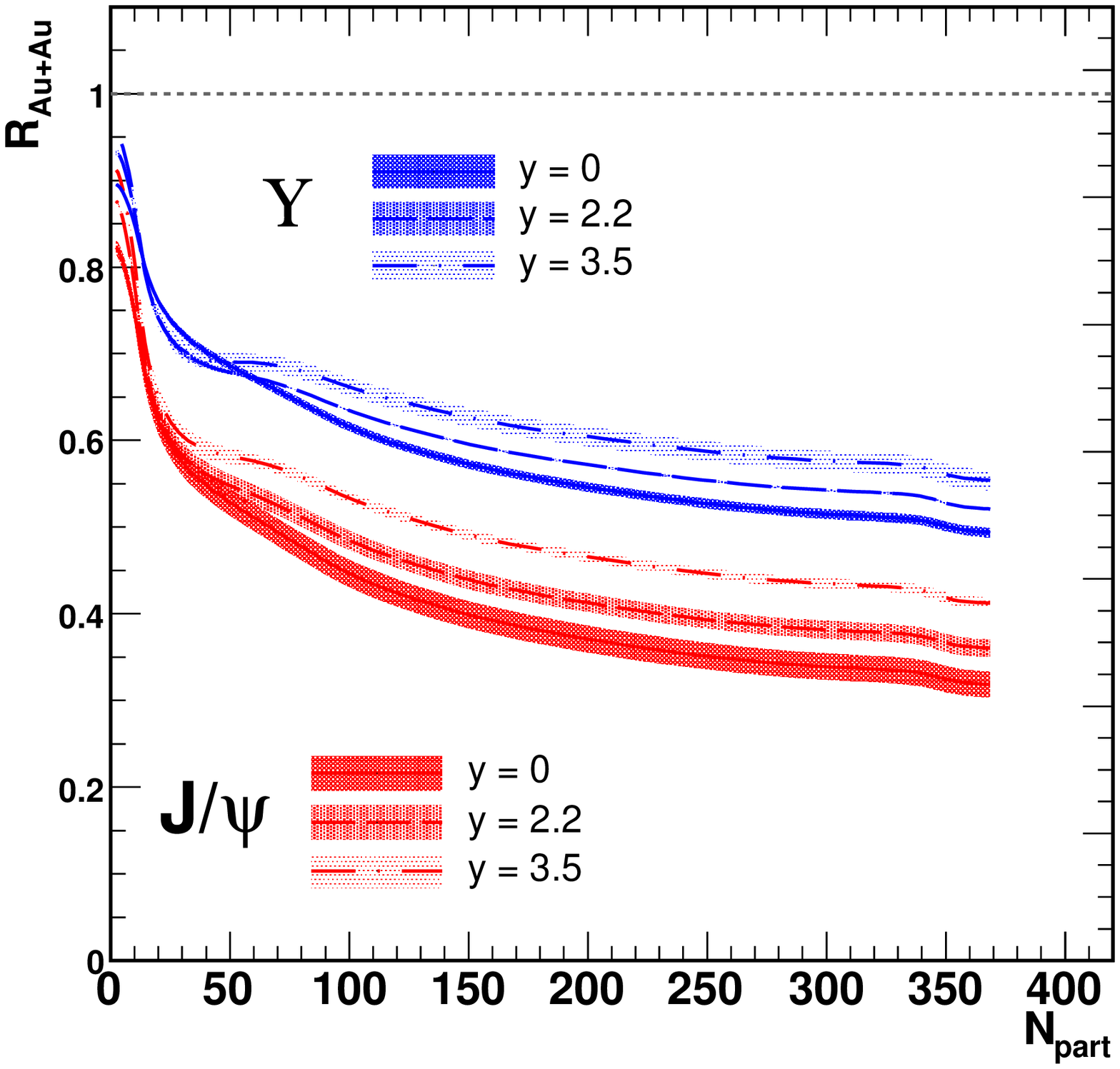}\hfill \includegraphics[width=6cm]{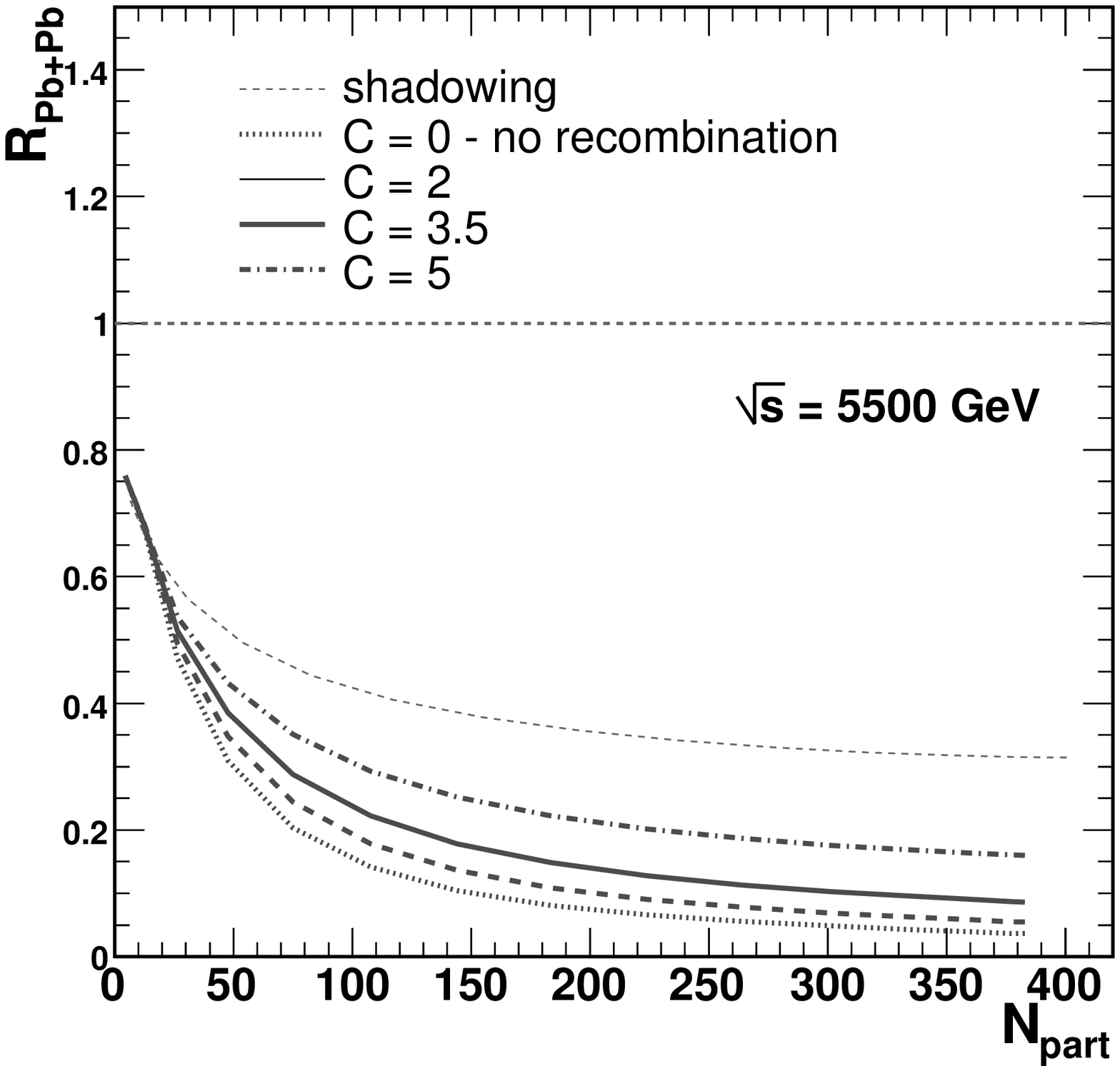}
\end{center}
\caption{Left: Nuclear modification factor of $J/\psi$ and $\Upsilon$ versus the number of participant nucleons for PbPb collisions at the LHC with only cold nuclear matter effects (shadowing) and at different rapidities, from Bravina et al., with the bands reflecting different parametrizations for the gluon densities and different kinematics. Figure taken from
$^{35}$.
Right: Nuclear modification factor of $J/\psi$ versus the number of participant nucleons for PbPb collisions at the LHC, with only the effects of shadowing and with comover suppression for different recombination strengths ranging from none to high ($C=0\div 5$), from Capella et al. Figure taken from
$^{196}$.}
\label{fig29}
\end{figure}

The only prediction which could be - to my knowledge - directly compared to data is that from \cite{Capella:2007jv}. It illustrates the different effects that enter in the calculation: no absorption, strong nuclear shadowing (see Fig. \ref{fig29} left), $J/\psi$ suppression by comoving particles (for a charged multiplicity at mid-pseudorapidity around 1800, see Capella et al. in Subsection \ref{multi}) and recombination with different values of $C(y)=(dN_{c\bar{c}}^{pp}/dy)^2/dN_{J/\psi}^{pp}/dy$. The different effects are illustrated in Fig. \ref{fig29} right. The magnitude of the effect of recombination in this plot is much smaller than the one to be seen in Fig. \ref{fig28} left, although the parameter fixing the recombination varies in a similar range. This is due to the kinematic requirements imposed on the $c\bar{c}$ for recombining, which are different in both models.

Finally, let me indicate that heavy-flavor and quarkonium production can discriminate the different mechanisms of particle production. In fact, recombination can be formulated in the framework of statistical hadronization models. In the proposal by Rafelski et al. \cite{Abreu:2007kv,Kuznetsova:2006bh} a sudden hadronization of the QGP could lead to strangeness over-saturation, which would imply an enhancement in the relative production of charmed-strange mesons and baryons over non-strange ones, with the corresponding diminution of the recombination probability for $J/\psi$ formation.

\subsection{Photons and dileptons}
\label{photons}

Electromagnetic probes - photons and dileptons - lie at the core of the discussions on QGP formation, see the recent reviews \cite{Chatterjee:2009rs,Tserruya:2009zt}. In principle, thermal real and virtual photons are the golden signature of hot matter produced in the collisions. But they have to compete with many other sources - thus reducing the signal to background ratio - originating from both non-equilibrated partonic matter or the hadronic phase.

\begin{figure}
\begin{center}
\includegraphics[width=6cm,height=5cm]{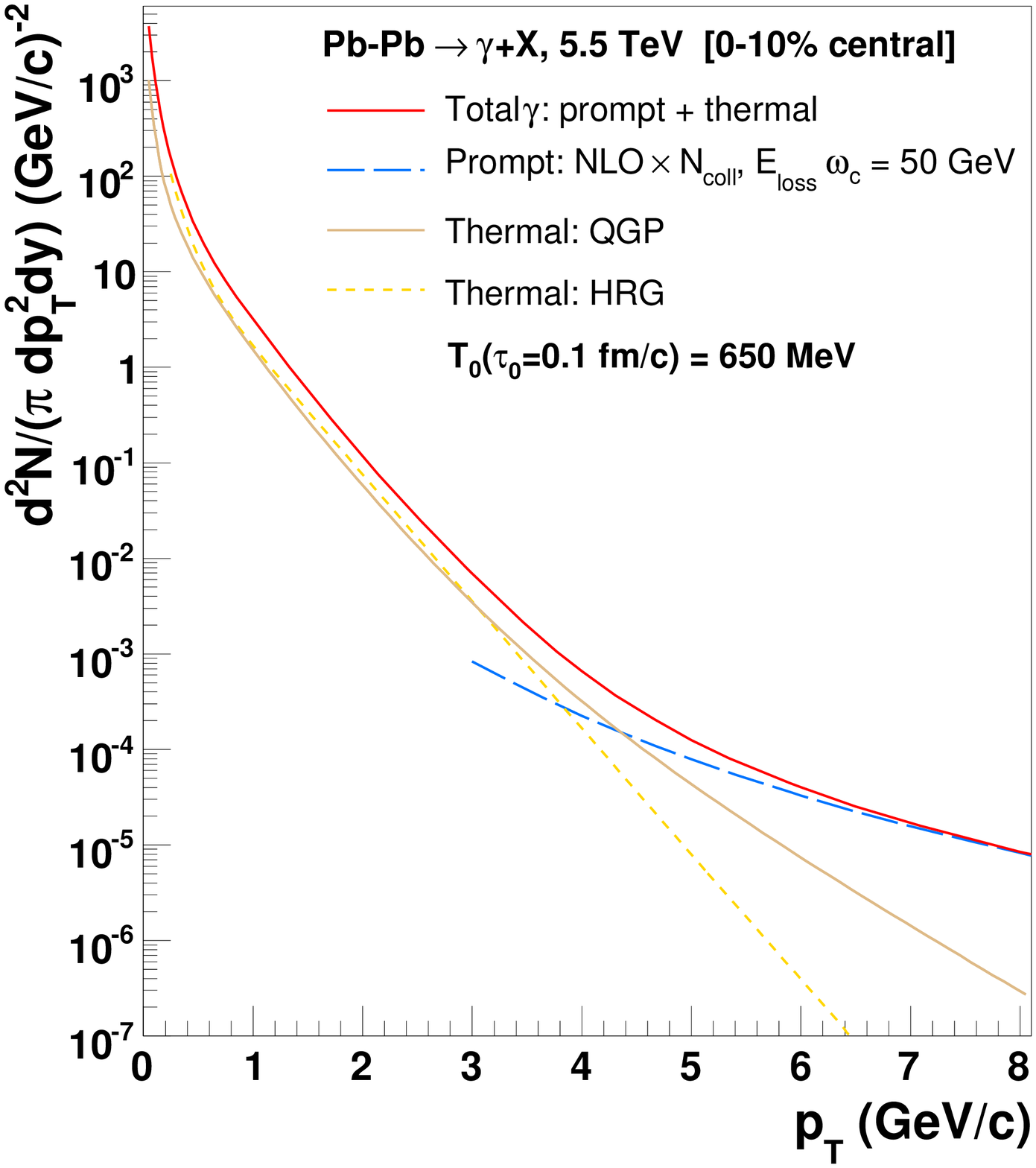}\hfill \includegraphics[width=6cm,height=5cm]{figivanphotons.eps}
\end{center}
\caption{Left: Photon yield versus transverse momentum at mid-rapidity from pQCD plus thermal emission from a QGP, from Arleo et al.
Right: Nuclear modification factor for photons in central dPb and PbPb collisions at the LHC with cold nuclear matter effects and initial and final state energy loss, from Vitev. Figures taken from
$^{35}$.}
\label{fig30}
\end{figure}

Photons at large and intermediate $p_T$ are the usual tool for calibration as they are not affected by the presence of a QGP and are well described by pQCD \cite{Adler:2005ig} (see also Rezaeian in \cite{Abreu:2007kv} for production mechanisms alternative to collinear factorization in pQCD). At low $p_T$, an excess is claimed \cite{:2008fqa} which is compatible with thermal emission from an equilibrated source with $T>300$ MeV at initial times $<0.6$ fm, see e.g. \cite{d'Enterria:2005vz}. Correspondingly, predictions exist for the LHC, see Fig. \ref{fig30} left in which the results of a hydrodynamical model coupled to a NLO pQCD calculation are shown, by Arleo et al. \cite{Abreu:2007kv}. The parameters used correspond to a multiplicity of 1300 charged particles at mid-rapidity. On the other hand, the description of photons in pQCD is far from being simple. There are contributions from the nuclear modification of parton densities, initial state energy loss, and final state energy loss which deviates the nuclear modification factor for direct photons (i.e. not coming from decays or conversions) from 1 - even in pA collisions, see Fig. \ref{fig30} right, Vitev \cite{Abreu:2007kv} and \cite{Vitev:2008vk,Arleo:2006xb}.

\begin{figure}
\begin{center}
\includegraphics[height=8cm,angle=270]{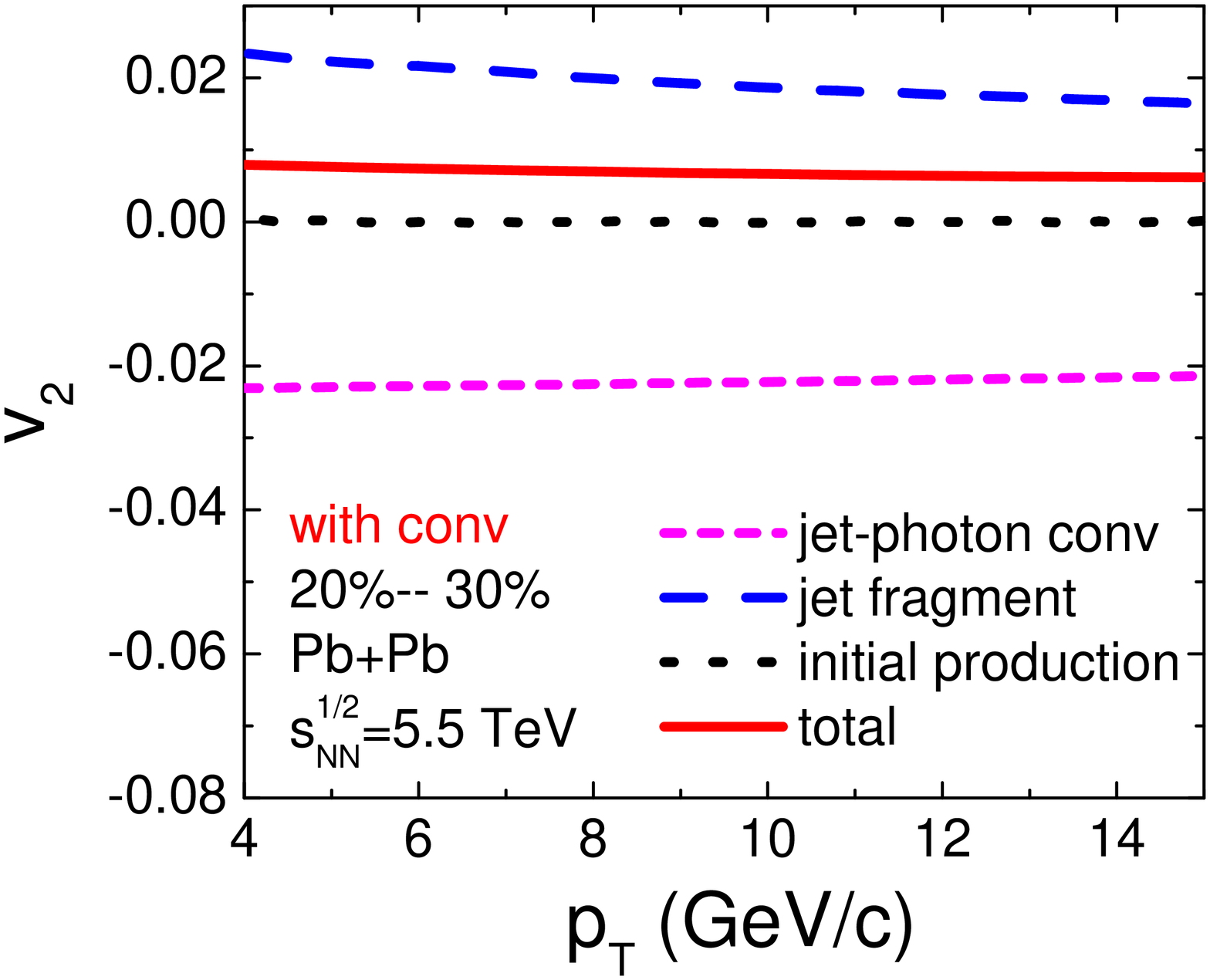}
\vskip 0.8cm
\includegraphics[width=7cm]{heinzphotons.eps}
\end{center}
\caption{Top: Photon $v_2$ versus transverse momentum from different sources, taking into account conversions. Figure taken from
$^{206}$.
Bottom: Photon $v_2$ versus transverse momentum within ideal hydrodynamics in AuAu at RHIC and PbPb at the LHC for $b=7$ fm, with separated contributions from QGP, from hadron matter and total, from Chatterjee et al. Figure taken from
$^{35}$.}
\label{fig31}
\end{figure}

The elliptic flow coefficient $v_2$ for photons is a most delicate measurement (a small signal affected by huge backgrounds) but offers great possibilities, as it is sensitive to the details of the mechanism of photon production. For example, the contributions from conversions (inverse Compton scattering, \cite{Fries:2002kt}) is negative, see Fig. \ref{fig31} top \cite{Liu:2008zb}. It is also very sensitive, within hydrodynamical calculations, to the initial thermalization time \cite{Chatterjee:2009ec,Chatterjee:2009ys,Chatterjee:2008tp}. In Fig. \ref{fig31} bottom the different contributions to the photon $v_2$ in the framework of ideal hydrodynamics are shown, from Chatterjee et al. \cite{Abreu:2007kv}. The contribution from the QGP phase with respect to the hadronic phase is larger at the LHC than at RHIC, as expected.

\begin{figure}
\begin{center}
\vskip 0.2cm
\includegraphics[width=7cm]{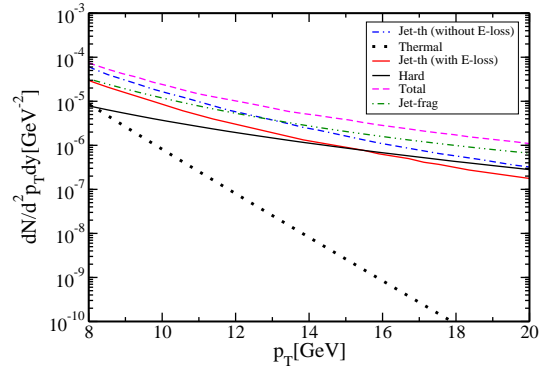}
\end{center}
\caption{Photon yield versus transverse momentum for 10\% central PbPb collisions at the LHC, for $T_i = 0.897$ GeV
and $\tau_i=0.073$ fm, with different contributions separated. Figure taken from
$^{210}$.}
\label{fig32}
\end{figure}

To conclude with photons, in Fig. \ref{fig32} the different contributions (hard, thermal-jet with and without energy loss, thermal, and fragmentation) to the transverse momentum spectrum of photons with $p_T>8$ GeV at the LHC, are shown \cite{Bhattacharya:2008kb} (see also \cite{Turbide:2005fk}). The disentanglement of a thermal component on the background looks defying and demands a very detailed understanding of the background sources.

Now I turn to dilepton production. Dileptons offer interesting information both in the low mass region $M<1$ GeV and in the intermediate mass region $1$ GeV $<M<M_{J/\psi}$, see e.g. 
\cite{Specht:2007ez,Ruppert:2007cr,vanHees:2007th} and references therein. In the former they are expected to reflect the changes of resonances (masses, widths) in the medium. In the latter a window of sizable thermal emission has been speculated.

\begin{figure}
\begin{center}
\includegraphics[width=7cm]{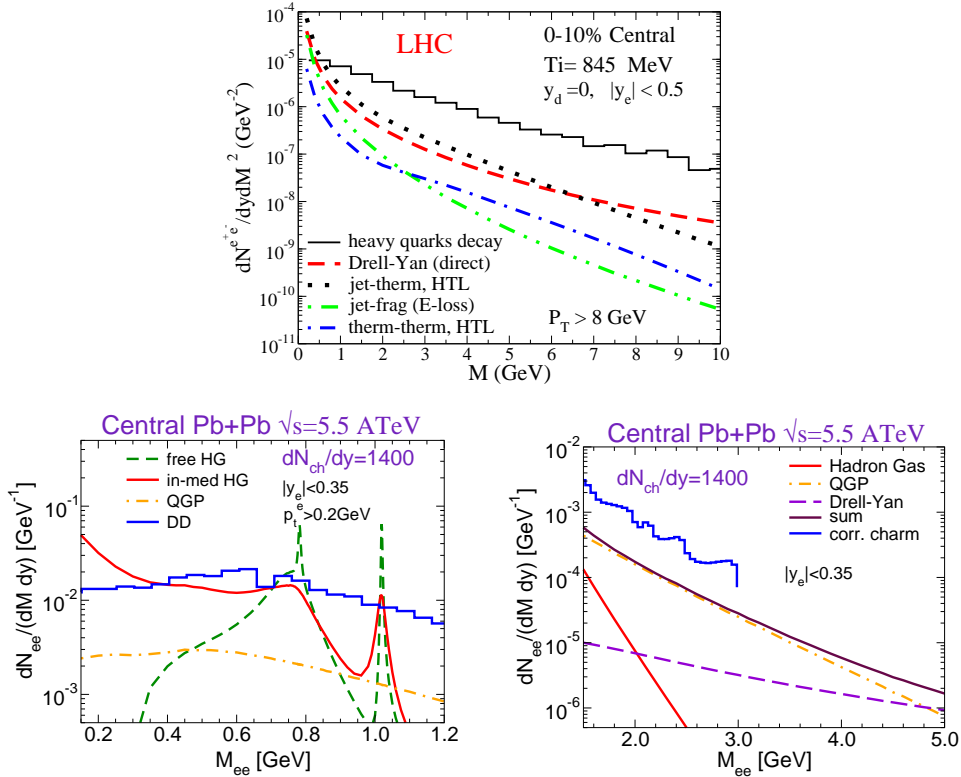}
\includegraphics[width=6cm]{vanheesdileptons1.eps}\hfill\includegraphics[width=6cm]{vanheesdileptons2.eps}
\end{center}
\caption{Top: Dilepton yield versus pair mass at mid-rapidity in central PbPb collisions at the LHC, for $p_T^{pair}>8$ GeV, with different sources indicated, for a scenario corresponding to $dN_{ch}/dy|_{y=0}=5625$, from Fries et al.
Bottom: Id. for $p_T^{pair}>0.2$ GeV, for the low (left) and intermediate (right) mass regions, for a scenario corresponding to $dN_{ch}/dy|_{y=0}=1400$, from van Hees et al. Figures taken from
$^{35}$.}
\label{fig33}
\end{figure}

In Fig. \ref{fig33} \cite{Turbide:2006mc,Abreu:2007kv} the different contributions to the dilepton spectra at large (top) and small (bottom) transverse momentum is shown. The contribution from heavy-quark decays seems to dominate all masses but $M<0.5$ GeV, where hadronic contributions are very large. Therefore, the identification of thermal sources looks defying. A larger multiplicity should be linked with a larger temperature but also with a larger hadronic background. It seems that only a larger light multiplicity - if originating from a larger temperature - linked with a smaller heavy-quark cross section - leading to a smaller background - would improve the situation for detection of thermal dileptons in the intermediate mass region.

\begin{figure}
\begin{center}
\vskip 0.2cm
\includegraphics[height=5.5cm]{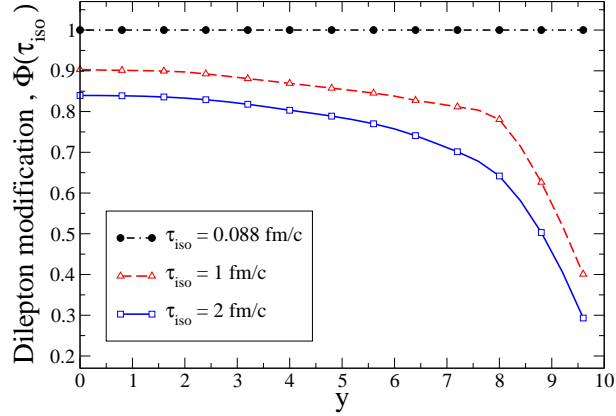}
\end{center}
\caption{Ratio of dilepton yield in the anisotropic situation over the isotropic one, versus rapidity, for $M\ge 2$ GeV and $p_T\ge 0.1$ GeV. The initial temperature is fixed to 845 MeV for $\tau_{iso}=0.088$ fm in the early isotropization scenario, and the parameters for the other situations are determined in order to keep the same final multiplicity. Figure taken from
$^{217}$.}
\label{fig34}
\end{figure}

All calculations shown until now for photons or dileptons in the low-$p_T$ region assume a very small thermalization time, $< 1$ fm, at which the system is isotropized/thermalized. The physical  mechanisms which could make such fast isotropization feasible, are not understood yet. Therefore some authors have studied the possibility of a later isotropization time and a previous evolution in an anisotropic stage. For example, the authors in \cite{Martinez:2008di,Martinez:2008mc} consider a model in which the system evolves from an early formation time $\sim 0.1$ fm in an anisotropic stage (a collisionally-broadened expansion) to an isotropization reached at 2 fm. They find a signal of such anisotropic behavior (depending on the kinematical cuts applied) in the enhancement of dileptons with large transverse momentum at $y=0$, and a suppression of the $p_T$-integrated yield (larger for forward rapidities), compared to the early isotropization scenario, see Fig. \ref{fig34}. Similar considerations for photons can be found in \cite{Bhattacharya:2008up}.

To conclude, much information can be obtained from real and virtual photons at the LHC but an accurate understanding of backgrounds is required\footnote{The ratio of real to virtual photons has been argued, Alam et al. in \cite{Abreu:2007kv} and \cite{Nayak:2007xv}, to develop a plateau at transverse momentum greater than $\sim 2$ GeV, quite insensitive to details of the model and reflecting the initial temperature of the system.}.

\section{pA collisions}
\label{pa}

While pA collisions will not take place until several successful data-taking heavy-ion runs have occurred, they offer a vast amount of information (see
\cite{Accardi:2004be} and references therein) which finally may turn out to be essential for the interpretation of the PbPb data, as it was the case with dAu collisions at RHIC. They should establish the benchmark for the cold nuclear matter effects on top of which the eventual signals of a dense partonic stage are to be searched. I do not intend to give a full overview of all the possibilities of the pA programme, but rather focus on some selected aspects.

\begin{figure}
\begin{center}
\includegraphics[height=5.8cm]{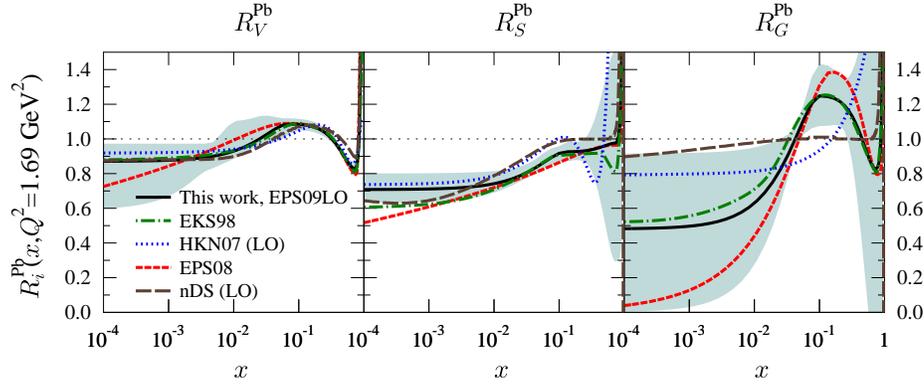}
\end{center}
\caption{Comparison of different parametrizations at lowest order, for the ratio of parton densities (valence, sea and gluons from left to right) per nucleon in Pb over that in p, at $Q^2=1.69$ GeV$^2$. The band represents the uncertainty extracted from the error analysis in the EPS09 parametrizations. Figure taken from
$^{223}$.}
\label{fig35}
\end{figure}

First, pA collisions offer the possibility of constraining the nuclear parton densities in kinematical regions, see Fig. \ref{fig2} right, which will not be explored in lepton-nucleus collisions unless future colliders \cite{eic,lhec} become eventually available. This is a key ingredient for hard probes, and the present situation of the parton densities in the $x$ region of interest for the LHC ($ 10^{-4}\lesssim x\lesssim 10^{-2}$) derived from DGLAP analysis (see e.g. the review \cite{Armesto:2006ph} or the recent work \cite{Eskola:2009uj} and references therein) is far from being satisfactory, see Fig. \ref{fig35} \cite{Eskola:2009uj}.
As evident from this figure, the nuclear gluon densities at a low virtuality for $x<0.05$ are very badly constrained\footnote{The situation is similar for NLO analysis, see \cite{Eskola:2009uj}. On the other hand, DGLAP evolution reduces the uncertainties at larger scales.}. The inclusion of pA data from the LHC in the fits can only  improve this situation.

\begin{figure}
\begin{center}
\includegraphics[height=7cm]{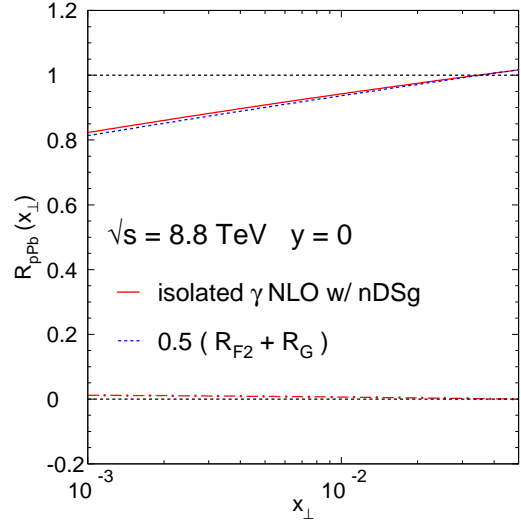}
\end{center}
\caption{$R_{pA}(y=0)$ (solid) for isolated photons in pPb collisions at $\sqrt{s_{NN}} = 8.8$ TeV, and $(R_{F_2}+R_{g})/2$ for Pb (dashed), versus $x_\perp=2p_T/\sqrt{s_{NN}}$. The lower (dashed-dotted)  curve corresponds to the difference between these two quantities divided by $R_{pA}$. Figure taken from
$^{35}$.}
\label{fig36}
\end{figure}

On the other hand, isolated photons offer the possibility of a direct access to the gluon distribution (Fig. \ref{fig36}), see Arleo in \cite{Abreu:2007kv} and \cite{Arleo:2007js}, where it is shown that the nuclear modification factor for isolated photons closely describes with the nuclear modifications factors for the gluon distribution and structure function $F_2$.

\begin{figure}
\begin{center}
\includegraphics[height=5cm]{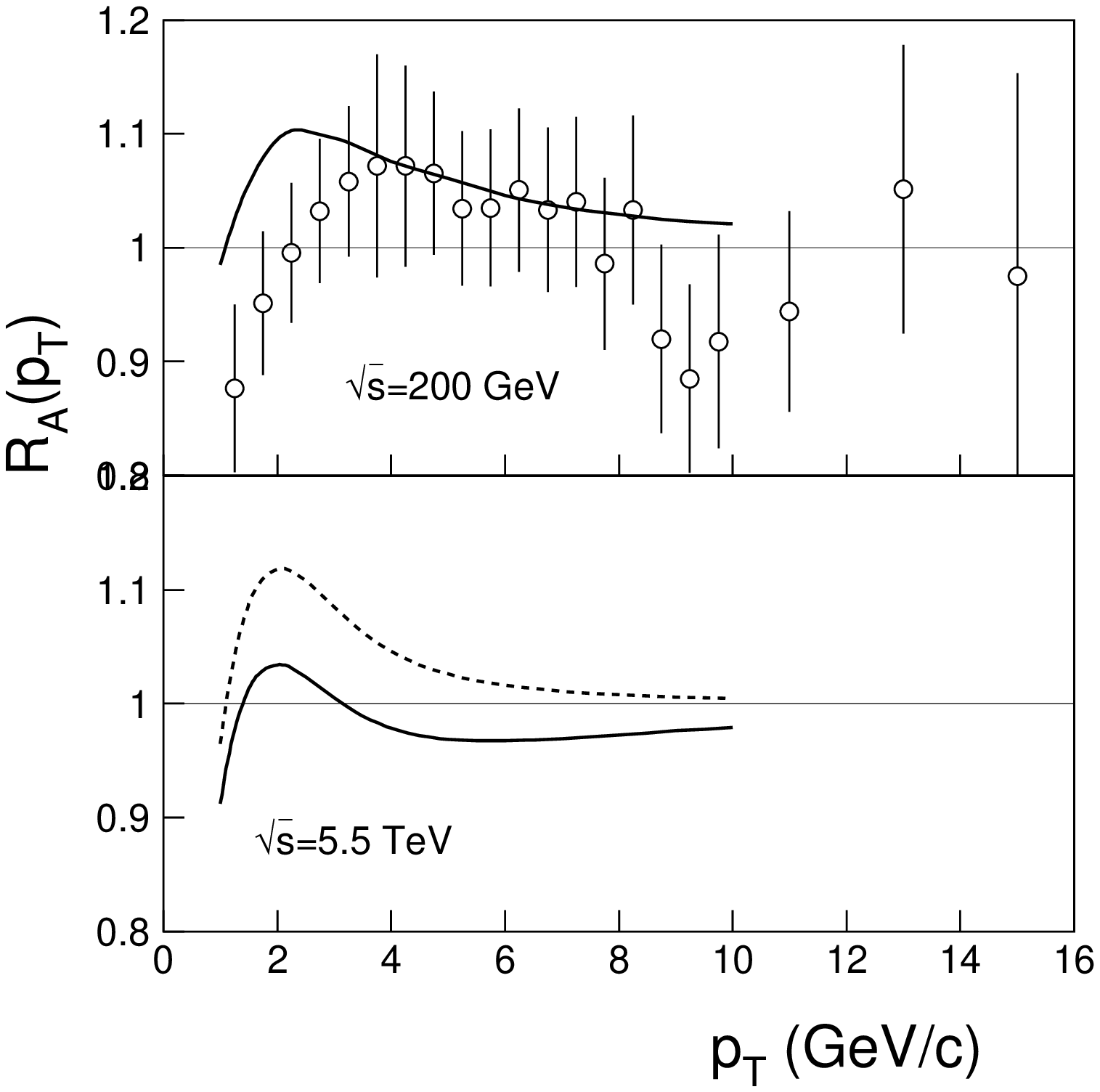}\hfill \includegraphics[height=6.4cm]{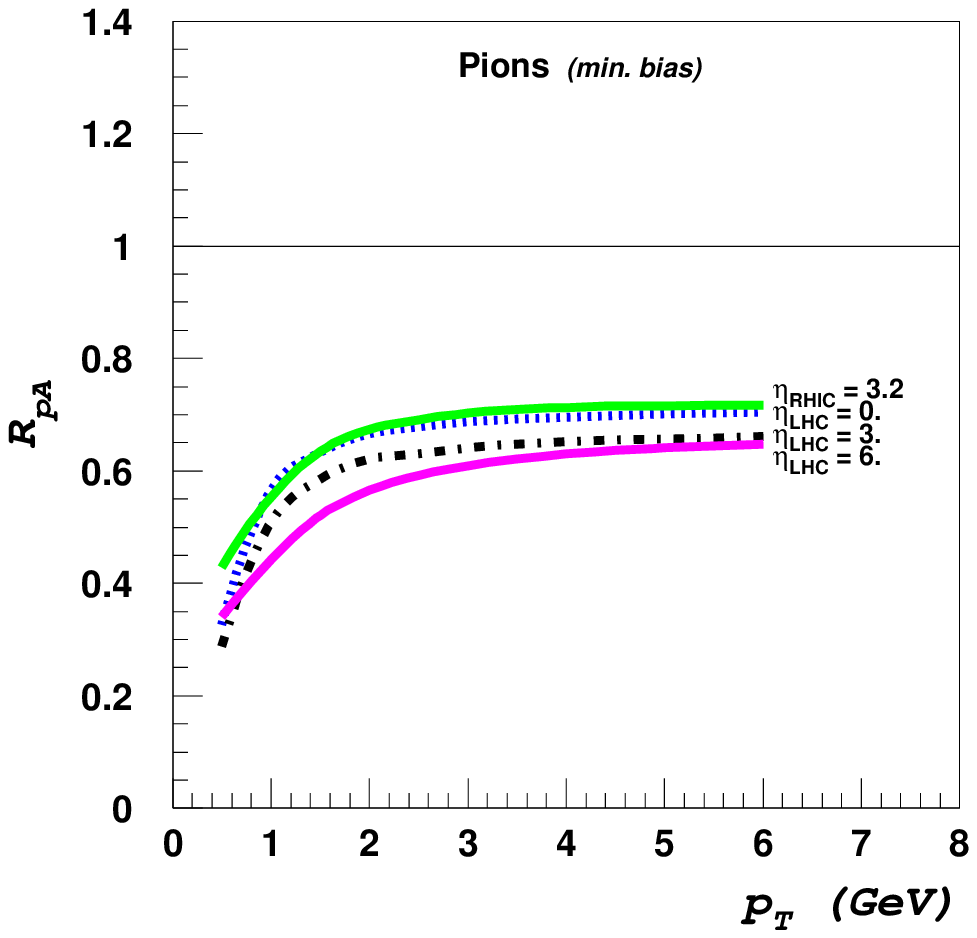}
\end{center}
\caption{Left: $R_{pA}(y=0)$ versus $p_T$ for pions in dAu collisions at RHIC (upper panel) and in pPb collisions at the LHC, by Kopeliovich et al. Solid and dashed lines correspond to the calculation with and without gluon shadowing respectively.
Right: $R_{pA}$ for pions versus $p_T$ in dAu collisions at RHIC and in pPb collisions at the LHC, for different $\eta$, from Tuchin.
Figures taken from
$^{35}$.}
\label{fig37}
\end{figure}

Another aspect which has raised large interest is the possibility to check the ideas of gluon saturation as proposed in the framework of the CGC, see the review in \cite{qgp3}. Generically, the saturation scale which characterizes the momentum below which the gluon densities are expected to be maximal, is expected to increase with increasing rapidity or energy, reaching values in the range from 1 to several GeV in heavy-ion collisions at the LHC. Therefore saturation effects should become visible in a region usually considered within the range of applicability of pQCD. Specifically, the CGC predicts \cite{Albacete:2003iq,JalilianMarian:2005jf} that the Cronin effect (the fact that the nuclear modification factor is larger than 1) observed at mid-rapidity in dAu collisions at RHIC disappears with increasing rapidity - as observed at RHIC, see \cite{Arsene:2004fa} - and increasing energy. While there is no consensus on this suppression at forward rapidities at RHIC being a clear signal of saturation in the CGC, see e.g.  \cite{Guzey:2004zp} or Bravina et al. in \cite{Abreu:2007kv} for an alternative approach to shadowing, pA collisions at the LHC offer the possibility of further tests. In Fig. 
\ref{fig37} left I show the predictions by Kopeliovich et al. 
\cite{Abreu:2007kv,Kopeliovich:2002yh} in which the Cronin effect at mid-rapidity is still present in pPb collisions at the LHC. On the other hand, in Fig. \ref{fig37} right predictions are shown within the CGC framework by Tuchin  \cite{Abreu:2007kv,Kharzeev:2004yx}. While these predictions are for light flavors (see also De Boer et al.  \cite{Abreu:2007kv}, or \cite{JalilianMarian:2004er,Betemps:2004xr,Betemps:2008yw}
for predictions for Drell-Yan and photons), Tuchin also provides predictions for heavy-flavor production with similar features, namely a marked suppression of ratios both at mid- and forward rapidities in pPb collisions at the LHC. Clearly different scenarios should be discriminated by LHC data.

\begin{figure}
\begin{center}
\includegraphics[width=6cm]{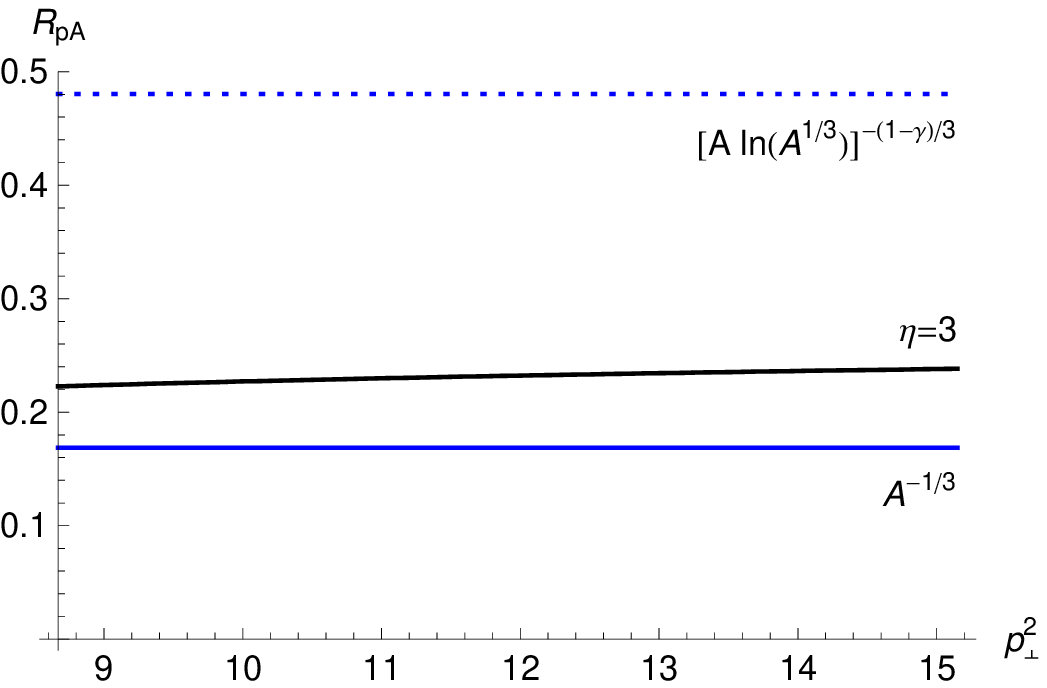}\hfill \includegraphics[width=6cm,height=4.3cm]{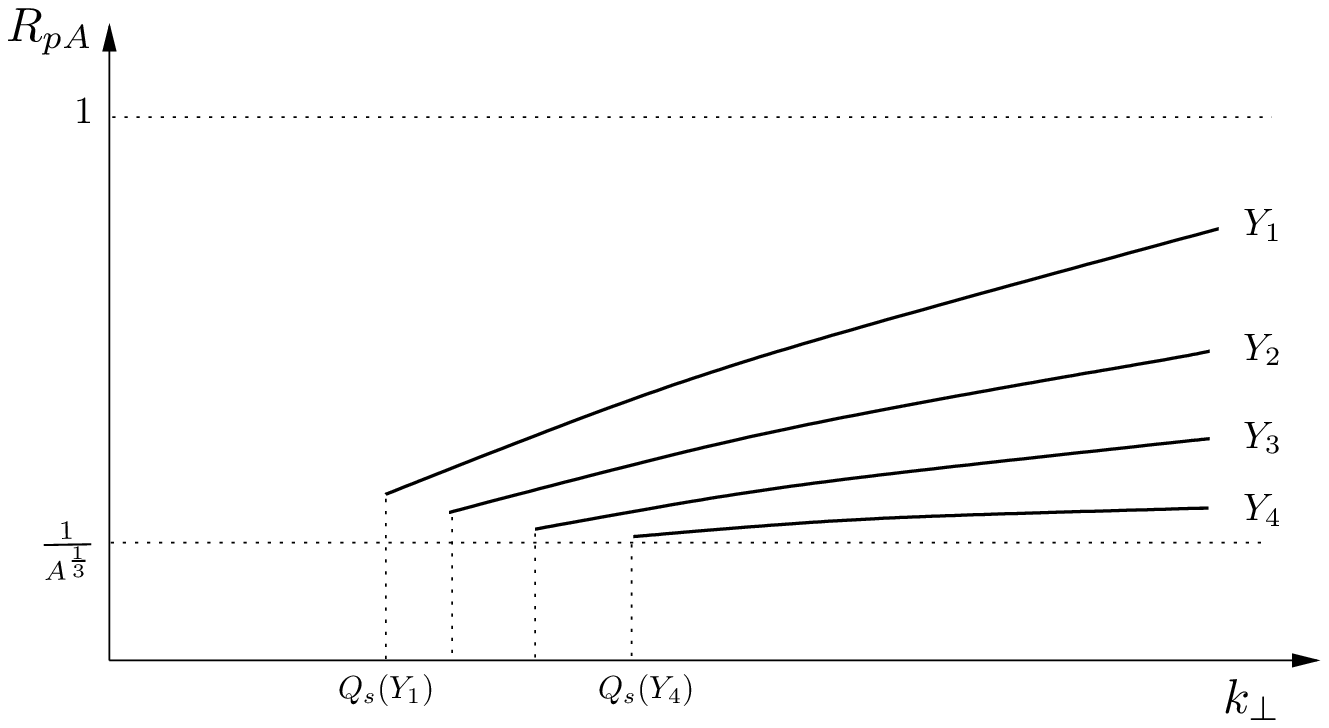}
\end{center}
\caption{Left:  $R_{pA}$ versus $p_T^2$ for pPb at the LHC at $\eta=3$ from small-$x$ evolution for fixed coupling (dashed) and running coupling (solid black), from Iancu et al. The lower, blue line corresponds to total gluon shadowing. Right: Schematic plot of $R_{pA}$ versus transverse momentum $k_\perp$ showing the increasing effect of fluctuations with increasing rapidities and the approach to total shadowing, by Kozlov et al.  Figures taken from
$^{35}$.}
\label{fig38}
\end{figure}

Finally, in the framework of the CGC, the nuclear modification factor in pPb collisions at the LHC offers the possibility of establishing the relevance of different effects. For example, considering a running coupling instead of a fixed coupling in the CGC evolution equations at high energies or small momentum fractions $x$ (the BK equation, see \cite{qgp3}) leads \cite{Iancu:2004bx} to a nuclear modification factor at very large rapidities which goes from $\left(A\ln A^{1/3}\right)^{-(1-\gamma)/3}$ ($\gamma\simeq 0.63$, fixed coupling) to $A^{-1/3}$ (running coupling, called total shadowing), with $A$ the mass number of the nucleus. The same total shadowing is achieved when fluctuations (or pomeron loops, see \cite{Kovner:2005pe,Triantafyllopoulos:2005cn}) are included \cite{Kozlov:2006qw}. Both effects are illustrated in Fig. \ref{fig38}.

\section{Summary and discussion}
\label{conclu}

In this work I have reviewed the predictions for the heavy-ion programme at the LHC, as available in early April 2009. After an introduction I have discussed some qualitative expectations with the aim of illustrating how a single observable, namely charged multiplicity at mid-rapidity, influences predictions for the energy density and other thermodynamical quantities, the evolution of the system, predictions for elliptic flow ($v_2$) or the nuclear modification factor ($R_{AA}$) in models of energy loss.

Then I have turned to a compilation of results (additional information can be found in \cite{Accardi:2004be,Accardi:2004gp,Bedjidian:2004gd,Arleo:2004gn,Armesto:2000xh,Borghini:2007ub,Abreu:2007kv,Armesto:2008fj}). Referring to PbPb collisions at the LHC and, otherwise stated, to observables at mid-rapidity, a summary of what was presented is:

\begin{enumerate}

\item In Subsection \ref{multi} I have discussed the predictions for charged multiplicity at mid-pseudorapidity. Most predictions (for $N_{part}\sim 350$, $\sim 10$ \% more central collisions) now lie below 2000 - a value sizably smaller than pre-RHIC predictions \cite{Armesto:2000xh}, and they include a large degree of coherence in particle production through saturation, strong gluon shadowing, strong color fields,$\dots$ On the other hand, the expectations for net protons at $\eta=0$ are systematically below 4.

\item In Subsection \ref{flow} I have analyzed the results for elliptic flow in several models. $p_T$-integrated $v_2$ increases in all models when going from RHIC to the LHC, but this increase is usually smaller in hydrodynamical models than in naive expectations, \cite{Borghini:2007ub} and Section \ref{qualitative}, and in some non-equilibrium, transport models. For $v_2(p_T)$, hydrodynamical models indicate a value for $p_T\lesssim 2$ GeV which is very close for pions, while a decrease is expected for protons. A strong decrease would be interpreted - once the initial conditions are settled - as an increase in viscous effects. On the other hand, non-equilibrium models generically result in an increase of $v_2(p_T)$.

\item In Subsection \ref{hadroch} predictions for hadrochemistry are reviewed. Different versions of the statistical models result in slightly different predictions, and non-equilibrium scenarios show distinctive features for resonance production. Hydrodynamical and recombination models predict large baryon-to-meson ratios at moderate $p_T$. Approaches with strong color fields or percolation show Cronin effect  for protons in central PbPb collisions at the LHC.

\item In Subsection \ref{correl} I have reviewed the predictions for correlations. HBT radii are expected to increase from RHIC to the LHC. The predictive power of ideal hydrodynamics is reduced by the limitations that appear in its description of RHIC data. The role of viscosity in the hydrodynamical descriptions of HBT radii is still to be clarified. On the other hand, correlations in rapidity are expected to extend along large intervals and offer additional possibilities  of constraining the multiparticle production mechanism.

\item In Subsection \ref{fluctu} I have shown the existing predictions for multiplicity fluctuations - few predictions are available as the evidence of a non-statistical or non-trivial origin of fluctuations at SPS and RHIC is still under debate. Fluctuations also hold discriminative power between different mechanisms of particle production e.g. different statistical ensembles.

\item In Subsection \ref{highpt} I have enumerated the predictions for the nuclear modification factor for high-$p_T$ charged particles or pions in central collisions. They generically lie, for radiative or collisional energy loss models, in the range $0.15\div 0.25$ at $p_T=20$ GeV and increasing with increasing $p_T$. Then I  have commented the possibilities of discriminating between the energy loss mechanism offered by jets, by hadrochemistry at large $p_T$ where several mechanisms like energy loss and parton conversions may be simultaneously at work, and by the study of correlations.

\item In Subsection \ref{heavy} results from different models with radiative or collisional energy loss for the nuclear modification factor of heavy flavors have been shown. They offer the possibility to further test the energy loss mechanism, as the energy diminution of a heavy quark traveling through the produced medium is different from that of a massless parton. On the other hand, predictions for quarkonium production are uncertain due to the lack of knowledge of both cold nuclear matter (nuclear parton densities and nuclear absorption) and hot nuclear matter (pattern of dissociation, recombination mechanism at work,$\dots$) effects. The identification of different quarkonium states and the large $p_T$ reach at the LHC, may help to settle the dissociation pattern and the role of recombination. But predictions for the nuclear modification factor of $J/\psi$ are plagued with uncertainties due to e.g. nuclear shadowing or the $c\bar{c}$ cross section for recombination.

\item In Subsection \ref{photons} I have reviewed the available predictions for photon and dilepton production. While the large initial temperature or energy density implies a large yield of thermal real and virtual photons, the huge backgrounds make the disentanglement of a thermal component in the final spectrum challenging - a very precise knowledge of the pp baseline will be required. Effects beyond the usual equilibrium scenarios like anisotropies in the pre-equilibrium stage may modify the yields with respect to the early thermalization expectations.

\item In Section \ref{pa} I have analyzed the usefulness of the pA programme at the LHC for the purpose of reducing the uncertainties in the nuclear parton distributions which weaken the capabilities of hard probes to characterize the medium produced in the collisions. I have also discussed the possibilities of studies of high gluon density QCD through measurements of the nuclear modification factor in pPb collisions in a large rapidity interval.

\end{enumerate}

To put in context the predictions with respect to our current interpretation of existing data, let me draw a set of rough predictions for the LHC 
with respect to every 'standard' claim based on the experimental findings at RHIC (see Section \ref{intro}):
\begin{center}
\begin{tabular}{c|c|c}
Finding at RHIC & 'Standard' interpretation & Prediction for the LHC \\ \hline
multiplicities  & highly coherent & \\
smaller than & particle production & $dN_{ch}^{PbPb}/d\eta|_{\eta=0}<2000$\\
expectations& (expected e.g. in CGC) & \\ \hline
$v_2$ in agreement &  quasi-ideal fluid & $v_2(p_T)$ for $p_T<2\ {\rm GeV}$ similar \\
with ideal hydro & (strongly coupled QGP) & or smaller than at RHIC \\ \hline
strong &  very opaque & $R_{AA}^{light}\sim 0.2$ at $p_T\sim 20$ GeV \\
jet quenching & medium & and increasing with $p_T$ \\ \hline
\end{tabular}
\end{center}

Obviously, neither the standard interpretations nor the predictions presented in this Table are free from problems and uncertainties, even more when the predictions tend to disagree with naive, data-driven expectations which would suggest multiplicities of order 1000, and sizably larger $v_2(p_T<2\ {\rm GeV})$  and smaller $R_{AA}$ than at RHIC.

Finally, I find it tempting to speculate on possible scenarios based on the first-day measurement of charged particle production at mid-pseudorapidity in central PbPb collisions. Without any intention beyond showing how our understanding may become affected by the very first data and having in mind the present experimental situation and its 'standard' interpretation, three rough possibilities can be discussed:

\begin{itemize}

\item A low multiplicity scenario, $dN_{ch}^{PbPb}/d\eta|_{\eta=0}<1000$, which would be close to the wounded nucleon model expectations and even smaller than most data-driven expectations. It would imply a extremely coherent particle production, difficult to describe even in saturation models. The conditions for collective flow would be relatively close to those at RHIC, and differentiating between naive extrapolations and hydrodynamical behaviors for $v_2$ more involved, as their predictions would be not so different. On high transverse momentum particle production, the fact that the densities are close to RHIC ones, would imply that the difference e.g. in $R_{AA}$ from RHIC would be driven by the different transverse momentum spectra - the trigger bias, so the expectation would be an $R_{AA}$ larger than at RHIC for the same large transverse momentum (e.g. of the order or greater than 20 GeV). The low multiplicity implies a small background for jet and correlation studies. A small light multiplicity could also be a good scenario for recombination models for quarkonia (for a fixed heavy quark cross section).

\item An intermediate multiplicity scenario, $1000<dN_{ch}^{PbPb}/d\eta|_{\eta=0}<2000$ as predicted by most models with a large degree of coherence and by data-driven extrapolations. The differences between naive predictions and results of hydrodynamical models for $v_2$ would be more noticeable. $R_{AA}$ should be more similar than at RHIC, for the same large transverse momentum, than in the previous scenario.

\item A large multiplicity scenario, $2000<dN_{ch}^{PbPb}/d\eta|_{\eta=0}$. This scenario would defy naive extrapolations based on logarithmic increases and limiting fragmentation, and would be very problematic for saturation physics. Discriminating between naive predictions and results of hydrodynamical models for $v_2$ should be relatively easy. In this case, a strong decrease of $v_2$ at fixed small $p_T$ with respect to RHIC, would strongly suggest viscous effects. $R_{AA}$ at large $p_T$, of the order or greater than 20 GeV, could be smaller than at RHIC for the same transverse momentum. Jet and correlation studies might be more defying due to the larger background. On the other hand, this scenario would imply larger temperatures and energy densities which may be welcome, even in spite of the larger background, for electromagnetic probes.

\end{itemize}

To conclude, the heavy-ion programme at the LHC will offer most valuable information for improving our understanding of high-density QCD matter - and, in a wider context, on the behavior of the strong interaction at high energies -  from the very first day of data taking. But it should be kept in mind that the usefulness of some observables will be restricted by our lack of knowledge of the pp and pA benchmarks, in particular to constrain the parton densities in nuclei. It seems plausible that a pA run will be needed - as it was the case at RHIC - in order to understand the effects of cold nuclear matter at LHC energies before strong conclusions about the heavy-ion programme can be drawn.

A large amount of work has already been done to extrapolate existing models to the LHC situation. Still much work is needed in order to deal with some observables e.g. viscous hydrodynamical calculations or transport models for collective flow, or Monte Carlo tools for jet analysis, just to mention two obvious ongoing developments. The first LHC data will reduce much of the available freedom in model parameters. The more restricted model predictions done after those very first data will indicate, when confronted with subsequent data on other observables, whether the physics at the LHC is qualitatively similar to that at the SPS and RHIC or, on the contrary, new aspects appear which will require new ideas. 

\section*{Acknowledgements}

I thank J. Albacete, J. Alvarez-Muniz, F. Bopp, W. Busza, R. Concei\c cao, L. Cunqueiro, A. Dainese, J. Dias de Deus, A. El, D. d'Enterria, K. Eskola, V. Gon\c calves, U. Heinz, C.-M. Ko, I. Lokhtin, M. Martinez, J. G. Milhano, C. Pajares, V. Pantuev, T. Renk, E. Sarkisyan, V. Topor Pop, K. Tywoniuk, R. Venugopalan, I. Vitev, X. N. Wang, K. Werner and G. Wolschin for information on their predictions, P. Brogueira, J. Dias de Deus and J. G. Milhano for an updated version of a figure in their paper \cite{Brogueira:2007ub}, and U. Heinz, A. Mischke,  E. Saridakis,  H. Song and D. Srivastava for useful comments. Special thanks are due to David d'Enterria, Guilherme Milhano, Carlos Pajares, Carlos Salgado and Konrad Tywoniuk for a critical reading of the manuscript. 
This work has been supported by Ministerio de Ciencia e Innovaci\'on of Spain under projects FPA2005-01963,  FPA2008-01177 and a contract Ram\'on y Cajal, by Xunta de Galicia (Conseller\'{\i}a de Educaci\'on) and through grant PGIDIT07PXIB206126PR, and by the Spanish Consolider-Ingenio 2010 Programme CPAN (CSD2007-00042).


\begin{thebibliography}{999}

\bibitem{Heinz:2000bk}
  U.~W.~Heinz and M.~Jacob,
 arXiv:nucl-th/0002042.

\bibitem{rhic}
K.~Adcox {\it et al.}  [PHENIX Collaboration],
Nucl.\ Phys.\ A {\bf 757} (2005) 184.

\bibitem{Back:2004je}
B.~B.~Back {\it et al.} [PHOBOS Collaboration],
Nucl.\ Phys.\ A {\bf 757} (2005) 28.

\bibitem{Arsene:2004fa}
I.~Arsene {\it et al.}  [BRAHMS Collaboration],
Nucl.\ Phys.\ A {\bf 757} (2005) 1.

\bibitem{Adams:2005dq}
J.~Adams {\it et al.}  [STAR Collaboration],
Nucl.\ Phys.\ A {\bf 757}  (2005) 102.

\bibitem{rhic2} Mid-term plan for RHIC:\\
 \texttt{http://www.bnl.gov/HENP/docs/RHICplanning/RHIC\_Mid-termplan\_print.pdf}.

\bibitem{sps2} J. Bartke et al. [NA49-future Collaboration],  ``A New Experimental Programme with Nuclei and
Proton Beams at the CERN SPS", preprint CERN-SPSC-2003-038 (SPSC-EOI-01).

\bibitem{Jowett:2008hb}
  J.~M.~Jowett,
  J.\ Phys.\ G {\bf 35} (2008) 104028.

\bibitem{Accardi:2004be}
  A. Accardi {\it et al.},
 arXiv:hep-ph/0308248.

\bibitem{lhc}
  F.~Carminati {\it et al.}  [ALICE Collaboration],
  J.\ Phys.\ G {\bf 30}  (2004) 1517.

\bibitem{Alessandro:2006yt}
  B.~Alessandro {\it et al.}  [ALICE Collaboration],
  J.\ Phys.\ G {\bf 32} (2006) 1295.

\bibitem{D'Enterria:2007xr}
  D.~G.~d'Enterria {\it et al.}  [CMS Collaboration],
  J.\ Phys.\ G {\bf 34} (2007) 2307.

\bibitem{Steinberg:2007nm}
  P.~Steinberg  [ATLAS Collaboration],
  J.\ Phys.\ G {\bf 34} (2007) S527.

\bibitem{Accardi:2004gp}
  A. Accardi {\it et al.},
 arXiv:hep-ph/0310274.

\bibitem{Bedjidian:2004gd}
  M. Bedjidian {\it et al.},
  arXiv:hep-ph/0311048.

\bibitem{Arleo:2004gn}
  F. Arleo {\it et al.},
 arXiv:hep-ph/0311131.

\bibitem{d'Enterria:2008ge}
  D.~d'Enterria  [CMS Collaboration],
  J.\ Phys.\ G {\bf 35} (2008) 104039.
    
\bibitem{Cortese:2008zza}
  P.~Cortese {\it et al.}  [ALICE Collaboration],
  ``ALICE electromagnetic calorimeter technical design report'', preprint CERN-LHCC-2008-014/CERN-ALICE-TDR-014.

\bibitem{jets}
  S.~Salur  [for the STAR Collaboration],
arXiv:0809.1609 [nucl-ex].

\bibitem{Putschke:2008wn}
  J.~Putschke [for the STAR Collaboration],
  arXiv:0809.1419 [nucl-ex].
  
\bibitem{d'Enterria:2008is}
  D.~d'Enterria,
  AIP Conf.\ Proc.\  {\bf 1038} (2008) 95.

\bibitem{Gyulassy:2004zy}
  M.~Gyulassy and L.~McLerran,
  Nucl.\ Phys.\  A {\bf 750} (2005) 30.

\bibitem{Jacobs:2004qv}
  P.~Jacobs and X.~N.~Wang,
  Prog.\ Part.\ Nucl.\ Phys.\  {\bf 54} (2005) 443.

\bibitem{Muller:2007rs}
  B.~Muller,
  Acta Phys.\ Polon.\  B {\bf 38} (2007) 3705.
  
\bibitem{d'Enterria:2006su}
  D.~G.~d'Enterria,
  J.\ Phys.\ G {\bf 34} (2007) S53.
  
\bibitem{Bass:1999zq}
  S. A. Bass  {\it et al.},
  Nucl. Phys.  A {\bf 661} (1999) 205.

\bibitem{Armesto:2000xh}
  N. Armesto and C. Pajares,
  Int. J. Mod. Phys. A {\bf 15} (2000) 2019.

\bibitem{lattice}
  F.~Karsch,
  Lect.\ Notes Phys.\  {\bf 583} (2002) 209.

\bibitem{Cheng:2007jq}
  M.~Cheng {\it et al.},
  Phys.\ Rev.\  D {\bf 77} (2008) 014511.
  
\bibitem{Fodor:2007ue}
  Z.~Fodor,
  PoS C {\bf POD07} (2007) 027.
    
  \bibitem{qgp1} R. Hwa (ed.), {\it Quark-Gluon Plasma}, Vol. 1 (World Scientific, Singapore, 1990).
    
  \bibitem{qgp2}  R. Hwa (ed.), {\it Quark-Gluon Plasma}, Vol. 2 (World Scientific, Singapore, 1995).
  
  \bibitem{qgp3} R. Hwa and X. N. Wang (eds.), {\it Quark-Gluon Plasma}, Vol. 3 (World Scientific, Singapore, 2004);
  
  P.~F.~Kolb and U.~W.~Heinz,
  arXiv:nucl-th/0305084;
  
  A.~Kovner and U.~A.~Wiedemann,
  arXiv:hep-ph/0304151;

  M.~Gyulassy, I.~Vitev, X.~N.~Wang and B.~W.~Zhang,
  arXiv:nucl-th/0302077;

  E.~Iancu and R.~Venugopalan,
  arXiv:hep-ph/0303204.

\bibitem{Borghini:2007ub}
  N.~Borghini and U.~A.~Wiedemann,
  J.\ Phys.\ G {\bf 35} (2008) 023001.
  
\bibitem{Abreu:2007kv}
N.~Armesto {\it et al.},
J.\ Phys.\ G {\bf 35} (2008) 054001.

\bibitem{Armesto:2008fj}
  N.~Armesto,
  J.\ Phys.\ G {\bf 35} (2008) 104042.

\bibitem{upc}
  K.~Hencken {\it et al.},
  Phys.\ Rept.\  {\bf 458} (2008) 1.

\bibitem{d'Enterria:2008sh}
  D.~d'Enterria, M.~Klasen and K.~Piotrzkowski,
  Nucl.\ Phys.\ Proc.\ Suppl.\  B {\bf 179} (2008) 1.

\bibitem{Amelin:2001sk}
  N. S. Amelin, N. Armesto, C. Pajares and D. Sousa,
 Eur. Phys. J. C {\bf 22} (2001) 149.
  
\bibitem{Bialas:1976ed}
  A.~Bialas, M.~Bleszynski and W.~Czyz,
  Nucl.\ Phys.\  B {\bf 111} (1976) 461.
  
\bibitem{Collins:1985gm}
  J.~C.~Collins, D.~E.~Soper and G.~Sterman,
  Nucl.\ Phys.\  B {\bf 263} (1986) 37.

\bibitem{Collins:1985ue}
  J.~C.~Collins, D.~E.~Soper and G.~Sterman,
  Nucl.\ Phys.\  B {\bf 261} (1985) 104.

\bibitem{Capella:1999kv}
  A.~Capella, A.~Kaidalov and J.~Tran Thanh Van,
  Heavy Ion Phys.\  {\bf 9} (1999) 169.

\bibitem{Abramovsky:1973fm}
  V.~A.~Abramovsky, V.~N.~Gribov and O.~V.~Kancheli,
  Yad.\ Fiz.\  {\bf 18} (1973) 595
  [Sov.\ J.\ Nucl.\ Phys.\  {\bf 18} (1974)] 308.
  
\bibitem{Back:2004dy}
  B.~B.~Back {\it et al.}  [PHOBOS Collaboration],
  Phys.\ Rev.\  C {\bf 70} (2004) 021902.
  
\bibitem{Abe:1989td}
  F.~Abe {\it et al.}  [CDF Collaboration],
  Phys.\ Rev.\  D {\bf 41} (1990) 2330.

\bibitem{Armesto:2004ud}
  N.~Armesto, C.~A.~Salgado and U.~A.~Wiedemann,
  Phys.\ Rev.\ Lett.\  {\bf 94} (2005) 022002.
  
\bibitem{Sjostrand:2006za}
  T.~Sjostrand, S.~Mrenna and P.~Skands,
  JHEP {\bf 0605} (2006) 026.
    
  \bibitem{alicetp}
  N. Ahmad {\it et al.} [ALICE Collaboration], preprint CERN/LHCC 1995-071.

\bibitem{Bjorken:1982qr}
  J.~D.~Bjorken,
  Phys.\ Rev.\  D {\bf 27} (1983) 140.

\bibitem{Albajar:1989an}
  C.~Albajar {\it et al.}  [UA1 Collaboration],
  Nucl.\ Phys.\  B {\bf 335} (1990) 261.
  
\bibitem{Hirano:2008hy}
  T.~Hirano, N.~van der Kolk and A.~Bilandzic,
  arXiv:0808.2684 [nucl-th].

\bibitem{Niemi:2008ta}
  H.~Niemi, K.~J.~Eskola and P.~V.~Ruuskanen,
  arXiv:0806.1116 [hep-ph].

\bibitem{viscous}
  M.~Luzum and P.~Romatschke,
  arXiv:0901.4588 [nucl-th].
  
\bibitem{Chaudhuri:2008je}
  A.~K.~Chaudhuri,
  Phys.\ Lett.\  B {\bf 672} (2009) 126.

\bibitem{Song:2008hj}
  H.~Song and U.~W.~Heinz,
  arXiv:0812.4274 [nucl-th].
  
\bibitem{Romatschke:2009im}
  P.~Romatschke,
  arXiv:0902.3663 [hep-ph].

\bibitem{cgcini}
  D.~Kharzeev and K.~Tuchin,
  Nucl.\ Phys.\  A {\bf 753} (2005) 316.

\bibitem{Kharzeev:2000ph}
   D.~Kharzeev and M.~Nardi,
  Phys.\ Lett.\  B {\bf 507} (2001) 121.
  
\bibitem{Beuf:2008vd}
  G.~Beuf, R.~Peschanski and E.~N.~Saridakis,
  Phys.\ Rev.\  C {\bf 78} (2008) 064909.

\bibitem{Voloshin:1999gs}
  S.~A.~Voloshin and A.~M.~Poskanzer,
  Phys.\ Lett.\  B {\bf 474} (2000) 27.

\bibitem{Alt:2003ab}
  C.~Alt {\it et al.}  [NA49 Collaboration],
  Phys.\ Rev.\  C {\bf 68} (2003) 034903.

\bibitem{Adler:2002pu}
  C.~Adler {\it et al.}  [STAR Collaboration],
  Phys.\ Rev.\  C {\bf 66} (2002) 034904.
  
\bibitem{Ollitrault:1992bk}
  J.~Y.~Ollitrault,
  Phys.\ Rev.\  D {\bf 46} (1992) 229.

\bibitem{Kolb:1999it}
  P.~F.~Kolb, J.~Sollfrank and U.~W.~Heinz,
  Phys.\ Lett.\  B {\bf 459} (1999) 667.
  
\bibitem{Lappi:2006xc}
  T.~Lappi and R.~Venugopalan,
  Phys.\ Rev.\  C {\bf 74} (2006) 054905.
  
\bibitem{Kestin:2008bh}
  G.~Kestin and U.~W.~Heinz,
  arXiv:0806.4539 [nucl-th].
  
\bibitem{Baier:2001yt}
  R.~Baier, Y.~L.~Dokshitzer, A.~H.~Mueller and D.~Schiff,
  JHEP {\bf 0109} (2001) 033.
  
\bibitem{Salgado:2003gb}
  C.~A.~Salgado and U.~A.~Wiedemann,
  Phys.\ Rev.\  D {\bf 68} (2003) 014008.
  
\bibitem{ToporPop:2007hb}
  V.~Topor Pop, M.~Gyulassy, J.~Barrette, C.~Gale, S.~Jeon and R.~Bellwied,
  Phys.\ Rev.\  C {\bf 75} (2007) 014904.
  
\bibitem{Bopp:2004xn}
  F.~W.~Bopp, J.~Ranft, R.~Engel and S.~Roesler,
  arXiv:hep-ph/0403084.
  
\bibitem{Lin:2004en}
  Z.~W.~Lin, C.~M.~Ko, B.~A.~Li, B.~Zhang and S.~Pal,
  Phys.\ Rev.\  C {\bf 72} (2005) 064901.
  
\bibitem{Wang:1991hta}
  X.~N.~Wang and M.~Gyulassy,
  Phys.\ Rev.\  D {\bf 44} (1991) 3501.
  
\bibitem{Lokhtin:2009be}
  I.~P.~Lokhtin, L.~V.~Malinina, S.~V.~Petrushanko, A.~M.~Snigirev, I.~Arsene and K.~Tywoniuk,
  arXiv:0903.0525 [hep-ph].
  
\bibitem{Mitrovski:2008hb}
  M.~Mitrovski, T.~Schuster, G.~Graf, H.~Petersen and M.~Bleicher,
  arXiv:0812.2041 [hep-ph].
  
\bibitem{Drescher:2000ha}
  H.~J.~Drescher, M.~Hladik, S.~Ostapchenko, T.~Pierog and K.~Werner,
  Phys.\ Rept.\  {\bf 350} (2001) 93.
  
\bibitem{Sa:2008fw}
  B.~H.~Sa, D.~M.~Zhou, B.~G.~Dong, Y.~L.~Yan, H.~L.~Ma and X.~M.~Li,
  J.\ Phys.\ G {\bf 36} (2009) 025007.
  
\bibitem{DiasdeDeus:2007wb}
  J.~Dias de Deus and J.~G.~Milhano,
  Nucl.\ Phys.\  A {\bf 795} (2007) 98.
  
\bibitem{Albacete:2007sm}
  J.~L.~Albacete,
  Phys.\ Rev.\ Lett.\  {\bf 99} (2007) 262301.
  
\bibitem{Kharzeev:2004if}
  D.~Kharzeev, E.~Levin and M.~Nardi,
  Nucl.\ Phys.\  A {\bf 747} (2005) 609.
  
\bibitem{Bzdak:2008gw}
  A.~Bzdak,
  Acta Phys.\ Polon.\  B {\bf 39} (2008) 1977.
  
\bibitem{El:2007vg}
  A.~El, Z.~Xu and C.~Greiner,
  Nucl.\ Phys.\  A {\bf 806} (2008) 287.
  
\bibitem{Humanic:2008nt}
  T.~J.~Humanic,
  arXiv:0810.0621 [nucl-th].
  
\bibitem{Sarkisyan:2005rt}
  E.~K.~G.~Sarkisyan and A.~S.~Sakharov,
  AIP Conf.\ Proc.\  {\bf 828} (2006) 35.
  
\bibitem{Wong:2008ta}
  C.~Y.~Wong,
  arXiv:0809.0517 [nucl-th].
  
\bibitem{Kuiper:2006si}
  R.~Kuiper and G.~Wolschin,
  Annalen Phys.\  {\bf 16} (2007) 67.
  
\bibitem{Bearden:2003hx}
  I.~G.~Bearden {\it et al.}  [BRAHMS Collaboration],
  Phys.\ Rev.\ Lett.\  {\bf 93} (2004) 102301.
  
\bibitem{MehtarTani:2008qg}
  Y.~Mehtar-Tani and G.~Wolschin,
  arXiv:0811.1721 [hep-ph].
      
\bibitem{Arsene:2009fn}
  I.~C.~Arsene  [BRAHMS Collaboration],
  arXiv:0901.0872 [nucl-ex].

\bibitem{AlvarezMuniz:2009kh}
  J.~Alvarez-Muniz, R.~Conceicao, J.~D.~de Deus, M.~C.~E.~Santo, J.~G.~Milhano and M.~Pimenta,
  arXiv:0903.0957 [hep-ph].

\bibitem{Gubser:2009sx}
  S.~S.~Gubser, S.~S.~Pufu and A.~Yarom,
  arXiv:0902.4062 [hep-th].
  
\bibitem{Torrieri:2009ms}
  G.~Torrieri, I.~Mishustin and B.~Tomasik,
  arXiv:0901.0226 [nucl-th].
  
\bibitem{Denicol:2009am}
  G.~S.~Denicol, T.~Kodama, T.~Koide and Ph.~Mota,
  arXiv:0903.3595 [hep-ph].

\bibitem{Monnai:2009ad}
  A.~Monnai and T.~Hirano,
  arXiv:0903.4436 [nucl-th].
  
\bibitem{Drescher:2007cd}
  H.~J.~Drescher, A.~Dumitru, C.~Gombeaud and J.~Y.~Ollitrault,
  Phys.\ Rev.\  C {\bf 76} (2007) 024905.
  
\bibitem{Back:2004mh}
  B.~B.~Back {\it et al.}  [PHOBOS Collaboration],
  Phys.\ Rev.\  C {\bf 72} (2005) 051901.
  
\bibitem{Chojnacki:2007rq}
  M.~Chojnacki, W.~Florkowski, W.~Broniowski and A.~Kisiel,
  Phys.\ Rev.\  C {\bf 78} (2008) 014905.

\bibitem{Petersen:2009vx}
  H.~Petersen and M.~Bleicher,
  arXiv:0901.3821 [nucl-th].
  
\bibitem{Molnar:2001ux}
  D.~Molnar and M.~Gyulassy,
  Nucl.\ Phys.\  A {\bf 697} (2002) 495
  [Erratum-ibid.\  A {\bf 703} (2002) 893].

\bibitem{Policastro:2001yc}
  G.~Policastro, D.~T.~Son and A.~O.~Starinets,
  Phys.\ Rev.\ Lett.\  {\bf 87} (2001) 081601.

\bibitem{Arnold:2000dr}
  P.~Arnold, G.~D.~Moore and L.~G.~Yaffe,
  JHEP {\bf 0011} (2000) 001.
  
\bibitem{Kraus:2009xa}
  I.~Kraus, J.~Cleymans, H.~Oeschler, K.~Redlich and S.~Wheaton,
  arXiv:0902.0873 [hep-ph].
  
\bibitem{Kraus:2007hf}
  I.~Kraus, J.~Cleymans, H.~Oeschler, K.~Redlich and S.~Wheaton,
  Phys.\ Rev.\  C {\bf 76} (2007) 064903.
  
  \bibitem{Rafelski:2005jc}
  J.~Rafelski and J.~Letessier,
  Eur.\ Phys.\ J.\  C {\bf 45} (2006) 61.
  
\bibitem{Albacete:2003iq}
  J.~L.~Albacete, N.~Armesto, A.~Kovner, C.~A.~Salgado and U.~A.~Wiedemann,
  Phys.\ Rev.\ Lett.\  {\bf 92} (2004) 082001.
  
\bibitem{JalilianMarian:2005jf}
  J.~Jalilian-Marian and Y.~V.~Kovchegov,
  Prog.\ Part.\ Nucl.\ Phys.\  {\bf 56} (2006) 104.
  
\bibitem{Lisa:2008gf}
  M.~A.~Lisa and S.~Pratt,
  arXiv:0811.1352 [nucl-ex].
  
  \bibitem{Kisiel:2008ws}
  A.~Kisiel, W.~Broniowski, M.~Chojnacki and W.~Florkowski,
  Phys.\ Rev.\  C {\bf 79} (2009) 014902.
  
\bibitem{Muronga:2004sf}
  A.~Muronga and D.~H.~Rischke,
  arXiv:nucl-th/0407114.
  
\bibitem{Pratt:2008qv}
  S.~Pratt,
  arXiv:0811.3363 [nucl-th].
  
\bibitem{Gombeaud:2009fk}
  C.~Gombeaud, T.~Lappi and J.~Y.~Ollitrault,
  arXiv:0901.4908 [nucl-th].
  
  \bibitem{Capella:1978rg}
  A.~Capella and A.~Krzywicki,
  Phys.\ Rev.\  D {\bf 18} (1978) 4120.
  
\bibitem{Brogueira:2007ub}
  P.~Brogueira, J.~Dias de Deus and J.~G.~Milhano,
  Phys.\ Rev.\  C {\bf 76} (2007) 064901.
  
\bibitem{Srivastava:2007ei}
  B.~K.~Srivastava  [STAR Collaboration],
  Int.\ J.\ Mod.\ Phys.\  E {\bf 16} (2008) 3371.
  
\bibitem{Brogueira:2009nj}
  P.~Brogueira, J.~D.~de Deus and C.~Pajares,
  arXiv:0901.0997 [hep-ph].

\bibitem{Armesto:2006bv}
  N.~Armesto, L.~McLerran and C.~Pajares,
  Nucl.\ Phys.\  A {\bf 781} (2007) 201.

\bibitem{vanLeeuwen:2008pn}
  M.~van Leeuwen  [STAR collaboration],
  arXiv:0808.4096 [nucl-ex].

\bibitem{Dumitru:2008wn}
  A.~Dumitru, F.~Gelis, L.~McLerran and R.~Venugopalan,
  Nucl.\ Phys.\  A {\bf 810} (2008) 91.
  
\bibitem{Vogt:2007aw}
  R.~Vogt,
  Eur.\ Phys.\ J.\ ST {\bf 155} (2008) 213.
  
\bibitem{rel}
  R.~Baier, D.~Schiff and B.~G.~Zakharov,
  Ann.\ Rev.\ Nucl.\ Part.\ Sci.\  {\bf 50} (2000) 37.

\bibitem{d'Enterria:2009am}
  D.~d'Enterria,
  arXiv:0902.2011 [nucl-ex].

\bibitem{CasalderreySolana:2007zz}
J.~Casalderrey-Solana and C.~A.~Salgado,
  Acta Phys.\ Polon.\  B {\bf 38} (2007) 3731.

\bibitem{Majumder:2007iu}
  A.~Majumder,
  J.\ Phys.\ G {\bf 34} (2007) S377.
  
\bibitem{Gubser:2009sn}
  S.~S.~Gubser, S.~S.~Pufu, F.~D.~Rocha and A.~Yarom,
  arXiv:0902.4041 [hep-th].
  
\bibitem{Baier:2002tc}
  R.~Baier,
  Nucl.\ Phys.\  A {\bf 715} (2003) 209.
  
\bibitem{Liu:2006ug}
  H.~Liu, K.~Rajagopal and U.~A.~Wiedemann,
  Phys.\ Rev.\ Lett.\  {\bf 97} (2006) 182301.
  
\bibitem{CasalderreySolana:2007sw}
  J.~Casalderrey-Solana and X.~N.~Wang,
  Phys.\ Rev.\  C {\bf 77} (2008) 024902.

\bibitem{Antonov:2007sh}
  D.~Antonov and H.~J.~Pirner,
  Eur.\ Phys.\ J.\  C {\bf 55} (2008) 439.
  
\bibitem{Arleo:2007bg}
  F.~Arleo,
  JHEP {\bf 0707} (2007) 032.
  
\bibitem{Dainese:2004te}
  A.~Dainese, C.~Loizides and G.~Paic,
  Eur.\ Phys.\ J.\  C {\bf 38} (2005) 461.

\bibitem{Renk:2006pk}
  T.~Renk and K.~Eskola,
  Phys.\ Rev.\  C {\bf 75} (2007) 054910.
  
\bibitem{Jeon:2002dv}
  S.~Jeon, J.~Jalilian-Marian and I.~Sarcevic,
  Phys.\ Lett.\  B {\bf 562} (2003) 45.
  
\bibitem{Vitev:2005he}
  I.~Vitev,
  Phys.\ Lett.\  B {\bf 639} (2006) 38.

\bibitem{Zhang:2007ja}
  H.~Zhang, J.~F.~Owens, E.~Wang and X.~N.~Wang,
  Phys.\ Rev.\ Lett.\  {\bf 98} (2007) 212301.
  
\bibitem{Qin:2007zz}
  G.~Y.~Qin, J.~Ruppert, S.~Turbide, C.~Gale, C.~Nonaka and S.~A.~Bass,
  Phys.\ Rev.\  C {\bf 76} (2007) 064907.
  
\bibitem{Wicks:2007am}
  S.~Wicks, W.~Horowitz, M.~Djordjevic and M.~Gyulassy,
  Nucl.\ Phys.\  A {\bf 783} (2007) 493.
  
\bibitem{Lokhtin:2005px}
  I.~P.~Lokhtin and A.~M.~Snigirev,
  Eur.\ Phys.\ J.\  C {\bf 45} (2006) 211.
    
\bibitem{Zakharov:2008kt}
  B.~G.~Zakharov,
  arXiv:0811.0445 [hep-ph].
  
\bibitem{Liu:2006sf}
  W.~Liu, C.~M.~Ko and B.~W.~Zhang,
  Phys.\ Rev.\  C {\bf 75} (2007) 051901.
  
\bibitem{Capella:2006fw}
  A.~Capella and E.~G.~Ferreiro,
  Phys.\ Rev.\  C {\bf 75} (2007) 024905.
  
\bibitem{Cunqueiro:2007fn}
  L.~Cunqueiro, J.~Dias de Deus, E.~G.~Ferreiro and C.~Pajares,
  Eur.\ Phys.\ J.\  C {\bf 53} (2008) 585.
  
\bibitem{Kopeliovich:2007yv}
  B.~Z.~Kopeliovich, H.~J.~Pirner, I.~K.~Potashnikova and I.~Schmidt,
  Phys.\ Lett.\  B {\bf 662} (2008) 117.
    
\bibitem{Pantuev:2008zz}
  V.~S.~Pantuev,
  Phys.\ Atom.\ Nucl.\  {\bf 71} (2008) 1625.
  
\bibitem{Cacciari:2007fd}
  M.~Cacciari and G.~P.~Salam,
  Phys.\ Lett.\  B {\bf 659} (2008) 119.
  
\bibitem{Borghini:2005em}
  N.~Borghini and U.~A.~Wiedemann,
  arXiv:hep-ph/0506218.

\bibitem{Borghini:2009eq}
  N.~Borghini,
  arXiv:0902.2951 [hep-ph].
  
\bibitem{Dremin:2006da}
  I.~M.~Dremin and O.~S.~Shadrin,
  J.\ Phys.\ G {\bf 32} (2006) 963.

\bibitem{Armesto:2008qe}
  N.~Armesto, C.~Pajares and P.~Quiroga-Arias,
  arXiv:0809.4428 [hep-ph].
  
\bibitem{Ramos:2008qb}
  R.~P.~Ramos,
  arXiv:0811.2418 [hep-ph].
  
\bibitem{Armesto:2007dt}
  N.~Armesto, L.~Cunqueiro, C.~A.~Salgado and W.~C.~Xiang,
  JHEP {\bf 0802} (2008) 048.
  
\bibitem{Majumder:2009zu}
  A.~Majumder,
  arXiv:0901.4516 [nucl-th].
  
\bibitem{Domdey:2008gp}
  S.~Domdey, G.~Ingelman, H.~J.~Pirner, J.~Rathsman, J.~Stachel and K.~Zapp,
  Nucl.\ Phys.\  A {\bf 808} (2008) 178.
  
\bibitem{Salgado:2003rv}
  C.~A.~Salgado and U.~A.~Wiedemann,
  Phys.\ Rev.\ Lett.\  {\bf 93} (2004) 042301.
  
\bibitem{Vitev:2008rz}
  I.~Vitev, S.~Wicks and B.~W.~Zhang,
  JHEP {\bf 0811} (2008) 093.
  
\bibitem{Armesto:2008qh}
  N.~Armesto, L.~Cunqueiro and C.~A.~Salgado,
  arXiv:0809.4433 [hep-ph].
  
\bibitem{Zapp:2008gi}
  K.~Zapp, G.~Ingelman, J.~Rathsman, J.~Stachel and U.~A.~Wiedemann,
  arXiv:0804.3568 [hep-ph].
  
\bibitem{Renk:2008pp}
  T.~Renk,
  Phys.\ Rev.\  C {\bf 78} (2008) 034908.
  
  \bibitem{Sapeta:2007ad}
  S.~Sapeta and U.~A.~Wiedemann,
  Eur.\ Phys.\ J.\  C {\bf 55} (2008) 293.
  
\bibitem{:2008afa}
  S.~Albino {\it et al.},
  arXiv:0804.2021 [hep-ph].
  
\bibitem{Hwa:2006zq}
  R.~C.~Hwa and C.~B.~Yang,
  Phys.\ Rev.\ Lett.\  {\bf 97} (2006) 042301.
  
  \bibitem{Renk:2007rn}
  T.~Renk and K.~J.~Eskola,
  Phys.\ Rev.\  C {\bf 77} (2008) 044905.
  
\bibitem{:2008nd}
  B.~I.~Abelev {\it et al.}  [STAR Collaboration],
  arXiv:0805.0622 [nucl-ex].
  
\bibitem{Rapp:2009my}
  R.~Rapp and H.~van Hees,
  arXiv:0903.1096 [hep-ph].
  
\bibitem{Dokshitzer:2001zm}
  Y.~L.~Dokshitzer and D.~E.~Kharzeev,
  Phys.\ Lett.\  B {\bf 519} (2001) 199.
  
\bibitem{Armesto:2005iq}
  N.~Armesto, A.~Dainese, C.~A.~Salgado and U.~A.~Wiedemann,
  Phys.\ Rev.\  D {\bf 71} (2005) 054027.
  
\bibitem{Mustafa:2004dr}
  M.~G.~Mustafa,
  Phys.\ Rev.\  C {\bf 72} (2005) 014905.
  
\bibitem{Djordjevic:2008iz}
  M.~Djordjevic and U.~W.~Heinz,
  Phys.\ Rev.\ Lett.\  {\bf 101} (2008) 022302.
  
\bibitem{Adare:2006nq}
  A.~Adare {\it et al.}  [PHENIX Collaboration],
  Phys.\ Rev.\ Lett.\  {\bf 98} (2007) 172301.
  
\bibitem{Abelev:2006db}
  B.~I.~Abelev {\it et al.}  [STAR Collaboration],
  Phys.\ Rev.\ Lett.\  {\bf 98} (2007) 192301.
  
\bibitem{Mischke:2008qj}
  A.~Mischke  [STAR Collaboration],
  J.\ Phys.\ G {\bf 35} (2008) 104117.
  
\bibitem{Mischke:2008af}
  A.~Mischke,
  Phys.\ Lett.\  B {\bf 671} (2009) 361.
  
\bibitem{Adare:2009ic}
  A.~Adare {\it et al.} [PHENIX Collaboration],
  arXiv:0903.4851 [hep-ex].
  
\bibitem{Armesto:2005mz}
  N.~Armesto, M.~Cacciari, A.~Dainese, C.~A.~Salgado and U.~A.~Wiedemann,
  Phys.\ Lett.\  B {\bf 637} (2006) 362.

\bibitem{Adil:2006ra}
  A.~Adil and I.~Vitev,
  Phys.\ Lett.\  B {\bf 649} (2007) 139.
  
\bibitem{Sharma:2009hn}
  R.~Sharma, I.~Vitev and B.~W.~Zhang,
  arXiv:0904.0032 [hep-ph].

\bibitem{Rapp:2005at}
  R.~Rapp, V.~Greco and H.~van Hees,
  Nucl.\ Phys.\  A {\bf 774} (2006) 685.

\bibitem{Gossiaux:2009mk}
  P.~B.~Gossiaux, R.~Bierkandt and J.~Aichelin,
  arXiv:0901.0946 [hep-ph].
  
\bibitem{ConesadelValle:2007sw}
  Z.~Conesa del Valle, A.~Dainese, H.~T.~Ding, G.~Martinez Garcia and D.~C.~Zhou,
  Phys.\ Lett.\  B {\bf 663} (2008) 202.
  
  \bibitem{Horowitz:2008zz}
  W.~A.~Horowitz and M.~Gyulassy,
  Phys.\ Lett.\  B {\bf 666} (2008) 320.
  
\bibitem{Liu:2008bw}
  W.~Liu and R.~J.~Fries,
  Phys.\ Rev.\  C {\bf 78} (2008) 037902.
  
\bibitem{Zhu:2007ne}
  X.~Zhu, N.~Xu and P.~Zhuang,
  Phys.\ Rev.\ Lett.\  {\bf 100} (2008) 152301.
  
\bibitem{Tsiledakis:2009da}
  G.~Tsiledakis and K.~Schweda,
  arXiv:0901.4296 [nucl-ex].
  
\bibitem{Pop:2009sd}
  V.~T.~Pop, J.~Barrette and M.~Gyulassy,
  arXiv:0902.4028 [hep-ph].
  
\bibitem{Matsui:1986dk}
  T.~Matsui and H.~Satz,
  Phys.\ Lett.\  B {\bf 178} (1986) 416.

\bibitem{Arleo:2006qk}
  F.~Arleo and V.~N.~Tram,
  Eur.\ Phys.\ J.\  C {\bf 55} (2008) 449.

\bibitem{Lourenco:2009sk}
  C.~Lourenco, R.~Vogt and H.~K.~Woehri,
  JHEP {\bf 0902} (2009) 014.

\bibitem{Braun:1997qw}
  M.~A.~Braun, C.~Pajares, C.~A.~Salgado, N.~Armesto and A.~Capella,
  Nucl.\ Phys.\  B {\bf 509} (1998) 357.
  
\bibitem{Kopeliovich:2001ee}
  B.~Kopeliovich, A.~Tarasov and J.~Hufner,
  Nucl.\ Phys.\  A {\bf 696} (2001) 669.
  
\bibitem{Capella:2006mb}
  A.~Capella and E.~G.~Ferreiro,
  Phys.\ Rev.\  C {\bf 76} (2007) 064906.

\bibitem{Kaczmarek:2004gv}
  O.~Kaczmarek, F.~Karsch, F.~Zantow and P.~Petreczky,
  Phys.\ Rev.\  D {\bf 70} (2004) 074505
  [Erratum-ibid.\  D {\bf 72} (2005) 059903].

\bibitem{Mocsy:2008eg}
  A.~Mocsy,
  arXiv:0811.0337 [hep-ph].
  
\bibitem{Liu:2006nn}
  H.~Liu, K.~Rajagopal and U.~A.~Wiedemann,
  Phys.\ Rev.\ Lett.\  {\bf 98} (2007) 182301.
  
\bibitem{Adare:2006ns}
  A.~Adare {\it et al.}  [PHENIX Collaboration],
  Phys.\ Rev.\ Lett.\  {\bf 98} (2007) 232301.
  
\bibitem{Andronic:2006ky}
  A.~Andronic, P.~Braun-Munzinger, K.~Redlich and J.~Stachel,
  Nucl.\ Phys.\  A {\bf 789} (2007) 334.
  
\bibitem{Thews:2005vj}
  R.~L.~Thews and M.~L.~Mangano,
  Phys.\ Rev.\  C {\bf 73} (2006) 014904.
  
\bibitem{Capella:2007jv}
  A.~Capella, L.~Bravina, E.~G.~Ferreiro, A.~B.~Kaidalov, K.~Tywoniuk and E.~Zabrodin,
  Eur.\ Phys.\ J.\  C {\bf 58} (2008) 437.
  
\bibitem{Kuznetsova:2006bh}
  I.~Kuznetsova and J.~Rafelski,
  Eur.\ Phys.\ J.\  C {\bf 51} (2007) 113.
 
\bibitem{Chatterjee:2009rs}
  R.~Chatterjee, L.~Bhattacharya and D.~K.~Srivastava,
  arXiv:0901.3610 [nucl-th].
  
\bibitem{Tserruya:2009zt}
  I.~Tserruya,
  arXiv:0903.0415 [nucl-ex].
  
\bibitem{Adler:2005ig}
  S.~S.~Adler {\it et al.}  [PHENIX Collaboration],
  Phys.\ Rev.\ Lett.\  {\bf 94} (2005) 232301.
  
\bibitem{:2008fqa}
  A.~Adare {\it et al.}  [PHENIX Collaboration],
  arXiv:0804.4168 [nucl-ex].
  
\bibitem{d'Enterria:2005vz}
  D.~G.~d'Enterria and D.~Peressounko,
  Eur.\ Phys.\ J.\  C {\bf 46} (2006) 451.
  
\bibitem{Vitev:2008vk}
  I.~Vitev and B.~W.~Zhang,
  Phys.\ Lett.\  B {\bf 669} (2008) 337.
  
\bibitem{Arleo:2006xb}
  F.~Arleo,
  JHEP {\bf 0609} (2006) 015.
  
\bibitem{Fries:2002kt}
  R.~J.~Fries, B.~Muller and D.~K.~Srivastava,
  Phys.\ Rev.\ Lett.\  {\bf 90} (2003) 132301.
  
\bibitem{Liu:2008zb}
  W.~Liu and R.~J.~Fries,
  Phys.\ Rev.\  C {\bf 77} (2008) 054902.
   
\bibitem{Chatterjee:2009ec}
  R.~Chatterjee, D.~K.~Srivastava and S.~Jeon,
  arXiv:0902.1036 [nucl-th].

\bibitem{Chatterjee:2009ys}
  R.~Chatterjee, D.~K.~Srivastava and U.~Heinz,
  arXiv:0901.3270 [nucl-th].

\bibitem{Chatterjee:2008tp}
  R.~Chatterjee and D.~K.~Srivastava,
  Phys.\ Rev.\  C {\bf 79} (2009) 021901.
  
\bibitem{Bhattacharya:2008kb}
  L.~Bhattacharya and P.~Roy,
  arXiv:0806.4033 [hep-ph].
  
\bibitem{Turbide:2005fk}
  S.~Turbide, C.~Gale, S.~Jeon and G.~D.~Moore,
  Phys.\ Rev.\  C {\bf 72} (2005) 014906.
  
\bibitem{Specht:2007ez}
  H.~J.~Specht,
  Nucl.\ Phys.\  A {\bf 805} (2008) 338.
  
\bibitem{Ruppert:2007cr}
  J.~Ruppert, C.~Gale, T.~Renk, P.~Lichard and J.~I.~Kapusta,
  Phys.\ Rev.\ Lett.\  {\bf 100} (2008) 162301.

\bibitem{vanHees:2007th}
  H.~van Hees and R.~Rapp,
  Nucl.\ Phys.\  A {\bf 806} (2008) 339.
  
\bibitem{Turbide:2006mc}
  S.~Turbide, C.~Gale, D.~K.~Srivastava and R.~J.~Fries,
  Phys.\ Rev.\  C {\bf 74} (2006) 014903.
  
\bibitem{Martinez:2008di}
  M.~Martinez and M.~Strickland,
  Phys.\ Rev.\  C {\bf 78} (2008) 034917.
  
\bibitem{Martinez:2008mc}
  M.~Martinez and M.~Strickland,
  arXiv:0808.3969 [hep-ph].
   
\bibitem{Bhattacharya:2008up}
  L.~Bhattacharya and P.~Roy,
  arXiv:0809.4596 [hep-ph].
  
\bibitem{Nayak:2007xv}
  J.~K.~Nayak, J.~Alam, S.~Sarkar and B.~Sinha,
  Phys.\ Rev.\  C {\bf 78} (2008) 034903.
  
  \bibitem{eic}
The Electron Ion Collider Working Group Collaboration, C.~Aidala {\em
  et~al.}, {\it {A High Luminosity, High Energy Electron Ion Collider}},
  \texttt{http://web.mit.edu/eicc/}.

\bibitem{lhec} M.~Klein {\em et.~al.}, {\it {Prospects for a Large
      Hadron Electron Collider (LHeC) at the LHC}}, EPAC'08, 11th
  European Particle Accelerator Conference, 23- 27 June 2008, Genoa,
  Italy.
  
\bibitem{Armesto:2006ph}
  N.~Armesto,
  J.\ Phys.\ G {\bf 32} (2006) R367.
  
\bibitem{Eskola:2009uj}
  K.~J.~Eskola, H.~Paukkunen and C.~A.~Salgado,
  arXiv:0902.4154 [hep-ph].
  
\bibitem{Arleo:2007js}
  F.~Arleo and T.~Gousset,
  Phys.\ Lett.\  B {\bf 660} (2008) 181.
  
\bibitem{Guzey:2004zp}
  V.~Guzey, M.~Strikman and W.~Vogelsang,
  Phys.\ Lett.\  B {\bf 603} (2004) 173.
  
\bibitem{Kopeliovich:2002yh}
  B.~Z.~Kopeliovich, J.~Nemchik, A.~Schafer and A.~V.~Tarasov,
  Phys.\ Rev.\ Lett.\  {\bf 88} (2002) 232303.
  
\bibitem{Kharzeev:2004yx}
  D.~Kharzeev, Y.~V.~Kovchegov and K.~Tuchin,
  Phys.\ Lett.\  B {\bf 599} (2004) 23.
  
\bibitem{JalilianMarian:2004er}
  J.~Jalilian-Marian,
  Nucl.\ Phys.\  A {\bf 739} (2004) 319.
  
\bibitem{Betemps:2004xr}
  M.~A.~Betemps and M.~B.~Gay Ducati,
  Phys.\ Rev.\  D {\bf 70} (2004) 116005.
  
\bibitem{Betemps:2008yw}
  M.~A.~Betemps and V.~P.~Goncalves,
  JHEP {\bf 0809} (2008) 019.

\bibitem{Iancu:2004bx}
  E.~Iancu, K.~Itakura and D.~N.~Triantafyllopoulos,
  Nucl.\ Phys.\  A {\bf 742} (2004) 182.
  
\bibitem{Kovner:2005pe}
  A.~Kovner,
  Acta Phys.\ Polon.\  B {\bf 36} (2005) 3551.

\bibitem{Triantafyllopoulos:2005cn}
  D.~N.~Triantafyllopoulos,
  Acta Phys.\ Polon.\  B {\bf 36} (2005) 3593.

\bibitem{Kozlov:2006qw}
  M.~Kozlov, A.~I.~Shoshi and B.~W.~Xiao,
  Nucl.\ Phys.\  A {\bf 792} (2007) 170.

\end{thebibliography}
\end{document}